\documentclass[11pt]{article}

\usepackage{amsfonts,amssymb,chet,dsfont,graphicx,mathrsfs,pgfplots,tikz}
\allowdisplaybreaks

\newcommand{\1}{\mathds{1}}

\newcommand{\Op}[2]{\mathcal{O}_{#1}(\eta_{#2})}

\newcommand{\ee}[3]{(\eta_{#1}\cdot\eta_{#2})^{#3}}

\newcommand{\D}{\mathcal{D}}

\newcommand{\cOPE}[4]{{}_{#1}c_{#2#3}^{\phantom{#2#3}#4}}
\newcommand{\DOPE}[4]{{}_{#1}\D_{#2#3}^{\phantom{#2#3}#4}}
\newcommand{\tOPE}[6]{{}_{#1}t_{#2#3}^{#5#6#4}}

\title{Six-Point Conformal Blocks\\in the Snowflake Channel}

\author{Jean-Fran\c{c}ois Fortin$^{\ast,}$\email{jean-francois.fortin@phy.ulaval.ca}, Wen-Jie Ma$^{\ast,}$\email{wenjie.ma.1@ulaval.ca} and Witold Skiba$^{\dagger,}$\email{witold.skiba@yale.edu}}

\affiliation{
$^\ast$D\'epartement de Physique, de G\'enie Physique et d'Optique\\Universit\'e Laval, Qu\'ebec, QC G1V 0A6, Canada\\
$^\dagger$Department of Physics, Yale University, New Haven, CT 06520, USA
}%Choices for affiliations $^{\ast,\dagger,\$,\S,\ddag,}$

\abstract{We compute $d$-dimensional scalar six-point conformal blocks in the two possible topologies allowed by the operator product expansion.  Our computation is a simple application of the embedding space operator product expansion formalism developed recently.  Scalar six-point conformal blocks in the comb channel have been determined not long ago, and we present here the first explicit computation of the scalar six-point conformal blocks in the remaining inequivalent topology.  For obvious reason, we dub the other topology the snowflake channel.  The scalar conformal blocks, with scalar external and exchange operators, are presented as a power series expansion in the conformal cross-ratios, where the coefficients of the power series are given as a double sum of the hypergeometric type.  In the comb channel, the double sum is expressible as a product of two ${}_3F_2$-hypergeometric functions.  In the snowflake channel, the double sum is expressible as a Kamp\'e de F\'eriet function where both sums are intertwined and cannot be factorized.  We check our results by verifying their consistency under symmetries and by taking several limits reducing to known results, mostly to scalar five-point conformal blocks in arbitrary spacetime dimensions.}

\date{September 2020} %Uncomment this line for month to be fixed

\begin{document}

\maketitle

%\toc

%%%%%%%%%%%%%%%%%%%%%%%%%%%%%%%%%%%%%%%%%%%%%%%%%%
%%%%%%%%%%%%%%%%%%%%%%%%%%%%%%%%%%%%%%%%%%%%%%%%%%

\section{Introduction}\label{SecIntro}

The study of higher-point conformal blocks in conformal field theory (CFT) is a complicated subject without many explicit results.  In a CFT, correlation functions, which are the natural observables of the theory, are given in terms of the CFT data and the conformal blocks.  The CFT data, which consist of the spectrum of quasi-primary operators as well as the operator product expansion (OPE) coefficients, completely determine all correlation functions with up to three points.  For higher-point correlation functions, the appearance of conformal cross-ratios, which are invariant under conformal transformations, leads to conformal blocks.  The conformal blocks are functions of the conformal cross-ratios which are in principle fully constrained by conformal invariance.

Although conformal blocks are fixed by conformal invariance, they are notoriously difficult to compute in all generality.  Several techniques have been developed over the years for the computation of four-point conformal blocks, which are the simplest blocks.  For example, various methods use Casimir equations \cite{Dolan:2003hv,Dolan:2011dv,Kravchuk:2017dzd}, the shadow formalism \cite{Ferrara:1972xe,Ferrara:1972uq,SimmonsDuffin:2012uy}, the weight-shifting formalism \cite{Karateev:2017jgd,Costa:2018mcg}, integrability \cite{Isachenkov:2016gim,Schomerus:2016epl,Schomerus:2017eny,Isachenkov:2017qgn,Buric:2019dfk}, 
AdS/CFT \cite{Hijano:2015zsa,Nishida:2016vds,Castro:2017hpx,Dyer:2017zef,Chen:2017yia,Sleight:2017fpc}, and the OPE \cite{Ferrara:1971vh,Ferrara:1971zy,Ferrara:1972cq,Ferrara:1973eg,Ferrara:1973vz,Ferrara:1974nf,Dolan:2000ut,Fortin:2016lmf,Fortin:2016dlj,Comeau:2019xco,Fortin:2019fvx,Fortin:2019dnq,Fortin:2019xyr,Fortin:2019pep,Fortin:2019gck,Fortin:2020ncr}.  Another important reason why four-point conformal blocks have been studied extensively is the conformal bootstrap \cite{Ferrara:1973yt,Polyakov:1974gs}, a way of constraining the CFT data solely from consistency of correlation functions under associativity.  Indeed, it is known that four-point conformal blocks are sufficient to implement the full conformal bootstrap.

Conformal blocks with more than four points have not been studied in great detail as of now.  Until very recently, the only results were for scalar $M$-point blocks in one and two spacetime dimensions as well as scalar five-point blocks in any spacetime dimensions \cite{Alkalaev:2015fbw,Rosenhaus:2018zqn,Goncalves:2019znr,Parikh:2019ygo,Jepsen:2019svc}.  Last year, the scalar $M$-point conformal blocks in the so-called comb channel were presented in \cite{Parikh:2019dvm,Fortin:2019zkm}.  They showed that the scalar $M$-point conformal blocks in the comb channel can be expressed as a power series expansion in the conformal cross-ratios with the coefficients containing a product of $M-4$ ${}_3F_2$-hypergeometric functions.  Although the techniques employed in the two references were different, AdS/CFT versus OPE, the two results have very similar forms even though the basis of conformal cross-ratios were distinct.

An interesting feature of higher-point correlation functions is that there exist several inequivalent topologies.  Indeed, starting at six points, the use of OPE among pairs of operators in different sequences of operator pairings leads to different topologies, see Figure \ref{Fig6pt}.  The comb channel \cite{Rosenhaus:2018zqn} is only one of the many topologies that are possible for higher-point correlation functions.

One possible advantage of the OPE formalism used in \cite{Fortin:2019zkm} is that it is not limited to the comb channel.  Indeed, it can be applied to any pair of operators in the $M$-point correlation function in any channel of interest.  The net effect of the OPE is ``adding" an operator to an $(M-1)$-point correlation function in a specific place of choice.  This allows generation of possible multiple topologies with $M$ points from one specific diagram with $M-1$ points.  After all, the existence of the many topologies is a consequence of the OPE, as was directly observed above.  More importantly, the choice in the pairs of quasi-primary operators does not lead to any new computational complications.

With the OPE formalism \cite{Fortin:2016lmf,Fortin:2019fvx,Fortin:2019dnq}, the OPE differential operator is applied on the known $(M-1)$-point correlation functions to generate $M$-point correlation functions.  Since the action of the OPE differential operator on arbitrary products of conformal cross-ratios has been determined \cite{Fortin:2019fvx,Fortin:2019dnq}, it is intuitively straightforward to proceed with the computation.  It can however be technically laborious to express the final results in the most convenient way possible.  Indeed, starting at five points, it is necessary to re-express the conformal cross-ratios of the original correlation function in terms of the conformal cross-ratios appropriate for the OPE differential operator, leading to several superfluous sums.  Re-summations must be performed to recover a simple result for the final correlation function.  For the computation of the scalar $M$-point conformal blocks, we found that, with an appropriate basis of conformal cross-ratios, all re-summations were easy ${}_2F_1$-hypergeometric function re-summations \cite{Fortin:2019zkm}.  For other channels, the necessary computations within the OPE formalism are exactly the same at the technical level, hence it should be straightforward to compute higher-point correlation functions in all topologies using the OPE formalism.

In this paper, we begin the investigation of higher-point correlation functions in all topologies by computing the scalar six-point conformal blocks in the remaining channel, see the bottom part of Figure \ref{Fig6pt}, which for obvious reason, we call the snowflake channel.\footnote{In \cite{Jepsen:2019svc,Parikh:2019dvm}, the snowflake channel is called the OPE channel.}  We show that the superfluous sums can be taken care of with simple ${}_2F_1$-hypergeometric function re-summations (as for the comb channel) and some identities for ${}_3F_2$-hypergeometric functions (absent in the comb channel).

This paper is organized as follows: Section \ref{SecCB} presents a summary of the embedding space OPE formalism.  The OPE is reviewed and the action of the scalar OPE differential operator is discussed.  A general form for the contribution to the correlation functions of external scalars with scalar exchanges is presented, as well as the associated scalar conformal blocks.  Moreover, the recurrence relation taking $(M-1)$-point correlation functions to $M$-point correlation functions is introduced, and, after re-considering scalar $M$-point conformal blocks in the comb channel, the scalar six-point conformal block in the snowflake channel is given.  Section \ref{SecChecks} provides several consistency checks of the snowflake result.  First, interesting symmetry properties of the scalar conformal blocks in the snowflake channel are proven.  Then, the OPE limit and the limit of unit operator are used to ascertain that the snowflake result passes those tests.  Finally, we conclude in Section \ref{SecConc} while several appendices present most of the more technical proofs.  Appendix \ref{SAppCB5} contains derivations of the scalar five-point conformal blocks of \cite{Parikh:2019dvm} directly from the OPE, showing that the results of \cite{Parikh:2019dvm} and \cite{Fortin:2019zkm} are equivalent for five points.  Appendix \ref{SAppCB6} demonstrates how the snowflake result is obtained.  An equivalent result is also presented and the proof of their equivalence is shown.  Appendix \ref{SAppSym} proves the identities that the scalar six-point conformal blocks in the snowflake channel verify under the symmetry group of the snowflake.

%%%%%%%%%%%%%%%%%%%%%%%%%%%%%%%%%%%%%%%%%%%%%%%%%%
%%%%%%%%%%%%%%%%%%%%%%%%%%%%%%%%%%%%%%%%%%%%%%%%%%

\section{Scalar Six-Point Conformal Blocks}\label{SecCB}

With the knowledge of the OPE and its explicit action on any function of conformal cross-ratios, one can in principle compute any correlation function starting from the known two-point functions or, for that matter, the only non-trivial one-point function.  With this technique, starting at five points and above, it is necessary to re-express the initial correlation function in terms of the conformal cross-ratios appropriate for the OPE differential operator, and then rewrite the solution in terms of the most convenient conformal cross-ratios by re-summing as many superfluous sums as possible.  This method is quite powerful, allowing the computation of conformal blocks in any channel.  Although straightforward, it is however not always clear \textit{a priori} what is the best choice of conformal cross-ratios that will lead to the simplest final answer.  In this section, we quickly review the OPE and then sketch the derivation and state the result for the scalar six-point conformal blocks in both channels.  Concrete proofs are left for the appendices.

%%%%%%%%%%%%%%%%%%%%%%%%%%%%%%%%%%%%%%%%%%%%%%%%%%

\subsection{\texorpdfstring{$M$}{M}-Point Correlation Functions from the OPE}

The embedding space OPE introduced in \cite{Fortin:2019fvx,Fortin:2019dnq} states that the product of two quasi-primary operators can be expressed as
\eqn{
\begin{gathered}
\Op{i}{1}\Op{j}{2}=\sum_k\sum_{a=1}^{N_{ijk}}\cOPE{a}{i}{j}{k}\DOPE{a}{i}{j}{k}(\eta_1,\eta_2)\Op{k}{2},\\
\DOPE{a}{i}{j}{k}(\eta_1,\eta_2)=\frac{1}{\ee{1}{2}{p_{ijk}}}(\mathcal{T}_{12}^{\boldsymbol{N}_i}\Gamma)(\mathcal{T}_{21}^{\boldsymbol{N}_j}\Gamma)\cdot\tOPE{a}{i}{j}{k}{1}{2}\cdot\D_{12}^{(d,h_{ijk}-n_a/2,n_a)}(\mathcal{T}_{12\boldsymbol{N}_k}\Gamma)*,\\
p_{ijk}=\frac{1}{2}(\tau_i+\tau_j-\tau_k),\qquad h_{ijk}=-\frac{1}{2}(\chi_i-\chi_j+\chi_k),\\
\tau_{\mathcal{O}}=\Delta_{\mathcal{O}}-S_{\mathcal{O}},\qquad\chi_{\mathcal{O}}=\Delta_{\mathcal{O}}-\xi_{\mathcal{O}},\qquad\xi_{\mathcal{O}}=S_{\mathcal{O}}-\lfloor S_{\mathcal{O}}\rfloor,
\end{gathered}
}[EqOPE]
where the OPE differential operator and the remaining quantities (half-projectors, tensor structures, \textit{etc.}) are introduced and detailed in \cite{Fortin:2019dnq}.  Since all the calculations involve only scalar operators, the OPE can be simplified significantly as the tensor structures and half-projectors are trivial in that case.  Considering an OPE of two scalar operators and neglecting any operators 
with spin on the right-hand side leads to
\eqn{
\begin{gathered}
\Op{i}{1}\Op{j}{2}=\cOPE{}{i}{j}{k}\frac{1}{\ee{1}{2}{p_{ijk}}}\D_{12}^{(d,h_{ijk},0)}\Op{k}{2}+\cdots,\\
p_{ijk}=\frac{1}{2}(\Delta_i+\Delta_j-\Delta_k),\qquad h_{ijk}=-\frac{1}{2}(\Delta_i-\Delta_j+\Delta_k),
\end{gathered}
}[EqOPE-scalar]
where $\Delta_i$ is the scaling dimension of $\mathcal{O}_i$ and $\cOPE{}{i}{j}{k}$ is the OPE coefficient.  The coordinates $\eta_i$ are the embedding space coordinates in $d+2$ dimensions that are constrained to the light cone $\eta_i\cdot\eta_i=0$ and are projectively identified $\eta_i\sim\lambda\eta_i$ for $\lambda>0$.

For scalar operators another simplification occurs in the differential operator that has a particularly simple form when the Lorentz indices are absent, indicated by $0$ in the third argument
\eqn{\D_{ij}^{(d,h,0)}=\left[\D_{ij}^2\right]^h\equiv\left[\ee{i}{j}{} \partial_j^2-(d+2\,\eta_j\cdot\partial_j)\eta_i\cdot\partial_j\right]^h,}
where $\partial_{jA}=\frac{\partial}{\partial\eta_j^A}$.  It is more practical to rescale the scalar differential operator, $\D_{ij}^2$, and introduce an operator that is homogeneous of degree $0$ with respect to all the coordinates
\eqn{\bar{\D}_{ij;kl;m}^2=\frac{\eta_{ij}\eta_{kl}}{\eta_{ik}\eta_{il}}\D_{ij}^2,}
with $\eta_{ij}\equiv\eta_i\cdot\eta_j$ for brevity.  The action of $\bar{\D}_{ij;kl;m}^2$ on the conformal cross-ratios
\eqn{
\begin{gathered}
x_m=\frac{\eta_{ij}\eta_{kl}\eta_{im}}{\eta_{ik}\eta_{il}\eta_{jm}},\\
y_a=1-\frac{\eta_{im}\eta_{ja}}{\eta_{ia}\eta_{jm}},\qquad1\leq a\leq M,\qquad a\neq i,j,m,\\
z_{ab}=\frac{\eta_{ik}\eta_{il}\eta_{ab}}{\eta_{kl}\eta_{ia}\eta_{ib}},\qquad1\leq a<b\leq M,\qquad a,b\neq i,j,
\end{gathered}
}[EqCROPE]
has been obtained explicitly\footnote{We note that the OPE differential operator used here is a simple rescaling of the one defined in \cite{Fortin:2019dnq}.}
\eqna{
&\bar{\D}_{ij;kl;m}^{2h}x_m^{\bar{q}}\prod_{\substack{1\leq a\leq M\\a\neq i,j,m}}(1-y_a)^{-q_a}\\
&\qquad=x_m^{\bar{q}+h}\sum_{\{n_a,n_{am},n_{ab}\}\geq0}\frac{(-h)_{\bar{n}_m+\bar{\bar{n}}}(q_m)_{\bar{n}_m}(\bar{q}+h)_{\bar{n}-\bar{\bar{n}}}}{(\bar{q})_{\bar{n}+\bar{n}_m}(\bar{q}+1-d/2)_{\bar{n}_m+\bar{\bar{n}}}}\\
&\qquad\phantom{=}\qquad\times\prod_{\substack{1\leq a\leq M\\a\neq i,j,m}}\frac{(q_a)_{n_a}}{n_{am}!(n_a-n_{am}-\bar{n}_a)!}y_a^{n_a}\left(\frac{x_mz_{am}}{y_a}\right)^{n_{am}}\prod_{\substack{1\leq a<b\leq M\\a,b\neq i,j,m}}\frac{1}{n_{ab}!}\left(\frac{x_mz_{ab}}{y_ay_b}\right)^{n_{ab}},
}[EqD]
where we have defined $\bar{q}=\sum_{\substack{1\leq a\leq M\\a\neq i,j}}q_a$ as well as
\eqn{
\begin{gathered}
\bar{n}=\sum_{\substack{1\leq a\leq M\\a\neq i,j,m}}n_a,\qquad\qquad\bar{n}_m=\sum_{\substack{1\leq a\leq M\\a\neq i,j,m}}n_{am},\\
\bar{n}_a=\sum_{\substack{1\leq b\leq M\\b\neq i,j,m,a}}n_{ab},\qquad\qquad\bar{\bar{n}}=\sum_{\substack{1\leq a<b\leq M\\a,b\neq i,j,m}}n_{ab}.
\end{gathered}
}

The OPE \eqref{EqOPE} was used to determine two-, three-, and four-points in \cite{Fortin:2019xyr,Fortin:2019pep,Fortin:2019gck}.  Very specific rules, somewhat reminiscent of Feynman rules, were also developed in \cite{Fortin:2020ncr} to compute all necessary ingredients to implement the full conformal bootstrap at the level of four-point correlation functions.

%%%%%%%%%%%%%%%%%%%%%%%%%%%%%%%%%%%%%%%%%%%%%%%%%%

\subsection{Scalar \texorpdfstring{$M$}{M}-Point Correlation Functions}

In general, it is possible to write the contribution to scalar $M$-point correlation functions, \textit{i.e.} with scalar external and exchanged quasi-primary operators, from a specific channel as
\eqna{
\left.I_{M(\Delta_{k_1},\ldots,\Delta_{k_{M-3}})}^{(\Delta_{i_2},\ldots,\Delta_{i_M},\Delta_{i_1})}\right|_{\text{channel}}&=L_{M|\text{channel}}^{(\Delta_{i_2},\ldots,\Delta_{i_M},\Delta_{i_1})}\left[\prod_{1\leq a\leq M-3}(u_a^M)^{\frac{\Delta_{k_a}}{2}}\right]G_{M|\text{channel}}^{(d,\boldsymbol{h};\boldsymbol{p})}(\boldsymbol{u}^M,\textbf{v}^M).
}[EqI]
The scalar $M$-point conformal block is given by
\eqna{
G_{M|\text{channel}}^{(d,\boldsymbol{h};\boldsymbol{p})}(\boldsymbol{u}^M,\textbf{v}^M)&=\sum_{\{m_a,m_{ab}\}\geq0}C_{M|\text{channel}}^{(d,\boldsymbol{h};\boldsymbol{p})}(\boldsymbol{m},\textbf{m})F_{M|\text{channel}}^{(d,\boldsymbol{h};\boldsymbol{p})}(\boldsymbol{m},\textbf{m})\\
&\phantom{=}\qquad\times\prod_{1\leq a\leq M-3}\frac{(u_a^M)^{m_a}}{m_a!}\prod_{1\leq a\leq b\leq M-3}\frac{(1-v_{ab}^M)^{m_{ab}}}{m_{ab}!},
}[EqG]
where the cross-ratios $u_a$ and $v_{ab}$ are defined below in \eqref{EqCBCRcomb}.

The complete $M$-point correlation functions are sums of the different $I_M$, including exchanges of operators in non-trivial representations which are not discussed here.

In \eqref{EqI} the conformal dimensions of the external scalar quasi-primary operators are $\Delta_{i_a}$ while the conformal dimensions of the exchanged scalar quasi-primary operators are $\Delta_{k_a}$.  Moreover, the legs $L_M$ are products of embedding space coordinates necessary to satisfy covariance under scale transformations while the conformal cross-ratios are denoted by the vector $\boldsymbol{u}^M$ of $u_a^M$ and the matrix $\textbf{v}^M$ of $v_{ab}^M$.  The scalar $M$-point conformal blocks \eqref{EqG} are written as sums over powers of conformal cross-ratios, with extra sums denoted by the function $F_M$.\footnote{In general, $F_M$ could be a function of both the vector $\boldsymbol{m}=(m_a)$ and the matrix $\textbf{m}=(m_{ab})$.  Thus the separation between $C_M$ and $F_M$ is somewhat arbitrary in \eqref{EqG}.  Here, $F_6$ in the snowflake channel is only a function of the vector $\boldsymbol{m}$, as is the case for $F_M$ in the comb channel.  That statement seems to generalize to all $F_M$, thus we conjecture that $F_M$ can always be chosen such that it is a function of the vector $\boldsymbol{m}$ only.  Moreover, we construct $F_{6|\text{snowflake}}$ such that it has some interesting symmetry properties.}

Finally, in the scalar $M$-point conformal blocks \eqref{EqG} the vectors $\boldsymbol{h}$ and $\boldsymbol{p}$ are generated by the action of the OPE \eqref{EqOPE} on the pairs of quasi-primary operators relevant to the channel of interest.  This statement translates into
\eqn{\left.I_{M(\Delta_{k_1},\ldots,\Delta_{k_{M-3}})}^{(\Delta_{i_2},\ldots,\Delta_{i_M},\Delta_{i_1})}\right|_{\text{channel}}=\frac{1}{\eta_{1M}^{p_M}}\left(\frac{\eta_{kM}\eta_{lM}}{\eta_{1M}\eta_{kl}}\right)^{h_{M-1}}\bar{\D}_{M1;kl;m}^{2h_{M-1}}\left.I_{M-1(\Delta_{k_1},\ldots,\Delta_{k_{M-4}})}^{(\Delta_{i_2},\ldots,\Delta_{i_{M-1}},\Delta_{k_{M-3}})}\right|_{\text{channel}},}[EqIfromOPE]
for some convenient choice of $k$, $l$ and $m$ with the appropriate channel on the RHS to generate the desired channel on the LHS.  For future convenience, we also define $\bar{p}_a=\sum_{b=2}^ap_b$ and $\bar{h}_a=\sum_{b=2}^ah_b$.  The above equation is the essence of the OPE approach to computing correlation functions.  Two operators in an $M$-point function are replaced by one operator appearing on the right-hand side of the OPE.  This reduces the $M$-point function to an $(M-1)$-point function.  Reading the equation in the other direction, the differential operator present in the OPE generates the expression for the $M$-point function when it acts on a previously computed $(M-1)$-point function.

In this context, the OPE \eqref{EqOPE} was also put to work in the computation of scalar $M$-point correlation functions in the comb channel in \cite{Fortin:2019zkm}, to which we now turn.

%%%%%%%%%%%%%%%%%%%%%%%%%%%%%%%%%%%%%%%%%%%%%%%%%%

\subsection{Scalar \texorpdfstring{$M$}{M}-Point Correlation Functions in the Comb Channel}

Acting repetitively with the OPE as in \eqref{EqIfromOPE}, we found in \cite{Fortin:2019zkm} the following scalar $M$-point conformal blocks \eqref{EqI} and \eqref{EqG} in the comb channel,
\eqn{
\begin{gathered}
L_{M|\text{comb}}^{(\Delta_{i_2},\ldots,\Delta_{i_M},\Delta_{i_1})}=\left(\frac{\eta_{34}}{\eta_{23}\eta_{24}}\right)^{\frac{\Delta_{i_2}}{2}}\left[\prod_{1\leq a\leq M-2}\left(\frac{\eta_{a+1,a+3}}{\eta_{a+1,a+2}\eta_{a+2,a+3}}\right)^{\frac{\Delta_{i_{a+2}}}{2}}\right]\left(\frac{\eta_{M-1,M}}{\eta_{1,M-1}\eta_{1M}}\right)^{\frac{\Delta_{i_1}}{2}},\\
u_a^M=\frac{\eta_{1+a,2+a}\eta_{3+a,4+a}}{\eta_{1+a,3+a}\eta_{2+a,4+a}},\qquad v_{ab}^M=\frac{\eta_{2-a+b,4+b}}{\eta_{2+b,4+b}}\prod_{1\leq c\leq a}\frac{\eta_{3+b-c,4+b-c}}{\eta_{2+b-c,4+b-c}},
\end{gathered}
}[EqCBCRcomb]
with
\eqn{
\begin{gathered}
C_{M|\text{comb}}^{(d,\boldsymbol{h};\boldsymbol{p})}=\frac{(p_3)_{m_1+\text{tr}_0\textbf{m}}(p_2+h_2)_{m_1+\text{tr}_1\textbf{m}}}{(p_3)_{m_1+\text{tr}_1\textbf{m}}}\left[\prod_{1\leq a\leq M-3}\frac{(\bar{p}_{a+2}+\bar{h}_{a+2})_{m_a+m_{a+1}+\bar{m}_a+\bar{\bar{m}}_a}}{(\bar{p}_{a+2}+\bar{h}_{a+1})_{2m_a+\bar{m}_{a-1}+\bar{m}_a+\bar{\bar{m}}_a}}\right.\\
\phantom{=}\qquad\left.\times(p_{a+2}-m_{a-1})_{m_a+\text{tr}_a\textbf{m}}\frac{(-h_{a+2})_{m_a}(-h_{a+2}+m_a-m_{a+1})_{\bar{m}_{a-1}}}{(\bar{p}_{a+2}+\bar{h}_{a+1}+1-d/2)_{m_a}}\right],\\
\text{tr}_a\textbf{m}=\sum_bm_{b,a+b},\qquad\qquad\bar{m}_a=\sum_{b\leq a}m_{ba},\qquad\qquad\bar{\bar{m}}_a=\sum_{b>a}(\bar{m}_b-\text{tr}_b\textbf{m}),
\end{gathered}
}[EqCBCcomb]
and
\eqn{F_{M|\text{comb}}^{(d,\boldsymbol{h};\boldsymbol{p})}(\boldsymbol{m})=\prod_{1\leq a\leq M-4}{}_3F_2\left[\begin{array}{c}-m_a,-m_{a+1},-\bar{p}_{a+2}-\bar{h}_{a+1}+d/2-m_a\\p_{a+3}-m_a,h_{a+2}+1-m_a\end{array};1\right],}[EqCBFcomb]
and finally
\eqn{
\begin{gathered}
2h_2=\Delta_{k_1}-\Delta_{i_2}-\Delta_{i_3},\qquad2h_a=\Delta_{k_{a-1}}-\Delta_{k_{a-2}}-\Delta_{i_{a+1}},\\
p_2=\Delta_{i_3},\qquad2p_3=\Delta_{i_2}+\Delta_{k_1}-\Delta_{i_3},\qquad2p_a=\Delta_{i_a}+\Delta_{k_{a-2}}-\Delta_{k_{a-3}},
\end{gathered}
}[EqCBhpcomb]
where $k_{M-2}\equiv i_1$.

The computations were straightforward yet somewhat tedious.  The choice of conformal cross-ratios was based on the OPE limit, a limit we use again later to check the validity of the scalar six-point conformal blocks in the snowflake channel discussed in the next subsection.

%%%%%%%%%%%%%%%%%%%%%%%%%%%%%%%%%%%%%%%%%%%%%%%%%%

\subsection{Scalar Six-Point Correlation Functions in the Snowflake Channel}

As already mentioned, the OPE \eqref{EqOPE} or \eqref{EqOPE-scalar} can be applied on any pair of quasi-primary operators in the $M$-point correlation function, allowing the $M$-point correlation function to be expressed in terms of the derivative operator acting on the $(M-1)$-point function in the corresponding channel.  Hence, with the knowledge of the only five-point correlation function (that is in the comb channel), appropriately choosing the pair of quasi-primary operators on which the OPE acts leads to six-point correlation function in all channels.  The desired channel follows from the choice of the OPE pair in the six-point function.

To reach the scalar six-point conformal blocks in the snowflake channel, we start from the four-point correlation functions and use the OPE appropriately.  For $M=4$, there is only one channel with the topology of the comb, as shown in Figure \ref{Fig4pt}.
\begin{figure}[t]
\centering
\resizebox{10cm}{!}{%
\begin{tikzpicture}[thick]
\begin{scope}
\node at (-2.2,0) {$\left.I_{4(\Delta_{k_1})}^{(\Delta_{i_2},\Delta_{i_3},\Delta_{i_4},\Delta_{i_1})}\right|_{\text{comb}}$};
\node at (0,0) {$=$};
\node at (1,0) {$\mathcal{O}_{i_2}$};
\draw[-] (1.5,0)--(4.5,0);
\node at (5.1,0) {$\mathcal{O}_{i_1}$};
\draw[-] (2.5,0)--(2.5,1) node[above]{$\mathcal{O}_{i_3}$};
\draw[-] (3.5,0)--(3.5,1) node[above]{$\mathcal{O}_{i_4}$};
\node at (3,-0.5) {$\mathcal{O}_{k_1}$};
\end{scope}
\end{tikzpicture}
}
\caption{Scalar four-point conformal blocks.}
\label{Fig4pt}
\end{figure}
One possible form for the contribution to scalar four-point correlation functions \eqref{EqI} and for scalar four-point conformal blocks \eqref{EqG} leads to \cite{Ferrara:1974nf,Dolan:2000ut,Dolan:2003hv,Dolan:2011dv}
\eqn{
\begin{gathered}
L_{4|\text{comb}}^{(\Delta_{i_2},\ldots,\Delta_{i_4},\Delta_{i_1})}=\left(\frac{\eta_{13}}{\eta_{12}\eta_{23}}\right)^{\frac{\Delta_{i_2}}{2}}\left(\frac{\eta_{12}}{\eta_{13}\eta_{23}}\right)^{\frac{\Delta_{i_3}}{2}}\left(\frac{\eta_{13}}{\eta_{14}\eta_{34}}\right)^{\frac{\Delta_{i_4}}{2}}\left(\frac{\eta_{34}}{\eta_{13}\eta_{14}}\right)^{\frac{\Delta_{i_1}}{2}},\\
u_1^4=\frac{\eta_{14}\eta_{23}}{\eta_{12}\eta_{34}},\qquad v_{11}^4=\frac{\eta_{13}\eta_{24}}{\eta_{12}\eta_{34}},\\
C_{4|\text{comb}}^{(d,\boldsymbol{h};\boldsymbol{p})}=\frac{(-h_2)_{m_1+m_{11}}(p_2+h_3)_{m_1}(-h_3)_{m_1+m_{11}}(p_2+h_2)_{m_1}}{(p_2)_{2m_1+m_{11}}(p_2+1-d/2)_{m_1}},\\
F_{4|\text{comb}}^{(d,\boldsymbol{h};\boldsymbol{p})}=1,
\end{gathered}
}[EqCB4]
\eqn{
\begin{gathered}
2h_2=\Delta_{i_3}-\Delta_{i_2}-\Delta_{k_1},\qquad2h_3=\Delta_{i_1}-\Delta_{i_4}-\Delta_{k_1},\\
p_2=\Delta_{k_1},\qquad2p_3=\Delta_{i_2}+\Delta_{i_3}-\Delta_{k_1},\qquad2p_4=\Delta_{i_4}+\Delta_{i_1}-\Delta_{k_1}.
\end{gathered}
}

It is clear from \eqref{EqIfromOPE} and Figure \ref{Fig4pt} that there is only one channel for five-point conformal blocks since all external quasi-primary operators in four-point conformal blocks are topologically equivalent.  This channel has the topology of the comb and is shown in Figure \ref{Fig5pt}.
\begin{figure}[t]
\centering
\resizebox{11cm}{!}{%
\begin{tikzpicture}[thick]
\begin{scope}
\node at (-2.2,0) {$\left.I_{5(\Delta_{k_1},\Delta_{k_2})}^{(\Delta_{i_3},\Delta_{i_4},\Delta_{i_2},\Delta_{i_5},\Delta_{i_1})}\right|_{\text{comb}}$};
\node at (0,0) {$=$};
\node at (1,0) {$\mathcal{O}_{i_3}$};
\draw[-] (1.5,0)--(5.5,0);
\node at (6.1,0) {$\mathcal{O}_{i_1}$};
\draw[-] (2.5,0)--(2.5,1) node[above]{$\mathcal{O}_{i_4}$};
\draw[-] (3.5,0)--(3.5,1) node[above]{$\mathcal{O}_{i_2}$};
\draw[-] (4.5,0)--(4.5,1) node[above]{$\mathcal{O}_{i_5}$};
\node at (3,-0.5) {$\mathcal{O}_{k_1}$};
\node at (4,-0.5) {$\mathcal{O}_{k_2}$};
\end{scope}
\end{tikzpicture}
}
\caption{Scalar five-point conformal blocks.}
\label{Fig5pt}
\end{figure}
It can be obtained from \eqref{EqIfromOPE} with $k=3$, $l=4$ and $m=4$, using the four-point correlation functions in the comb channel \eqref{EqCB4} where we first shifted $\mathcal{O}_{i_a}(\eta_a)\to\mathcal{O}_{i_{a-1}}(\eta_{a-1})$ with $\mathcal{O}_{i_0}(\eta_0)\equiv\mathcal{O}_{i_4}(\eta_4)$.  The scalar five-point conformal blocks can be expressed as in \eqref{EqI} and \eqref{EqG} with the help of
\eqn{
\begin{gathered}
L_{5|\text{comb}}^{(\Delta_{i_3},\Delta_{i_4},\Delta_{i_2},\Delta_{i_5},\Delta_{i_1})}=\left(\frac{\eta_{14}}{\eta_{12}\eta_{24}}\right)^{\frac{\Delta_{i_2}}{2}}\left(\frac{\eta_{24}}{\eta_{23}\eta_{34}}\right)^{\frac{\Delta_{i_3}}{2}}\left(\frac{\eta_{23}}{\eta_{24}\eta_{34}}\right)^{\frac{\Delta_{i_4}}{2}}\left(\frac{\eta_{14}}{\eta_{15}\eta_{45}}\right)^{\frac{\Delta_{i_5}}{2}}\left(\frac{\eta_{45}}{\eta_{14}\eta_{15}}\right)^{\frac{\Delta_{i_1}}{2}},\\
u_1^5=\frac{\eta_{12}\eta_{34}}{\eta_{14}\eta_{23}},\qquad u_2^5=\frac{\eta_{15}\eta_{24}}{\eta_{12}\eta_{45}},\qquad v_{11}^5=\frac{\eta_{13}\eta_{24}}{\eta_{14}\eta_{23}},\qquad v_{12}^5=\frac{\eta_{14}\eta_{25}}{\eta_{12}\eta_{45}},\qquad v_{22}^5=\frac{\eta_{24}\eta_{35}}{\eta_{23}\eta_{45}},\\
C_{5|\text{comb}}^{(d,\boldsymbol{h};\boldsymbol{p})}=(-h_2)_{m_1+m_2+m_{11}+m_{22}}\frac{(-h_3)_{m_1+m_{11}+m_{22}}(p_2+h_2)_{m_1}(p_2+h_3)_{m_1}}{(p_2)_{2m_1+m_{11}+m_{22}}(p_2+1-d/2)_{m_1}}\\
\qquad\qquad\times\frac{(-h_4)_{m_2+m_{12}+m_{22}}(p_3-h_2+h_4)_{m_2+m_{11}}(p_3-m_1)_{m_2+m_{12}}}{(p_3-h_2)_{2m_2+m_{11}+m_{12}+m_{22}}(p_3-h_2+1-d/2)_{m_2}},\\
F_{5|\text{comb}}^{(d,\boldsymbol{h};\boldsymbol{p})}={}_3F_2\left[\begin{array}{cc}-m_1,-m_2,-p_2+d/2-m_1\\1-p_2-h_2-m_1,p_3-m_1\end{array};1\right],\\
2h_2=\Delta_{i_2}-\Delta_{k_2}-\Delta_{k_1},\qquad2h_3=\Delta_{i_4}-\Delta_{i_3}-\Delta_{k_1},\qquad2h_4=\Delta_{i_1}-\Delta_{i_5}-\Delta_{k_2},\\
p_2=\Delta_{k_1},\qquad2p_3=\Delta_{k_2}+\Delta_{i_2}-\Delta_{k_1},\qquad2p_4=\Delta_{i_3}+\Delta_{i_4}-\Delta_{k_1},\qquad2p_5=\Delta_{i_5}+\Delta_{i_1}-\Delta_{k_2}.
\end{gathered}
}[EqCB5]
We note here that the result \eqref{EqCB5} for the scalar five-point conformal blocks is equivalent to but different than the results \eqref{EqCBCRcomb}, \eqref{EqCBCcomb}, \eqref{EqCBFcomb}, and \eqref{EqCBhpcomb} obtained in \cite{Fortin:2019zkm}.\footnote{It is also different than the results of \cite{Rosenhaus:2018zqn} and \cite{Parikh:2019dvm}.}

For $M=6$, there is now two possible topologies that can be obtained from five-point correlation functions.  Indeed, from Figure \ref{Fig5pt}, it is clear that the external quasi-primary operator $\mathcal{O}_{i_2}$ is different.  Transforming the external quasi-primary operators $\mathcal{O}_{i_1}$, $\mathcal{O}_{i_3}$, $\mathcal{O}_{i_4}$, or $\mathcal{O}_{i_5}$ into exchanged quasi-primary operators by appending two new external quasi-primary operators leads to the scalar six-point correlation function in the comb channel (see previous subsection).  Doing the same with the quasi-primary operator $\mathcal{O}_{i_2}$ gives instead the scalar six-point correlation in the snowflake channel.  The difference can be seen in Figure \ref{Fig6pt}.
\begin{figure}[t]
\centering
\resizebox{12cm}{!}{%
\begin{tikzpicture}[thick]
\begin{scope}
\node at (-2.2,0) {$\left.I_{6(\Delta_{k_1},\Delta_{k_2},\Delta_{k_3})}^{(\Delta_{i_2},\ldots,\Delta_{i_6},\Delta_{i_1})}\right|_{\text{comb}}$};
\node at (0,0) {$=$};
\node at (1,0) {$\mathcal{O}_{i_2}$};
\draw[-] (1.5,0)--(6.5,0);
\node at (7.1,0) {$\mathcal{O}_{i_1}$};
\draw[-] (2.5,0)--(2.5,1) node[above]{$\mathcal{O}_{i_3}$};
\draw[-] (3.5,0)--(3.5,1) node[above]{$\mathcal{O}_{i_4}$};
\draw[-] (4.5,0)--(4.5,1) node[above]{$\mathcal{O}_{i_5}$};
\draw[-] (5.5,0)--(5.5,1) node[above]{$\mathcal{O}_{i_6}$};
\node at (3,-0.5) {$\mathcal{O}_{k_1}$};
\node at (4,-0.5) {$\mathcal{O}_{k_2}$};
\node at (5,-0.5) {$\mathcal{O}_{k_3}$};
\end{scope}
\begin{scope}[yshift=-4cm]
\node at (-2.4,0) {$\left.I_{6(\Delta_{k_1},\Delta_{k_2},\Delta_{k_3})}^{(\Delta_{i_2},\ldots,\Delta_{i_6},\Delta_{i_1})}\right|_{\text{snowflake}}$};
\node at (0,0) {$=$};
\draw[-] (4,0)--+(-150:1) node[pos=0.6,above]{$\mathcal{O}_{k_1}$};
\draw[-] (4,0)++(-150:1)--+(-90:1) node[below]{$\mathcal{O}_{i_2}$};
\draw[-] (4,0)++(-150:1)--+(150:1) node[left]{$\mathcal{O}_{i_3}$};
\draw[-] (4,0)--+(90:1) node[pos=0.5,right]{$\mathcal{O}_{k_2}$};
\draw[-] (4,0)++(90:1)--+(30:1) node[above]{$\mathcal{O}_{i_5}$};
\draw[-] (4,0)++(90:1)--+(150:1) node[above]{$\mathcal{O}_{i_4}$};
\draw[-] (4,0)--+(-30:1) node[pos=0.4,below]{$\mathcal{O}_{k_3}$};;
\draw[-] (4,0)++(-30:1)--+(30:1) node[right]{$\mathcal{O}_{i_6}$};
\draw[-] (4,0)++(-30:1)--+(-90:1) node[below]{$\mathcal{O}_{i_1}$};
\end{scope}
\end{tikzpicture}
}
\caption{Scalar six-point conformal blocks in the comb (top) and snowflake (bottom) channels.}
\label{Fig6pt}
\end{figure}
For the scalar six-point correlation functions in the snowflake channel, we start from the scalar five-point conformal blocks \eqref{EqCB5} and shift the quasi-primary operators such that $\mathcal{O}_{i_a}(\eta_a)\to\mathcal{O}_{i_{a-1}}(\eta_{a-1})$ with the understanding that $\mathcal{O}_{i_0}(\eta_0)\equiv\mathcal{O}_{i_5}(\eta_5)$, and then use \eqref{EqIfromOPE} with $k=4$, $l=5$ and $m=5$, to get
\eqn{
\begin{gathered}
L_{6|\text{snowflake}}^{(\Delta_{i_2},\ldots,\Delta_{i_6},\Delta_{i_1})}=\left(\frac{\eta_{13}}{\eta_{12}\eta_{23}}\right)^{\frac{\Delta_{i_2}}{2}}\left(\frac{\eta_{12}}{\eta_{13}\eta_{23}}\right)^{\frac{\Delta_{i_3}}{2}}\left(\frac{\eta_{35}}{\eta_{34}\eta_{45}}\right)^{\frac{\Delta_{i_4}}{2}}\\
\qquad\qquad\qquad\qquad\qquad\qquad\times\left(\frac{\eta_{34}}{\eta_{35}\eta_{45}}\right)^{\frac{\Delta_{i_5}}{2}}\left(\frac{\eta_{15}}{\eta_{16}\eta_{56}}\right)^{\frac{\Delta_{i_6}}{2}}\left(\frac{\eta_{56}}{\eta_{15}\eta_{16}}\right)^{\frac{\Delta_{i_1}}{2}},\\
u_1^6=\frac{\eta_{15}\eta_{23}}{\eta_{12}\eta_{35}},\qquad u_2^6=\frac{\eta_{13}\eta_{45}}{\eta_{15}\eta_{34}},\qquad u_3^6=\frac{\eta_{16}\eta_{35}}{\eta_{13}\eta_{56}},\\
v_{11}^6=\frac{\eta_{13}\eta_{25}}{\eta_{12}\eta_{35}},\qquad v_{12}^6=\frac{\eta_{14}\eta_{35}}{\eta_{15}\eta_{34}},\qquad v_{22}^6=\frac{\eta_{13}\eta_{24}}{\eta_{12}\eta_{34}},\\
v_{13}^6=\frac{\eta_{15}\eta_{26}}{\eta_{12}\eta_{56}},\qquad v_{23}^6=\frac{\eta_{15}\eta_{36}}{\eta_{13}\eta_{56}},\qquad v_{33}^6=\frac{\eta_{35}\eta_{46}}{\eta_{34}\eta_{56}},
\end{gathered}
}[EqCB6CR]
with
\eqna{
C_{6|\text{snowflake}}^{(d,\boldsymbol{h};\boldsymbol{p})}&=\frac{(p_2+h_3)_{m_1+m_{23}}(p_3)_{-m_1+m_2+m_3+m_{12}+m_{33}}(-h_3)_{m_1+m_{11}+m_{22}+m_{13}}}{(p_2)_{2m_1+m_{11}+m_{13}+m_{22}+m_{23}}(p_2+1-d/2)_{m_1}}\\
&\phantom{=}\qquad\times\frac{(p_3-h_2+h_4)_{m_2+m_{11}}(p_2+h_2)_{m_1-m_2+m_3+m_{13}+m_{23}}(-h_4)_{m_2+m_{12}+m_{22}+m_{33}}}{(p_3-h_2)_{2m_2+m_{11}+m_{12}+m_{22}+m_{33}}(p_3-h_2+1-d/2)_{m_2}}\\
&\phantom{=}\qquad\times\frac{(\bar{p}_3+h_2+h_5)_{m_3+m_{12}}(-h_2)_{m_1+m_2-m_3+m_{11}+m_{22}}(-h_5)_{m_3+m_{13}+m_{23}+m_{33}}}{(\bar{p}_3+h_2)_{2m_3+m_{12}+m_{13}+m_{23}+m_{33}}(\bar{p}_3+h_2+1-d/2)_{m_3}},
}[EqCB6C]
as well as
\eqna{
F_{6|\text{snowflake}}^{(d,\boldsymbol{h};\boldsymbol{p})}(\boldsymbol{m})&=\frac{(-p_3+h_2+d/2-m_2)_{m_2}}{(p_3)_{-m_1}(-h_2+m_1)_{-m_3}}\sum_{t_1,t_2\geq0}\frac{(-m_1)_{t_2}(-m_2)_{t_1}(-m_3)_{t_2}}{(p_3-h_2+1-d/2)_{t_1}}\\
&\phantom{=}\qquad\times\frac{(p_3)_{t_1}(\bar{p}_3-d/2)_{t_1}(-p_2+d/2-m_1)_{t_2}(1+h_2-m_1+m_3)_{-t_1}}{(1+h_2-m_1)_{-t_1+t_2}(p_3-m_1)_{t_1+t_2}t_1!t_2!}\\
&=\frac{(-p_3+h_2+d/2-m_2)_{m_2}}{(p_3)_{-m_1}(-h_2)_{-m_3}}\\
&\phantom{=}\qquad\times F_{2,1,0}^{1,3,2}\left[\left.\begin{array}{c}\bar{p}_3-d/2;-m_2,-h_2,p_3;-m_1,-m_3\\-h_2-m_3,p_3-m_1;p_3-h_2+1-d/2;-\end{array}\right|1,1\right],
}[EqCB6F]
and finally
\eqn{
\begin{gathered}
2h_2=\Delta_{k_3}-\Delta_{k_2}-\Delta_{k_1},\qquad2h_3=\Delta_{i_3}-\Delta_{i_2}-\Delta_{k_1},\\
2h_4=\Delta_{i_5}-\Delta_{i_4}-\Delta_{k_2},\qquad2h_5=\Delta_{i_1}-\Delta_{i_6}-\Delta_{k_3},\\
p_2=\Delta_{k_1},\qquad2p_3=\Delta_{k_2}+\Delta_{k_3}-\Delta_{k_1},\qquad2p_4=\Delta_{i_2}+\Delta_{i_3}-\Delta_{k_1},\\
2p_5=\Delta_{i_4}+\Delta_{i_5}-\Delta_{k_2},\qquad2p_6=\Delta_{i_6}+\Delta_{i_1}-\Delta_{k_3}.
\end{gathered}
}[EqCB6hp]
As mentioned previously, the function $F_6$ \eqref{EqCB6F} in the snowflake channel is also a double sum of the hypergeometric type, same as in the comb channel.  However, the snowflake double sum does not factorize into two hypergeometric functions, unlike the comb sum, compare \eqref{EqCBFcomb} with $M=6$.  It can however be written as a Kamp\'e de F\'eriet function $F_{2,1,0}^{1,3,2}$ as shown in Appendix \ref{SAppCB6} (another expression in terms of a different Kamp\'e de F\'eriet function $F_{1,1,1}^{2,1,1}$ is also given there).  See \eqref{EqKdF} and \cite{exton1976multiple,srivastava1985multiple} for its definition.

In the following section, we study the snowflake results \eqref{EqCB6CR}, \eqref{EqCB6C}, \eqref{EqCB6F}, and \eqref{EqCB6hp}.  We explicitly check the symmetry properties of $G_6$ and verify that it behaves properly under the OPE limit and the limit of unit operator.  The proof of the snowflake results as well as an alternative form for the snowflake are shown in Appendix \ref{SAppCB6}

%%%%%%%%%%%%%%%%%%%%%%%%%%%%%%%%%%%%%%%%%%%%%%%%%%
%%%%%%%%%%%%%%%%%%%%%%%%%%%%%%%%%%%%%%%%%%%%%%%%%%

\section{Sanity Checks}\label{SecChecks}

The scalar six-point conformal blocks obtained in the previous section must satisfy several properties.  This section investigates the identities of $G_6$ \eqref{EqG} in the snowflake channel from the symmetries of the associated snowflake diagram Figure \ref{Fig6pt}.  Then, the OPE limit and the limit of unit operator are taken to verify that the scalar six-point correlation functions reduce to the appropriate scalar five-point correlation functions.

%%%%%%%%%%%%%%%%%%%%%%%%%%%%%%%%%%%%%%%%%%%%%%%%%%

\subsection{Symmetry Properties}\label{SecSymmetry}

The scalar $M$-point conformal blocks must verify several identities where the conformal cross-ratios and the vectors $\boldsymbol{h}$ and $\boldsymbol{p}$ are transformed.  These identities are generated from the symmetries of the scalar $M$-point conformal blocks in the associated topology.  The symmetries of the scalar $M$-point conformal blocks in the comb channel are relatively trivial.
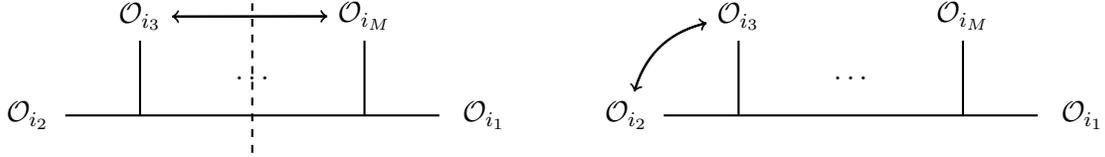
\begin{figure}[t]
\centering
\resizebox{15cm}{!}{%
\begin{tikzpicture}[thick]
\begin{scope}[xshift=0cm]
\node at (1,0) {$\mathcal{O}_{i_2}$};
\draw[-] (1.5,0)--(6.5,0);
\node at (7.1,0) {$\mathcal{O}_{i_1}$};
\draw[-] (2.5,0)--(2.5,1) node[above] (1) {$\mathcal{O}_{i_3}$};
\node at (4,0.5) {$\ldots$};
\draw[dashed] (4,-0.5)--(4,1.5);
\draw[-] (5.5,0)--(5.5,1) node[above] (2) {$\mathcal{O}_{i_M}$};
\path[->]
(1) edge node[left]{} (2)
(2) edge node[right]{} (1);
\end{scope}
\begin{scope}[xshift=8cm]
\node at (1,0) (1) {$\mathcal{O}_{i_2}$};
\draw[-] (1.5,0)--(6.5,0);
\node at (7.1,0) {$\mathcal{O}_{i_1}$};
\draw[-] (2.5,0)--(2.5,1) node[above] (2) {$\mathcal{O}_{i_3}$};
\node at (4,0.5) {$\ldots$};
\draw[-] (5.5,0)--(5.5,1) node[above] {$\mathcal{O}_{i_M}$};
\path[->]
(1) edge[bend left] node[right]{} (2)
(2) edge[bend right] node[left]{} (1);
\end{scope}
\end{tikzpicture}
}
\caption{Symmetries of the scalar $M$-point conformal blocks in the comb channel.  The figure shows the two generators, with reflections on the left and dendrite permutations on the right.}
\label{FigSymComb}
\end{figure}
They correspond to $(\mathbb{Z}_2)^2\rtimes\mathbb{Z}_2$, the semi-direct product of the direct product of two cyclic groups $\mathbb{Z}_2$ of order two (for OPE, or dendrite, permutations depicted on the right of Figure \ref{FigSymComb}) and the cyclic group $\mathbb{Z}_2$ of order two (for reflection shown in the left part of Figure \ref{FigSymComb}).  Here we present the identities of the scalar six-point conformal blocks in the snowflake channel.  The proofs are left to Appendix \ref{SAppSym}.

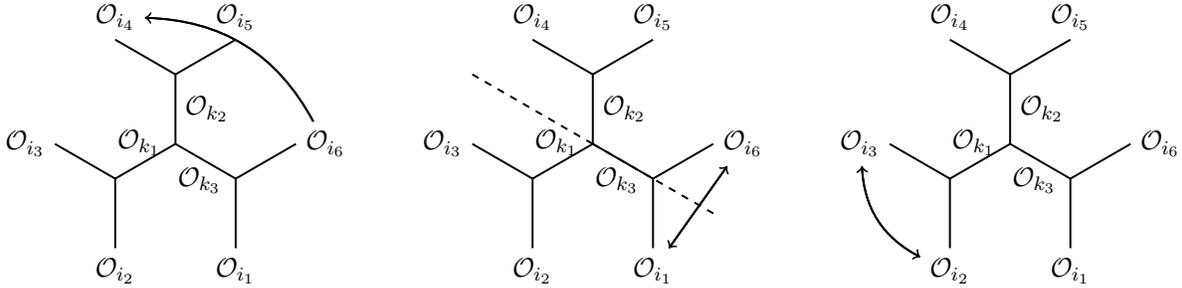
\begin{figure}[t]
\centering
\resizebox{16cm}{!}{%
\begin{tikzpicture}[thick]
\begin{scope}[xshift=0cm]
\draw[-] (4,0)--+(-150:1) node[pos=0.6,above]{$\mathcal{O}_{k_1}$};
\draw[-] (4,0)++(-150:1)--+(-90:1) node[below]{$\mathcal{O}_{i_2}$};
\draw[-] (4,0)++(-150:1)--+(150:1) node[left]{$\mathcal{O}_{i_3}$};
\draw[-] (4,0)--+(90:1) node[pos=0.5,right]{$\mathcal{O}_{k_2}$};
\draw[-] (4,0)++(90:1)--+(30:1) node[above]{$\mathcal{O}_{i_5}$};
\draw[-] (4,0)++(90:1)--+(150:1) node[above] (1) {$\mathcal{O}_{i_4}$};
\draw[-] (4,0)--+(-30:1) node[pos=0.4,below]{$\mathcal{O}_{k_3}$};;
\draw[-] (4,0)++(-30:1)--+(30:1) node[right] (2) {$\mathcal{O}_{i_6}$};
\draw[-] (4,0)++(-30:1)--+(-90:1) node[below]{$\mathcal{O}_{i_1}$};
\path[-]
(1) edge[bend left] node[above]{} (2)
(2) edge[->,bend right] node[right]{} (1);
\end{scope}
\begin{scope}[xshift=6cm]
\draw[-] (4,0)--+(-150:1) node[pos=0.6,above]{$\mathcal{O}_{k_1}$};
\draw[-] (4,0)++(-150:1)--+(-90:1) node[below]{$\mathcal{O}_{i_2}$};
\draw[-] (4,0)++(-150:1)--+(150:1) node[left]{$\mathcal{O}_{i_3}$};
\draw[-] (4,0)--+(90:1) node[pos=0.5,right]{$\mathcal{O}_{k_2}$};
\draw[-] (4,0)++(90:1)--+(30:1) node[above]{$\mathcal{O}_{i_5}$};
\draw[-] (4,0)++(90:1)--+(150:1) node[above]{$\mathcal{O}_{i_4}$};
\draw[-] (4,0)--+(-30:1) node[pos=0.4,below]{$\mathcal{O}_{k_3}$};;
\draw[-] (4,0)++(-30:1)--+(30:1) node[right] (1) {$\mathcal{O}_{i_6}$};
\draw[-] (4,0)++(-30:1)--+(-90:1) node[below] (2) {$\mathcal{O}_{i_1}$};
\draw[dashed] (4,0)--+(-30:2);
\draw[dashed] (4,0)--+(150:2);
\path[->]
(1) edge node[left]{} (2)
(2) edge node[right]{} (1);
\end{scope}
\begin{scope}[xshift=12cm]
\draw[-] (4,0)--+(-150:1) node[pos=0.6,above]{$\mathcal{O}_{k_1}$};
\draw[-] (4,0)++(-150:1)--+(-90:1) node[below] (1) {$\mathcal{O}_{i_2}$};
\draw[-] (4,0)++(-150:1)--+(150:1) node[left] (2) {$\mathcal{O}_{i_3}$};
\draw[-] (4,0)--+(90:1) node[pos=0.5,right]{$\mathcal{O}_{k_2}$};
\draw[-] (4,0)++(90:1)--+(30:1) node[above]{$\mathcal{O}_{i_5}$};
\draw[-] (4,0)++(90:1)--+(150:1) node[above]{$\mathcal{O}_{i_4}$};
\draw[-] (4,0)--+(-30:1) node[pos=0.4,below]{$\mathcal{O}_{k_3}$};;
\draw[-] (4,0)++(-30:1)--+(30:1) node[right]{$\mathcal{O}_{i_6}$};
\draw[-] (4,0)++(-30:1)--+(-90:1) node[below]{$\mathcal{O}_{i_1}$};
\path[->]
(1) edge[bend left] node[right]{} (2)
(2) edge[bend right] node[left]{} (1);
\end{scope}
\end{tikzpicture}
}
\caption{Symmetries of the scalar six-point conformal blocks in the snowflake channel.  The figure shows rotations by $2\pi/3$ (left), reflections (middle), and dendrite permutations (right).}
\label{FigSymSnowflake}
\end{figure}

The snowflake diagram Figure \ref{Fig6pt} is invariant under the symmetry group generated by the three transformations shown in Figure \ref{FigSymSnowflake}.
Since the scalar six-point correlation functions in the snowflake channel $I_6$ are the same under symmetry transformations generated by the rotations, reflections and permutations described in Figure \ref{FigSymSnowflake}, there are $47$ identities that the scalar six-point conformal blocks $G_6$ should satisfy.

Although we dubbed it the snowflake channel, at first glance the symmetry group generated by rotations and reflections is only the dihedral group of order six, $D_3$, which is the symmetry group of the triangle, not the hexagon expected for snowflakes.  Including the OPE permutations, \textit{i.e.} the permutations of the dendrites (or arms) of the snowflake, the full symmetry group of the snowflake diagram is however given by $(\mathbb{Z}_2)^3\rtimes D_3$ where each cyclic group of order two corresponds to dendrite permutations.  Since the order of this symmetry group is $|(\mathbb{Z}_2)^3\rtimes D_3|=48$, the snowflake diagram has a larger symmetry group than the hexagon, contrary to expectations.

Before proceeding, we verify that the transformations depicted in Figure~\ref{FigSymSnowflake} indeed generate $(\mathbb{Z}_2)^3\rtimes D_3$, the semi-direct product of $(\mathbb{Z}_2)^3$ (for dendrite permutations) and the dihedral group of order six.  By defining the action of the symmetry generators on the external quasi-primary operators as $R$ for the rotation (left diagram in Figure~\ref{FigSymSnowflake}), $S$ for the reflection (center), and $P$ for the permutation (right), it is easy to see that the dihedral part of symmetry group of the snowflake diagram has for presentation
\eqn{\langle r,s|r^3=s^2=(rs)^2=1\rangle,}[EqD3]
with $r=R^{-1}$ and $s=S$.  The presentation \eqref{EqD3} corresponds to $D_3$, with $r$ and $s$ representing rotations by $2\pi/3$ and reflections with respect to one of the three different axes, respectively.  The $(\mathbb{Z}_2)^3=\mathbb{Z}_2\times\mathbb{Z}_2\times\mathbb{Z}_2$ part of the symmetry group is generated by $p_1=P$, $p_2=R^{-1}PR$, and $p_3=RPR^{-1}$ with $p_i^2=1$ for all $i$.  It is trivial to check that the $p_i$'s commute and that they correspond to dendrite permutations.  To exclude the direct nature of the product, it suffices to observe that the generators $p_i$ do not commute with the generators $r$ and $s$.  Having excluded $(\mathbb{Z}_2)^3\times D_3$, it is easy to verify that the snowflake diagram has the symmetry group $(\mathbb{Z}_2)^3\rtimes D_3$ of order $48$.\footnote{The homomorphism from $D_3\simeq S_3$ to the automorphism group of $(\mathbb{Z}_2)^3=\mathbb{Z}_2\times\mathbb{Z}_2\times\mathbb{Z}_2$, $Aut[(\mathbb{Z}_2)^3]=S_3$, associated to the semi-direct nature of the product is simply given by $S_3$ permutations of the three $\mathbb{Z}_2$ factors.}

For rotations, we choose the generator where $\mathcal{O}_{i_a}(\eta_a)\to\mathcal{O}_{i_{a+2}}(\eta_{a+2})$ and $\Delta_{k_a}\to\Delta_{k_{a+1}}$ with $\mathcal{O}_{i_7}(\eta_7)\equiv\mathcal{O}_{i_1}(\eta_1)$, $\mathcal{O}_{i_8}(\eta_8)\equiv\mathcal{O}_{i_2}(\eta_2)$ as well as $\Delta_{k_4}\equiv\Delta_{k_1}$.  Under this transformation, the legs and conformal cross-ratios \eqref{EqCB6CR} transform as
\eqn{
\begin{gathered}
L_6\prod_{1\leq a\leq3}(u_a^6)^{\frac{\Delta_{k_a}}{2}}\to L_6\prod_{1\leq a\leq3}(u_a^6)^{\frac{\Delta_{k_a}}{2}},\\
u_1^6\to u_2^6,\qquad u_2^6\to u_3^6,\qquad u_3^6\to u_1^6,\\
v_{11}^6\to v_{12}^6,\qquad v_{12}^6\to v_{23}^6,\qquad v_{13}^6\to v_{22}^6,\\
v_{22}^6\to v_{33}^6,\qquad v_{23}^6\to v_{11}^6,\qquad v_{33}^6\to v_{13}^6,
\end{gathered}
}
which imply the following identity,
\eqna{
&G_{6|\text{snowflake}}^{(d,h_2,h_3,h_4,h_5;p_2,p_3,p_4,p_5,p_6)}(u_1^6,u_2^6,u_3^6;v_{11}^6,v_{12}^6,v_{13}^6,v_{22}^6,v_{23}^6,v_{33}^6)\\
&\qquad=G_{6|\text{snowflake}}^{(d,-p_3,h_4,h_5,h_3;p_3-h_2,p_2+h_2,p_5,p_6,p_4)}(u_2^6,u_3^6,u_1^6;v_{12}^6,v_{23}^6,v_{22}^6,v_{33}^6,v_{11}^6,v_{13}^6).
}[EqSymRot]
Using the decomposition \eqref{EqG} with \eqref{EqCB6C} and \eqref{EqCB6F}, it is easy to see that $C_6$ does not change under this rotation generator, resulting in a non-trivial identity for $F_6$ [see \eqref{EqSymRotF}].  This identity can be translated into the language of Kamp\'e de F\'eriet functions as discussed in the conclusion.

For reflections, we start with the generator acting as 
\eqn{\mathcal{O}_{i_2}(\eta_2)\leftrightarrow\mathcal{O}_{i_4}(\eta_4),\qquad\mathcal{O}_{i_3}(\eta_3)\leftrightarrow\mathcal{O}_{i_5}(\eta_5),\qquad\Delta_{k_1}\leftrightarrow\Delta_{k_2}.}
For this reflection generator, the legs and conformal cross-ratios \eqref{EqCB6CR} transform as
\eqn{
\begin{gathered}
L_6\prod_{1\leq a\leq3}(u_a^6)^{\frac{\Delta_{k_a}}{2}}\to(v_{11}^6)^{h_3}(v_{12}^6)^{h_4}(v_{23}^6)^{h_5}L_6\prod_{1\leq a\leq3}(u_a^6)^{\frac{\Delta_{k_a}}{2}},\\
u_1^6\to\frac{u_2^6}{v_{12}^6},\qquad u_2^6\to\frac{u_1^6}{v_{11}^6},\qquad u_3^6\to\frac{u_3^6}{v_{23}^6},\\
v_{11}^6\to\frac{1}{v_{12}^6},\qquad v_{12}^6\to\frac{1}{v_{11}^6},\qquad v_{13}^6\to\frac{v_{33}^6}{v_{12}^6v_{23}^6},\\
v_{22}^6\to\frac{v_{22}^6}{v_{11}^6v_{12}^6},\qquad v_{23}^6\to\frac{1}{v_{23}^6},\qquad v_{33}^6\to\frac{v_{13}^6}{v_{11}^6v_{23}^6},
\end{gathered}
}
and that observation translates into the identity
\eqna{
&G_{6|\text{snowflake}}^{(d,h_2,h_3,h_4,h_5;p_2,p_3,p_4,p_5,p_6)}(u_1^6,u_2^6,u_3^6;v_{11}^6,v_{12}^6,v_{13}^6,v_{22}^6,v_{23}^6,v_{33}^6)\\
&\qquad=(v_{11}^6)^{h_3}(v_{12}^6)^{h_4}(v_{23}^6)^{h_5}\\
&\qquad\phantom{=}\qquad\times G_{6|\text{snowflake}}^{(d,h_2,h_4,h_3,h_5;p_3-h_2,p_2+h_2,p_5,p_4,p_6)}\left(\frac{u_2^6}{v_{12}^6},\frac{u_1^6}{v_{11}^6},\frac{u_3^6}{v_{23}^6};\frac{1}{v_{12}^6},\frac{1}{v_{11}^6},\frac{v_{33}^6}{v_{12}^6v_{23}^6},\frac{v_{22}^6}{v_{11}^6v_{12}^6},\frac{1}{v_{23}^6},\frac{v_{13}^6}{v_{11}^6v_{23}^6}\right).
}[EqSymRef]
Again, from the decomposition \eqref{EqG} with \eqref{EqCB6C} and \eqref{EqCB6F}, we remark that $F_6$ in the form \eqref{EqF6} does not change under this reflection generator, implying an identity for $C_6$.

Finally, the generator $\mathcal{O}_{i_2}(\eta_2)\leftrightarrow\mathcal{O}_{i_3}(\eta_3)$ for dendrite permutations lead to
\eqn{
\begin{gathered}
L_6\prod_{1\leq a\leq3}(u_a^6)^{\frac{\Delta_{k_a}}{2}}\to(v_{11}^6)^{h_2-h_4}(v_{22}^6)^{h_4}L_6\prod_{1\leq a\leq3}(u_a^6)^{\frac{\Delta_{k_a}}{2}},\\
u_1^6\to\frac{u_1^6}{v_{11}^6},\qquad u_2^6\to\frac{u_2^6}{v_{22}^6},\qquad u_3^6\to u_3^6v_{11}^6,\\
v_{11}^6\to\frac{1}{v_{11}^6},\qquad v_{12}^6\to\frac{v_{11}^6v_{12}^6}{v_{22}^6},\qquad v_{13}^6\to v_{23}^6,\\
v_{22}^6\to\frac{1}{v_{22}^6},\qquad v_{23}^6\to v_{13}^6,\qquad v_{33}^6\to\frac{v_{11}^6v_{33}^6}{v_{22}^6},
\end{gathered}
}
for the legs and conformal cross-ratios \eqref{EqCB6CR}.  Thus, the corresponding identity is
\eqna{
&G_{6|\text{snowflake}}^{(d,h_2,h_3,h_4,h_5;p_2,p_3,p_4,p_5,p_6)}(u_1^6,u_2^6,u_3^6;v_{11}^6,v_{12}^6,v_{13}^6,v_{22}^6,v_{23}^6,v_{33}^6)\\
&\qquad=(v_{11}^6)^{h_2-h_4}(v_{22}^6)^{h_4}\\
&\qquad\phantom{=}\qquad\times G_{6|\text{snowflake}}^{(d,h_2,-p_2-h_3,h_4,h_5;p_2,p_3,p_4,p_5,p_6)}\left(\frac{u_1^6}{v_{11}^6},\frac{u_2^6}{v_{22}^6},u_3^6v_{11}^6;\frac{1}{v_{11}^6},\frac{v_{11}^6v_{12}^6}{v_{22}^6},v_{23}^6,\frac{1}{v_{22}^6},v_{13}^6,\frac{v_{11}^6v_{33}^6}{v_{22}^6}\right).
}[EqSymPerm]
Once again, the decomposition \eqref{EqG} with \eqref{EqCB6C} and \eqref{EqCB6F} shows that $F_6$ \eqref{EqCB6F} is invariant under this generator for dendrite permutations, resulting in a second identity for $C_6$.

Therefore, the three symmetry transformations of Figure \ref{FigSymSnowflake} generate the symmetry group $(\mathbb{Z}_2)^3\rtimes D_3$ of order $48$ (see Appendix \ref{SAppSym}).  Each generator has an associated identity for the scalar six-point conformal blocks \eqref{EqG} in the snowflake channel, with \eqref{EqSymRot}, \eqref{EqSymRef}, and \eqref{EqSymPerm} being the identities for rotations, reflections, and dendrite permutations, respectively.  From these three identities, it is straightforward to generate the remaining $45$ identities of the snowflake symmetry group by composition.  The proofs that our explicit solution \eqref{EqCB6C} and \eqref{EqCB6F} satisfies these symmetry transformations can be found in Appendix \ref{SAppSym}.

We now turn to the OPE limit and the limit of unit operator.  Since we have already demonstrated that the scalar six-point conformal blocks in the snowflake channel obey several identities originating from the symmetry group of the snowflake diagram, it is only necessary to check the two limits once, all the other cases are equivalent by symmetry.

%%%%%%%%%%%%%%%%%%%%%%%%%%%%%%%%%%%%%%%%%%%%%%%%%%

\subsection{OPE Limit}

The OPE limit is defined as having two embedding space coordinates coincide.  The two embedding space coordinates must correspond to an OPE in the associated topology.  In this limit, the original $M$-point correlation function reduces to the proper $(M-1)$-point correlation function with a pre-factor originating from the OPE \eqref{EqOPE}, as dictated by \eqref{EqIfromOPE} for the scalar case.

For scalar six-point correlation functions in the snowflake channel depicted in Figure \ref{Fig6pt}, the possible OPE limits are $\eta_{2}\to\eta_{3}$, $\eta_{4}\to\eta_{5}$, and $\eta_{6}\to\eta_{1}$.  However, since $I_6$ is invariant under rotations as discussed above, it is only necessary to assess the behavior of $I_6$ in one OPE limit.  Here, we check that in the limit $\eta_{2}\to\eta_{3}$, we have
\eqn{\left.I_{6(\Delta_{k_1},\Delta_{k_2},\Delta_{k_3})}^{(\Delta_{i_2},\Delta_{i_3},\Delta_{i_4},\Delta_{i_5},\Delta_{i_6},\Delta_{i_1})}\right|_{\text{snowflake}}\to\frac{1}{\eta_{23}^{\frac{1}{2}(\Delta_2+\Delta_3-\Delta_{k_1})}}\left.I_{5(\Delta_{k_2},\Delta_{k_3})}^{(\Delta_{i_5},\Delta_{i_4},\Delta_{i_3},\Delta_{i_6},\Delta_{i_1})}\right|_{\text{comb}}.}[EqOPELim]

For this proof, we start from the alternative form of the scalar six-point correlation functions in the snowflake channel given in Appendix \ref{SAppCB6}, which must lead in the OPE limit \eqref{EqOPELim} to the scalar five-point correlation functions in the comb channel of \cite{Parikh:2019dvm} discussed in Appendix \ref{SAppCB5}.  This choice is of no consequence since we prove that these are equal to the results of Section \ref{SecCB} in Appendices \ref{SAppCB6} and \ref{SAppCB5}, respectively [see \eqref{EqCB6CR}, \eqref{EqCB6C}, \eqref{EqCB6F}, and \eqref{EqCB6hp} for scalar six-point correlation functions in the snowflake channel and \eqref{EqCB5} for scalar five-point correlation functions in the comb channel].

In the OPE limit \eqref{EqOPELim}, we have
\eqn{
\begin{gathered}
L_6^*\prod_{1\leq a\leq3}(u_a^{*6})^{\frac{\Delta_{k_a}}{2}}\to\frac{1}{(\eta_{23})^{\frac{1}{2}(\Delta_2+\Delta_3-\Delta_{k_1})}}(v_{34}^P)^{h_5}L_5^P\prod_{1\leq a\leq2}(u_a^P)^{\frac{\Delta_{k_a}}{2}},\\
u_1^{*6}\to0,\qquad u_2^{*6}\to u_1^P,\qquad u_3^{*6}\to\frac{u_2^P}{v_{34}^P},\\
v_{1a}^{*6}\to1\qquad(1\leq a\leq 3),\\
v_{22}^{*6}\to v_{23}^P,\qquad v_{23}^{*6}\to\frac{v_{24}^P}{v_{34}^P},\qquad v_{33}^{*6}\to\frac{1}{v_{34}^P},
\end{gathered}
}
where the conformal dimensions on the RHS are the ones relevant for the five-point correlation functions, \textit{i.e.}
\eqn{
\begin{gathered}
p_3-h_2+h_4\to p_3,\qquad p_3\to\bar{p}_3+\bar{h}_3,\qquad p_2+h_2\to p_4,\qquad h_5\to h_4,\qquad h_2\to h_3,\\
\bar{p}_3+\bar{h}_2+h_5\to\bar{p}_4+\bar{h}_4,\qquad\bar{p}_3+h_2\to\bar{p}_4+\bar{h}_3,\qquad h_4\to-p_2-h_2,\qquad p_3-h_2\to\bar{p}_3+h_2.
\end{gathered}
}
Thus, the OPE limit \eqref{EqOPELim} corresponds to the identity
\eqn{(v_{34}^P)^{h_5}G_6^*=G_5^P,}
with the appropriate changes for the vectors $\boldsymbol{h}$ and $\boldsymbol{p}$.

Since
\eqn{F_6^*=F_6\to\frac{(p_2+h_2-m_2)_{m_2}}{(-h_2)_{-m_3}}{}_3F_{2}\left[\begin{array}{c}-m_2,-m_3,\bar{p}_3-d/2\\p_2+h_2-m_2,-h_2-m_3\end{array};1\right]=\frac{(p_2+h_2-m_2)_{m_2}}{(-h_2)_{-m_3}}F_5^P,}
in the OPE limit \eqref{EqOPELim}, we have
\eqna{
(v_{34}^P)^{h_5}G_6^*&=\sum\frac{(p_3)_{m_2+m_3+m_{23}+m_{33}}(-h_2-m_3)_{m_2+m_{22}}(-h_5)_{m_3+m_{23}+m_{33}}(\bar{p}_3+h_2+h_5)_{m_3}}{(\bar{p}_3+h_2)_{2m_3+m_{23}+m_{33}}(\bar{p}_3+h_2+1-d/2)_{m_3}}\\
&\phantom{=}\qquad\times\frac{(p_2+h_2-m_2)_{m_3}(-h_4)_{m_2+m_{22}+m_{23}}(p_3-h_2+h_4)_{m_2+m_{33}}}{(p_3-h_2)_{2m_2+m_{22}+m_{23}+m_{33}}(p_3-h_2+1-d/2)_{m_2}}\binom{m_{23}}{k_{23}}\binom{m_{33}}{k_{33}}\\
&\phantom{=}\qquad\times\binom{k_{23}}{m^{\prime}_{24}}\binom{h_5-m_3-k_{23}-k_{33}}{m^{\prime}_{34}}(-1)^{k_{23}+k_{33}+m^{\prime}_{24}+m^{\prime}_{34}}\\
&\phantom{=}\qquad\times\frac{(u_1^P)^{m_2}}{m_2!}\frac{(u_2^P)^{m_3}}{m_3!}\frac{(1-v_{23}^P)^{m_{22}}}{m_{22}!}\frac{(1-v_{24}^P)^{m^{\prime}_{24}}}{m_{23}!}\frac{(1-v_{34}^P)^{m^{\prime}_{34}}}{m_{33}!}F_5^P,
}
after expanding in the proper conformal cross-ratios $u_a^P$ and $(1-v_{ab}^P)$ of \cite{Parikh:2019dvm}.  Here the vectors $\boldsymbol{h}$ and $\boldsymbol{p}$ are still the original ones.  Thus to complete the proof, we need to evaluate the extra sums for which we did not explicitly write the indices of summation (to make the notation less cluttered) with the help of \eqref{Eq2F1}, and express the vectors $\boldsymbol{h}$ and $\boldsymbol{p}$ in terms of the five-point ones.

First, we sum over $k_{22}$ and then $k_{23}$ after changing the variable by $k_{23}\to k_{23}+m^{\prime}_{24}$, which lead to
\eqna{
(v_{34}^P)^{h_5}G_6^*&=\sum\frac{(p_3)_{m_2+m_3+m_{23}+m_{33}}(-h_2-m_3)_{m_2+m_{22}}(-h_5)_{m_3+m^{\prime}_{24}+m^{\prime}_{34}}(\bar{p}_3+h_2+h_5)_{m_3}}{(\bar{p}_3+h_2)_{2m_3+m_{23}+m_{33}}(\bar{p}_3+h_2+1-d/2)_{m_3}}\\
&\phantom{=}\qquad\times\frac{(p_2+h_2-m_2)_{m_3}(-h_4)_{m_2+m_{22}+m_{23}}(p_3-h_2+h_4)_{m_2+m_{33}}}{(p_3-h_2)_{2m_2+m_{22}+m_{23}+m_{33}}(p_3-h_2+1-d/2)_{m_2}}\frac{(-m^{\prime}_{34})_{m_{23}+m_{33}-m^{\prime}_{24}}}{m_{33}!(m_{23}-m^{\prime}_{24})!}\\
&\phantom{=}\qquad\times\frac{(u_1^P)^{m_2}}{m_2!}\frac{(u_2^P)^{m_3}}{m_3!}\frac{(1-v_{23}^P)^{m_{22}}}{m_{22}!}\frac{(1-v_{24}^P)^{m^{\prime}_{24}}}{m^{\prime}_{24}!}\frac{(1-v_{34}^P)^{m^{\prime}_{34}}}{m^{\prime}_{34}!}F_5^P.
}
We then shift $m_{23}$ by $m_{23}\to m_{23}+m^{\prime}_{24}$, and then redefine $m_{33}=m-m_{23}$.  With these changes, we can compute the sums over $m_{23}$ and then $m$, which give
\eqna{
(v_{34}^P)^{h_5}G_6^*&=\sum\frac{(p_3)_{m_2+m_3+m^{\prime}_{24}}(-h_2)_{m_2-m_3+m_{22}}(-h_5)_{m_3+m^{\prime}_{24}+m^{\prime}_{34}}(\bar{p}_3+h_2+h_5)_{m_3}}{(-h_2)_{-m_3}(\bar{p}_3+h_2)_{2m_3+m^{\prime}_{24}+m^{\prime}_{34}}(\bar{p}_3+h_2+1-d/2)_{m_3}}\\
&\phantom{=}\qquad\times\frac{(p_2+h_2)_{-m_2+m_3+m^{\prime}_{34}}(-h_4)_{m_2+m_{22}+m^{\prime}_{24}}(p_3-h_2+h_4)_{m_2}}{(p_2+h_2)_{-m_2}(p_3-h_2)_{2m_2+m_{22}+m^{\prime}_{24}}(p_3-h_2+1-d/2)_{m_2}}\\
&\phantom{=}\qquad\times\frac{(u_1^P)^{m_2}}{m_2!}\frac{(u_2^P)^{m_3}}{m_3!}\frac{(1-v_{23}^P)^{m_{22}}}{m_{22}!}\frac{(1-v_{24}^P)^{m^{\prime}_{24}}}{m^{\prime}_{24}!}\frac{(1-v_{34}^P)^{m^{\prime}_{34}}}{m^{\prime}_{34}!}F_5^P.
}
Finally, changing the vectors $\boldsymbol{h}$ and $\boldsymbol{p}$ by their five-point counterparts and renaming $m_2\to m_1$, $m_3\to m_1$, $m_{22}\to m_{23}$, $m^{\prime}_{24}\to m_{24}$, and $m^{\prime}_{34}\to m_{34}$, we have
\eqn{F_5^P={}_3F_{2}\left[\begin{array}{c}-m_2,-m_3,\bar{p}_3-d/2\\p_2+h_2-m_2,-h_2-m_3\end{array};1\right]\to{}_3F_{2}\left[\begin{array}{c}-m_1,-m_2,\bar{p}_4+h_2-d/2\\p_4-m_1,-h_3-m_2\end{array};1\right],}
as in \eqref{EqCB5P} (hence its name), which indeed proves that $(v_{34}^P)^{h_5}G_6^*=G_5^P$, as expected.

%%%%%%%%%%%%%%%%%%%%%%%%%%%%%%%%%%%%%%%%%%%%%%%%%%

\subsection{Limit of Unit Operator}

The limit of unit operator is defined by setting one external operator to the identity operator.  In this limit, a $M$-point correlation function directly becomes the corresponding $(M-1)$-point correlation function.

For scalar six-point correlation functions in the snowflake channel, the symmetry properties of $I_6$ liberate us to verify the limit of unit operator for just one quasi-primary operator.  We choose this quasi-primary operator to be $\mathcal{O}_{i_6}(\eta_6)\to\1$, for which $\Delta_{i_6}=0$, $\Delta_{k_3}=\Delta_{i_1}$, and
\eqn{\left.I_{6(\Delta_{k_1},\Delta_{k_2},\Delta_{k_3})}^{(\Delta_{i_2},\Delta_{i_3},\Delta_{i_4},\Delta_{i_5},\Delta_{i_6},\Delta_{i_1})}\right|_{\text{snowflake}}\to\left.I_{5(\Delta_{k_1},\Delta_{k_2})}^{(\Delta_{i_2},\Delta_{i_3},\Delta_{i_1},\Delta_{i_4},\Delta_{i_5})}\right|_{\text{comb}}.}[EqLimUnit]

Since $h_5=p_6=0$ and the remaining vector elements of $\boldsymbol{h}$ and $\boldsymbol{p}$ \eqref{EqCB6hp} translate directly into their five-point counterparts \eqref{EqCB5} in the limit of unit operator \eqref{EqLimUnit}, the sums over $m_3$, $m_{13}$, $m_{23}$, and $m_{33}$ in \eqref{EqCB6C} are trivial due to the Pochhammer symbol $(-h_5)_{m_3+m_{13}+m_{23}+m_{33}}$, forcing $m_3=m_{13}=m_{23}=m_{33}=0$.  Therefore, the conformal cross-ratios $u_3^6$, $v_{13}^6$, $v_{23}^6$, and $v_{33}^6$ disappear in the limit \eqref{EqLimUnit}.

The remaining conformal cross-ratios \eqref{EqCB6CR} relate to the conformal cross-ratios \eqref{EqCB5} as in
\eqn{u_a^6\to u_a^5\qquad(1\leq a\leq2),\qquad\qquad v_{ab}^6\to v_{ab}^5\qquad(1\leq a\leq b\leq2),}
and thus
\eqn{L_6\prod_{1\leq a\leq3}(u_a^6)^{\frac{\Delta_{k_a}}{2}}\to L_5\prod_{1\leq a\leq2}(u_a^5)^{\frac{\Delta_{k_a}}{2}}.}
These observations imply that $G_6\to G_5$ in the limit of unit operator \eqref{EqLimUnit}, which is straightforward to verify since $m_3=m_{13}=m_{23}=m_{33}=0$.

%%%%%%%%%%%%%%%%%%%%%%%%%%%%%%%%%%%%%%%%%%%%%%%%%%
%%%%%%%%%%%%%%%%%%%%%%%%%%%%%%%%%%%%%%%%%%%%%%%%%%

\section{Discussion and Conclusion}\label{SecConc}

The field of research on $d$-dimensional higher-point correlation functions in CFT is a relatively uncharted territory.  Although completely determined by conformal invariance, higher-point conformal blocks are notoriously difficult to compute in all generality.  Moreover, there exists several inequivalent topologies for higher-point correlation functions, each of them having their associated set of identities originating from their topology.  In this paper, we introduced all scalar six-point conformal blocks by computing the remaining topology, the scalar six-point conformal blocks in the snowflake channel.

Our results---presented in \eqref{EqCB6CR}, \eqref{EqCB6C}, \eqref{EqCB6F}, and \eqref{EqCB6hp}---are obtained with the help of the embedding space OPE formalism developed in \cite{Fortin:2019fvx,Fortin:2019dnq}.  They show that the embedding space OPE formalism is very powerful, leading to explicit results for any conformal higher-point correlation function of interest, after straightforward (yet somewhat tedious) re-summations of the hypergeometric type.

From the symmetry group of the snowflake diagram, we showed that scalar six-point conformal blocks in the snowflake channel have symmetry groups of order $48$, larger than the hexagon symmetry group, contrary to expectations.  We then showed that the result verifies both the OPE limit (in which two embedding space coordinates coincide) and the limit of unit operator (where one external operator is set to the identity).  Our snowflake result thus passes several non-trivial consistency checks, lending further credence to the embedding space OPE formalism.

Contrary to the scalar six-point conformal blocks in the comb channel, for which there are two extra sums that factorize into the product of two ${}_3F_2$-hypergeometric functions, the scalar six-point conformal blocks in the snowflake channel have two extra sums that do not factorize.  They can be written as a Kamp\'e de F\'eriet function, and the snowflake invariance under rotations implies \eqref{EqSymRotF}, that translates into the identities (for $m_1$, $m_2$, and $m_3$ non-negative integers and $a_1$, $a_2$, $b_1$, $b_2$, $d_1$, and $g_1$ arbitrary)
\eqna{
&F_{2,1,0}^{1,3,2}\left[\left.\begin{array}{c}a_1;-m_2,b_1+m_3,b_2+m_1;-m_1,-m_3\\b_1,b_2;d_1;-\end{array}\right|1,1\right]\\
&\qquad=\frac{(d_1-a_1)_{m_2}(b_1-a_1)_{m_3}}{(d_1)_{m_2}(b_1)_{m_3}}F_{2,1,0}^{1,3,2}\left[\left.\begin{array}{c}a_1;b_2+m_1,1+a_1-d_1,-m_3;-m_1,-m_2\\1+a_1-d_1-m_2,b_2;1+a_1-b_1-m_3;-\end{array}\right|1,1\right],\\
&F_{1,1,1}^{2,1,1}\left[\left.\begin{array}{c}a_1,a_2;-m_2;-m_3\\a_2-m_1;d_1;g_1\end{array}\right|1,1\right]\\
&\qquad=\frac{(a_2)_{-m_1}(d_1-a_1)_{m_2}}{(a_2-a_1)_{-m_1}(d_1)_{m_2}}F_{1,1,1}^{2,1,1}\left[\left.\begin{array}{c}a_1,1+a_1-d_1;-m_1;-m_3\\1+a_1-d_1-m_2,1+a_1-a_2;g_1\end{array}\right|1,1\right],
}
with
\eqn{F_{2,1,0}^{1,3,2}\left[\left.\begin{array}{c}a_1;-m_2,1+a_1-g_1,a_2;-m_1,-m_3\\1+a_1-g_1-m_3,a_2-m_1;d_1;-\end{array}\right|1,1\right]=\frac{(g_1)_{m_3}}{(g_1-a_1)_{m_3}}F_{1,1,1}^{2,1,1}\left[\left.\begin{array}{c}a_1,a_2;-m_2;-m_3\\a_2-m_1;d_1;g_1\end{array}\right|1,1\right],}
which are obtained from repeated use of well-known ${}_3F_2$-hypergeometric identities.\footnote{Kamp\'e de F\'eriet functions have not been studied as extensively as standard hypergeometric functions.}

It is obvious that the embedding space OPE formalism can be used to investigate higher-point correlation functions, including their symmetry groups.  For example, starting from the scalar three-point correlation function, which has symmetry group $D_3$, we conjecture that doubling the number of all external legs $N$ times with the help of the OPE gives scalar $(3\times2^N)$-point correlation functions with the symmetry groups $\left((\mathbb{Z}_2)^{3\times2^{N-1}}\rtimes\left(\cdots\rtimes\left((\mathbb{Z}_2)^3\rtimes D_3\right)\cdots\right)\right)$.\footnote{Although this is only a conjecture, the order of the symmetry groups should be correct.}  From the successive action of the OPE limit or the limit of unit operator, these maximally-symmetric topologies should lead to all topologies for smaller scalar higher-point correlation functions (starting from sufficiently large $N$).  In general, starting from a specific $M$-point topology with symmetry group $H_{M|\text{channel}}$, this procedure of doubling the number of all external legs $N$ times with the help of the OPE should generate $(M\times2^N)$-point topologies with symmetry groups $\left((\mathbb{Z}_2)^{M\times2^{N-1}}\rtimes\left(\cdots\rtimes\left((\mathbb{Z}_2)^M\rtimes H_{M|\text{channel}}\right)\cdots\right)\right)$.

It is less clear what happens to diagrams with fewer symmetries.  For example, scalar seven-point correlation functions with the topology resulting from scalar six-point correlation functions in the snowflake channel (called the extended snowflake channel below) should have symmetry group of order sixteen only, smaller than for the snowflake diagram but larger than the scalar seven-point correlation functions in the comb channel.  In fact, their symmetry group should be $\mathbb{Z}_2\times\left((\mathbb{Z}_2)^2\rtimes\mathbb{Z}_2\right)$, the direct product of the symmetry group for the extra dendrite permutations and the symmetry group of the six-point comb diagram from which it can also be built.  The symmetries are relatively easy to enumerate for a specific topology, but it is not clear how to write general expressions for the orders of the symmetry groups.  Nevertheless, studying the cosets $S_M/H_{M|\text{channel}}$, with $S_M$ the symmetric group of $M$ elements, leads to interesting consequences.

For example, two- and three-point correlation functions have symmetry groups $\mathbb{Z}_2$ and $D_3$, respectively, while four- and five-point correlation functions have symmetry groups $(\mathbb{Z}_2)^2\rtimes\mathbb{Z}_2$ (the topologies are unique for fewer than six points).  Since $\mathbb{Z}_2\simeq S_2$ and $D_3\simeq S_3$, the symmetry groups of two- and three-point conformal blocks correspond to the symmetry groups of the full two- and three-point correlation functions, and their cosets are trivial with only one element.  This observation implies that there is no non-trivial information that can be derived from bootstrap considerations starting with two- and three-point correlation functions, as expected.  This is not the case for $(M>3)$-point correlation functions.  Indeed, for four-point correlation functions, the associated coset is $S_4/[(\mathbb{Z}_2)^2\rtimes\mathbb{Z}_2]$ which has cardinality three.  There are thus three elements, corresponding to the well-known $s$-, $t$-, and $u$-channels.  For five-point correlation functions, the coset is $S_5/[(\mathbb{Z}_2)^2\rtimes\mathbb{Z}_2]$ with cardinality fifteen, hence there should be fifteen different ways of expressing five-point correlation functions.  For six-point correlation functions, the number of different expressions should depend on the channels.  For the comb channel, there should be $|S_6/[(\mathbb{Z}_2)^2\rtimes\mathbb{Z}_2]|=90$ different expressions while for the snowflake channel there should only be $|S_6/[(\mathbb{Z}_2)^3\rtimes D_6]|=15$, for a total of $105$.  In the case of the seven-point conformal bootstrap, the comb channel and the extended snowflake channel should have coset cardinalities $|S_7/[(\mathbb{Z}_2)^2\rtimes\mathbb{Z}_2]|=630$ and $|S_7/\left[\mathbb{Z}_2\times\left((\mathbb{Z}_2)^2\rtimes\mathbb{Z}_2\right)\right]|=315$, respectively.  Hence, there should be a total of $945$ different ways of expressing seven-point correlation functions.

These numbers can also be understood from the counting of topologies.  Each $M$-point topology is represented by an unrooted binary tree with $2M-2$ nodes, where $M$ nodes represent the external quasi-primary operators (vertices of degree one, which are the leaves of the tree) and $M-2$ nodes correspond to the OPEs (vertices of degree three).  It is transparent that unrooted binary trees with $2M-2$ nodes have a total of $2M-3$ edges, with $M$ edges connecting the OPE nodes to the leaves and $M-3$ edges representing the exchanged quasi-primary operators.  The number of $M$-point topologies is thus given by the number of unrooted binary trees with $M$ unlabeled leaves, denoted here by $T_0(M)$.  Unfortunately, there is no simple closed-form formula for $T_0(M)$, but the first few integers in the sequence are $(1,1,1,1,2,2,4,6,11,\ldots)$, starting at $M=2$.\footnote{See \textit{The On-line Encyclopedia of Integer Sequences} at \href{https://oeis.org/A000672}{https://oeis.org/A000672} and \href{https://oeis.org/A129860}{https://oeis.org/A129860} for more details.}  On the contrary, the number of different ways of expressing the same full $M$-point correlation functions is given by the number of unrooted binary trees with $M$ labeled leaves, denoted here by $T(M)$, which is given by $T(M)=(2M-5)!!$.  Comparing the symmetry groups and the number of topologies, we thus conclude that we should have the following identity for $M$-point correlation functions,
\eqn{\sum_{1\leq i\leq T_0(M)}\frac{1}{|H_{M|\text{channel $i$}}|}=\frac{T(M)}{|S_M|}=\frac{(2M-5)!!}{M!}.}
This identity is in agreement with the partial results obtained above with $M$ up to seven and we verified it for $M=8,9$ as well.  It can be used to explore the orders of the symmetry groups of the different topologies.  Moreover, the previous discussion implies that the number of ways of writing the same $M$-point correlation function is $T(M)$, although redundancies lead to a number of independent bootstrap equations which is much smaller, to which we now turn.

It is well known that the conformal bootstrap program can benefit from the knowledge of higher-point correlation functions.  Indeed, it has been argued that bootstrapping higher-point correlation functions with external quasi-primary operators in the trivial representation is equivalent to the usual full conformal bootstrap of four-point correlation functions.\footnote{This statement must be modified accordingly when there are fermions in the theory.}  We conjecture that the study of the symmetry groups of the different topologies of $M$-point conformal blocks also leads to interesting insights on the conformal bootstrap.  Once again, let us denote the symmetry group of the scalar $M$-point conformal blocks with some specific topology as $H_{M|\text{channel}}$ and the symmetry group of the full $M$-point correlation functions, including the contributions of the non-trivial representations, as $S_M$ (the symmetric group of $M$ elements).  As discussed above, the former depends on the topology of the particular channel under consideration while the latter corresponds to all the possible ways of re-arranging the $M$ external quasi-primary operators in the full correlation function.  We observe that the analysis of the symmetry groups $H_{M|\text{channel}}$ gives an intuitive picture of the $M$-point conformal bootstrap.

\begin{figure}[t!]
\centering
\resizebox{10cm}{!}{%
\begin{tikzpicture}[thick]
\begin{scope}
\draw[-] (-0.5,-0.5)--(0.5,0.5);
\draw[-] (-0.5,1.5)--(0.5,0.5);
\draw[-] (0.5,0.5)--(1.5,0.5);
\draw[-] (1.5,0.5)--(2.5,1.5);
\draw[-] (1.5,0.5)--(2.5,-0.5);
\node at (-1,1.5) {$\mathcal{O}_{i_1}$};
\node at (-1,-0.5) {$\mathcal{O}_{i_2}$};
\node at (3,1.5) {$\mathcal{O}_{i_3}$};
\node at (3,-0.5) {$\mathcal{O}_{i_4}$};
\node at (3.5,0.5) {$=$};
\draw[-] (4.5,-0.5)--(5.5,0.5);
\draw[-] (4.5,1.5)--(5.5,0.5);
\draw[-] (5.5,0.5)--(6.5,0.5);
\draw[-] (6.5,0.5)--(7.5,1.5);
\draw[-] (6.5,0.5)--(7.5,-0.5);
\node at (4,1.5) {$\mathcal{O}_{i_1}$};
\node at (4,-0.5) {$\mathcal{O}_{i_4}$};
\node at (8,1.5) {$\mathcal{O}_{i_3}$};
\node at (8,-0.5) {$\mathcal{O}_{i_2}$};
\end{scope}
\end{tikzpicture}
}
\caption{Four-point bootstrap equation that is equality between the $s$- and $t$-channels.}
\label{Fig4ptbootstrap}
\end{figure}
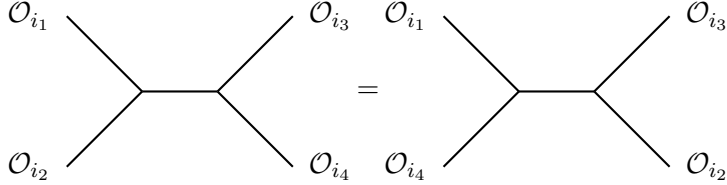
As already mentioned, there are $T(M)=(2M-5)!!$ ways of writing the same full $M$-point correlation function.  One can therefore write a total of $(2M-5)!!-1$ \textit{a priori} independent equations for the $M$-point conformal bootstrap, up to redundancies implied by symmetries of different channels.  In fact, it can be shown that most of the equations are redundant and one can choose specific equations to minimize the overall number of required conformal bootstrap equations.  The smallest number of independent bootstrap equations $N_B$ is equal to the greater of $1$ and $T_0(M)-1$.  Namely, with a unique topology ($M=4,5$) or with two topologies ($M=6,7$) it is sufficient to have only one bootstrap equation, while with more than two topologies, the smallest number of bootstrap equations is equal to the number of topologies minus one.

\begin{figure}[t!]
\centering
\resizebox{16cm}{!}{%
\begin{tikzpicture}[thick]
\begin{scope}
\draw[-] (-0.5,0)--(-1,-0.5) node[left] {$1$};
\draw[-] (-0.5,0)--(-1,0.5) node[left] {$2$};
\draw[-] (-0.5,0)--(5,0);
\draw[-] (5,0)--(5.5,0.5) node[right] {$M$-$1$};
\draw[-] (5,0)--(5.5,-0.5) node[right] {$M$};
\draw[-] (0,0)--(0,0.5);
\draw[-] (0,0.5)--(-0.5,1) node[above] {$3$};
\draw[-] (0,0.5)--(0.5,1) node[above] {$4$};
\node at (1.2,0.5) {$\cdots$};
\draw[-] (2.5,0)--(2.5,0.5);
\draw[-] (2.5,0.5)--(2,1) node[above] {$M$-$5$};
\draw[-] (2.5,0.5)--(3,1) node[above] {$M$-$4$};
\draw[-] (4.5,0)--(4.5,0.5);
\draw[-] (4.5,0.5)--(4,1) node[above] {$M$-$3$};
\draw[-] (4.5,0.5)--(5,1) node[above] {$M$-$2$};
\node at (7.0,0) {$=$};
\draw[-] (8.5,0)--(8,-0.5) node[left] {$2$};
\draw[-] (8.5,0)--(8,0.5) node[left] {$3$};
\draw[-] (8.5,0)--(14,0);
\draw[-] (14,0)--(14.5,0.5) node[right] {$M$-$2$};
\draw[-] (14,0)--(14.5,-0.5) node[right] {$M$-$1$};
\draw[-] (9,0)--(9,0.5);
\draw[-] (9,0.5)--(8.5,1) node[above] {$4$};
\draw[-] (9,0.5)--(9.5,1) node[above] {$5$};
\node at (10.2,0.5) {$\cdots$};
\draw[-] (11.5,0)--(11.5,0.5);
\draw[-] (11.5,0.5)--(11,1) node[above] {$M$-$4$};
\draw[-] (11.5,0.5)--(12,1) node[above] {$M$-$3$};
\draw[-] (12.7,0)--(12.7,1) node[above] {$1$};
\draw[-] (13.5,0)--(13.5,1) node[above] {$M$};
\end{scope}
\begin{scope}[yshift=-3cm]
\draw[-] (-0.5,0)--(-1,-0.5) node[left] {$1$};
\draw[-] (-0.5,0)--(-1,0.5) node[left] {$2$};
\draw[-] (-0.5,0)--(5,0);
\draw[-] (5,0)--(5.5,0.5) node[right] {$M$-$2$};
\draw[-] (5,0)--(5.5,-0.5) node[right] {$M$-$1$};
\draw[-] (0,0)--(0,0.5);
\draw[-] (0,0.5)--(-0.5,1) node[above] {$3$};
\draw[-] (0,0.5)--(0.5,1) node[above] {$4$};
\node at (1.5,0.5) {$\cdots$};
\draw[-] (3,0)--(3,0.5);
\draw[-] (3,0.5)--(2.5,1) node[above] {$M$-$4$};
\draw[-] (3,0.5)--(3.5,1) node[above] {$M$-$3$};
\draw[-] (4.5,0)--(4.5,	1)  node[above] {$M$};
\node at (7.0,0) {$=$};
\draw[-] (8.5,0)--(8,-0.5) node[left] {$2$};
\draw[-] (8.5,0)--(8,0.5) node[left] {$3$};
\draw[-] (8.5,0)--(14,0);
\draw[-] (14,0)--(14.5,0.5) node[right] {$M$-$1$};
\draw[-] (14,0)--(14.5,-0.5) node[right] {$M$};
\draw[-] (9,0)--(9,0.5);
\draw[-] (9,0.5)--(8.5,1) node[above] {$4$};
\draw[-] (9,0.5)--(9.5,1) node[above] {$5$};
\node at (10.2,0.5) {$\cdots$};
\draw[-] (13.5,0)--(13.5,0.5);
\draw[-] (13.5,0.5)--(13,1) node[above] {$M$-$3$};
\draw[-] (13.5,0.5)--(14,1) node[above] {$M$-$2$};
\draw[-] (12,0)--(12,	1)  node[above] {$1$};\end{scope}
\end{tikzpicture}
}
\caption{Bootstrap equations for even $M$ (top) and odd $M$ (bottom) that generate the full permutation group $S_M$.  For better readability, we denoted the external operators just by their subscripts, for example $1$ instead of $\mathcal{O}_{i_1}$.}
\label{FigMptbootstrap}
\end{figure}
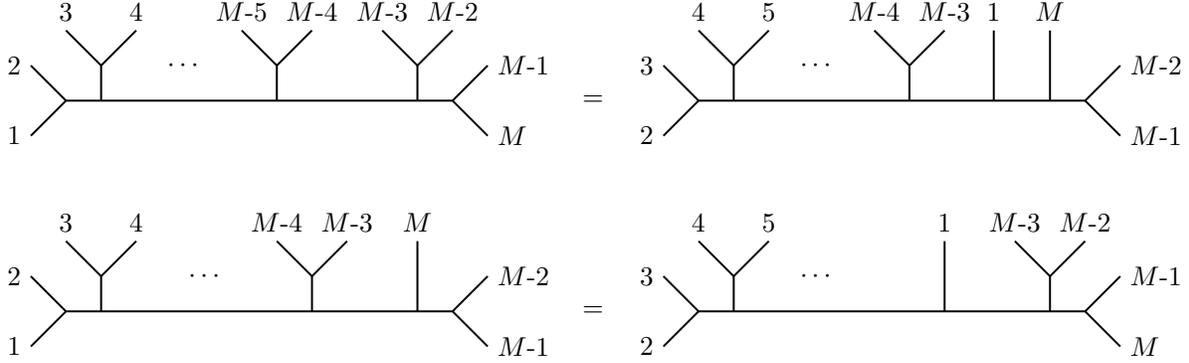
This counting statement is easy to prove using the symmetries of different topologies.  Let us start with the well-known case of four points.  The only required bootstrap equation is the equality of the $s$- and $t$-channels illustrated in Figure~\ref{Fig4ptbootstrap}.  The two diagrams identified by the bootstrap equality have the same symmetries $H_4=(\mathbb{Z}_2)^2\rtimes\mathbb{Z}_2$ since they belong to the same topology, but these symmetries are embedded differently in the full permutation group that acts on different operators in the two channels.  In the $s$-channel, the non-trivial elements of the three $\mathbb{Z}_2$'s in $(\mathbb{Z}_2)^2\rtimes\mathbb{Z}_2$ are $\sigma_{12}$, $\sigma_{34}$, and $\sigma_{13}\sigma_{24}$, where $\sigma_{ab}$ denotes an exchange of external operators $\mathcal{O}_{i_a}$ and $\mathcal{O}_{i_b}$.  The full $H_4$ symmetry is obtained by multiplying these elements.  By comparison, the non-trivial elements of the $t$-channel symmetry group are $\sigma_{14}$, $\sigma_{23}$, and $\sigma_{12}\sigma_{34}$.  Equating the $s$- and $t$-channels allows one to use the symmetries of both diagrams to get all other bootstrap equations related by symmetry.  After all, satisfying the equation means that all equations related by either set of symmetries of the diagram on the right-hand side or the left-hand side are also satisfied.  Picking the following symmetries $\sigma_{12}$, $\sigma_{23}$, and $\sigma_{34}$ from both channels, one generates the full permutation group $S_4$ since these choices comprise the generators of $S_4$.  Once we have the full permutation group, it is clear that one of these permutations includes the $u$-channel and there is no independent equation equating the $u$-channel to either the $s$- or $t$-channels.

Analogous use of symmetries allows us to prove the statement on the minimal number of independent bootstrap equations for any $M$.  The illustration for even $M\geq6 $ and odd $M\geq9$ is shown in Figure~\ref{FigMptbootstrap}.  In both the even and odd cases we start with a single bootstrap equation that identifies two channels of different topologies.  By the assignment of the external operators, the diagrams in the bootstrap equation contain $\sigma_{12},\sigma_{23},\cdots,\sigma_{M-1,M}$ (as well as other symmetries that are unimportant for the argument) that generate the full permutation group $S_M$.  Thus, any other diagram with different operator assignment of either of the two channels can be obtained by symmetries and automatically satisfies the bootstrap equations.  This includes equalities between diagrams of the two topologies chosen in Figure~\ref{FigMptbootstrap} and, by the transitive property, also all equalities between diagrams of the same topology---either the one on the left or on the right of the equation.  Diagrams in topologies that are not present in Figure~\ref{FigMptbootstrap} are independent and each additional topology must be included in the set of bootstrap equations.  However, since the original equation in Figure~\ref{FigMptbootstrap} guarantees equality of all permutations, any operator assignment for the additional topologies will do.  All other assignments will be equivalent by the action of $S_M$.  Hence the smallest number of independent bootstrap equations is $T_0(M)-1$.

\begin{figure}[t!]
\centering
\resizebox{10cm}{!}{%
\begin{tikzpicture}[thick]
\begin{scope}
\draw[-] (4,0)--(3.5,-0.5) node[left] {$1$};
\draw[-] (4,0)--(3.5,0.5) node[left] {$2$};
\draw[-] (4,0)--(5,0);
\draw[-] (5,0)--(5.5,0.5) node[right] {$3$};
\draw[-] (5,0)--(5.5,-0.5) node[right] {$4$};
\draw[-] (4.5,0)--(4.5,0.5) node[above] {$5$};
\node at (6.5,0) {$=$};
\draw[-] (8,0)--(7.5,-0.5) node[left] {$2$};
\draw[-] (8,0)--(7.5,0.5) node[left] {$3$};
\draw[-] (8,0)--(9,0);
\draw[-] (9,0)--(9.5,0.5) node[right] {$4$};
\draw[-] (9,0)--(9.5,-0.5) node[right] {$5$};
\draw[-] (8.5,0)--(8.5,0.5) node[above] {$1$};
\end{scope}
\begin{scope}[yshift=-3cm]
\draw[-] (2.5,0)--(2,-0.5) node[left] {$1$};
\draw[-] (2.5,0)--(2,0.5) node[left] {$2$};
\draw[-] (2.5,0)--(5,0);
\draw[-] (3,0)--(3.0,0.5);
\draw[-] (3.0,0.5)--(2.5,1) node[above] {3};
\draw[-] (3.0,0.5)--(3.5,1) node[above] {4};
\draw[-] (5,0)--(5.5,0.5) node[right] {$5$};
\draw[-] (5,0)--(5.5,-0.5) node[right] {$6$};
\draw[-] (4.5,0)--(4.5,1) node[above] {$7$};
\node at (6.5,0) {$=$};
\draw[-] (8,0)--(7.5,-0.5) node[left] {$2$};
\draw[-] (8,0)--(7.5,0.5) node[left] {$3$};
\draw[-] (8,0)--(10.5,0);
\draw[-] (10.5,0)--(11,0.5) node[right] {$4$};
\draw[-] (10.5,0)--(11,-0.5) node[right] {$5$};
\draw[-] (10,0)--(10,1) node[above] {$7$};
\draw[-] (8.5,0)--(8.5,1) node[above] {$6$};
\draw[-] (9.25,0)--(9.25,1) node[above] {$1$};
\end{scope}
\end{tikzpicture}
}
\caption{Bootstrap equations for five points (top) and seven points (bottom) that generate the full permutation groups $S_5$ and $S_7$, respectively.  For better readability, we denoted the external operators just by their subscripts, for example $1$ instead of $\mathcal{O}_{i_1}$.}
\label{Fig5-7ptbootstrap}
\end{figure}
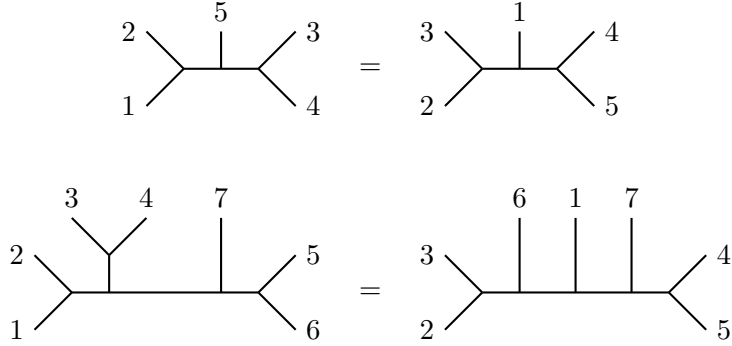
One needs to consider $M=5$, where there is only one topology, and $M=7$ with two topologies, separately.  This is because the choice of topologies for odd $M$ in the bottom drawing in Figure~\ref{FigMptbootstrap} requires topologies that are not present for $M=5$ or identical for $M=7$.  Suitable equations for these cases are illustrated in Figure~\ref{Fig5-7ptbootstrap}.  It is straightforward to check that the symmetries of these diagrams generate the full permutation groups $S_5$ and $S_7$, accordingly.  For $M=5$ the $\mathbb{Z}_2$'s of the dendrite permutations obviously generate $S_5$.  For $M=7$ the dendrite permutations generate $S_6$ acting on the set $\{1,2,\ldots,6\}$.  Additionally, the left-right reflection symmetry of the $M=7$ comb diagram is realized by $\sigma_{67}\sigma_{34}\sigma_{25}$.  This element can be combined with permutations contained in $S_6$ to yield $\sigma_{67}$, therefore generating the full permutation group $S_7$.  This completes the argument that one can satisfy all possible $M$-point bootstrap equations by suitably choosing only $N_B=\max\{1,T_0(M)-1\}$ independent equations.

Coming back to our main result, it is unclear for now what type of generalizations occurs for the extra sums, encoded here in our function $F_M$.  In the comb channel, we argued in \cite{Fortin:2019zkm} that the $M-4$ extra sums appearing in scalar $M$-point correlation functions were necessary for the limit of unit operator to make sense.  For $M=6$, this argument cannot be used for the snowflake diagram.  Nevertheless, \eqref{EqCB6F} shows that in the snowflake channel, the number of extra sums is the same, although these extra sums do not factorize as in the comb channel.  Are there three extra sums that do not factorize for the non-comb topology of the scalar seven-point correlation functions or does the number of sums depend on the topology?

An interesting avenue of research is to initiate the computation of higher-point correlation functions with spins, either for external quasi-primary operators or internal quasi-primary operators, or both.  Since the embedding space OPE formalism developed in \cite{Fortin:2019fvx,Fortin:2019dnq} treats all irreducible representations of the Lorentz group on the same footing, such computations should be feasible.

Finally, higher-point correlation functions can also be of use in the AdS/CFT correspondence.  Indeed, scalar six-point conformal blocks in the snowflake channel correspond to some geodesic Witten diagrams (see for example \cite{Jepsen:2019svc}) and their knowledge might elucidate some of the kinematics in AdS.

%%%%%%%%%%%%%%%%%%%%%%%%%%%%%%%%%%%%%%%%%%%%%%%%%%
%%%%%%%%%%%%%%%%%%%%%%%%%%%%%%%%%%%%%%%%%%%%%%%%%%

\ack{
The authors would like to thank Valentina Prilepina for useful discussions. The work of JFF is supported by NSERC and FRQNT.  The work of WJM is supported by the China Scholarship Council and in part by NSERC and FRQNT.
}

%%%%%%%%%%%%%%%%%%%%%%%%%%%%%%%%%%%%%%%%%%%%%%%%%%
%%%%%%%%%%%%%%%%%%%%%%%%%%%%%%%%%%%%%%%%%%%%%%%%%%

\setcounter{section}{0}
\renewcommand{\thesection}{\Alph{section}}

\section{Scalar Five-Point Conformal Blocks and the OPE}\label{SAppCB5}

The scalar higher-point correlation functions in the comb channel were obtained very recently in \cite{Parikh:2019dvm,Fortin:2019zkm}.  In this appendix, we show how to obtain the scalar five-point conformal blocks of \cite{Parikh:2019dvm} from the OPE approach of \cite{Fortin:2019zkm}.  The proof is a straightforward application of the re-summation formula
\eqn{{}_2F_1\left[\begin{array}{c}-n,b\\c\end{array};1\right]=\frac{(c-b)_n}{(c)_n},}[Eq2F1]
for $n$ a non-negative integer.  Other useful identities used in the proof are the binomial identity
\eqn{(1-v)^{a+b}=\sum_{i\geq0}(-1)^i\binom{a+b}{i}v^i=\sum_{i,j\geq0}(-1)^{i+j}\binom{a}{i}\binom{b}{j}v^{i+j},}[EqBinom]
and
\eqn{
\begin{gathered}
{}_3F_2\left[\begin{array}{c}-n,b,c\\d,e\end{array};1\right]=\frac{(d-b)_n}{(d)_n}{}_3F_2\left[\begin{array}{c}-n,b,e-c\\b-d-n+1,e\end{array};1\right],\\
{}_3F_2\left[\begin{array}{c}-n,b,c\\d,1+b+c-d-n\end{array};1\right]=\frac{(d-b)_n(d-c)_n}{(d)_n(d-b-c)_n}.
\end{gathered}
}[Eq3F2]
The scalar five-point conformal blocks \eqref{EqCB5}, which is our starting point to compute the scalar six-point conformal blocks in the snowflake channel, can be obtained similarly.

%%%%%%%%%%%%%%%%%%%%%%%%%%%%%%%%%%%%%%%%%%%%%%%%%%

\subsection{Proof of the Equivalence}

The comb channel of \cite{Parikh:2019dvm} is depicted in Figure \ref{Fig5ptParikh}.  To get Figure \ref{Fig5ptParikh}, we need to shift the quasi-primary operators in \cite{Parikh:2019dvm} such that $\mathcal{O}_{i_a}(\eta_a)\to\mathcal{O}_{i_{a+1}}(\eta_{a+1})$ with $\mathcal{O}_{i_6}(\eta_6)\equiv\mathcal{O}_{i_1}(\eta_1)$.
\begin{figure}[t]
\centering
\resizebox{11cm}{!}{%
\begin{tikzpicture}[thick]
\begin{scope}
\node at (-2.2,0) {$\left.I_{5(\Delta_{k_1},\Delta_{k_2})}^{(\Delta_{i_2},\ldots,\Delta_{i_5},\Delta_{i_1})}\right|_{\text{comb}}$};
\node at (0,0) {$=$};
\node at (1,0) {$\mathcal{O}_{i_2}$};
\draw[-] (1.5,0)--(5.5,0);
\node at (6.1,0) {$\mathcal{O}_{i_1}$};
\draw[-] (2.5,0)--(2.5,1) node[above]{$\mathcal{O}_{i_3}$};
\draw[-] (3.5,0)--(3.5,1) node[above]{$\mathcal{O}_{i_4}$};
\draw[-] (4.5,0)--(4.5,1) node[above]{$\mathcal{O}_{i_5}$};
\node at (3,-0.5) {$\mathcal{O}_{k_1}$};
\node at (4,-0.5) {$\mathcal{O}_{k_2}$};
\end{scope}
\end{tikzpicture}
}
\caption{Scalar five-point conformal blocks of \cite{Parikh:2019dvm}.}
\label{Fig5ptParikh}
\end{figure}
Using \eqref{EqIfromOPE} with $k=2$, $l=3$ and $m=2$ on the scalar four-point correlation functions of \cite{Fortin:2019zkm}, we obtain, after expanding with the binomial identity \eqref{EqBinom},
\eqna{
I_5&=L_5\left(\frac{\eta_{23}\eta_{45}}{\eta_{24}\eta_{35}}\right)^{\frac{\Delta_{k_1}}{2}}\left(\frac{\eta_{34}\eta_{15}}{\eta_{14}\eta_{35}}\right)^{\frac{\Delta_{k_2}}{2}}\left(\frac{\eta_{14}\eta_{25}}{\eta_{12}\eta_{45}}\right)^{\bar{p}_4+\bar{h}_4}\\
&\phantom{=}\qquad\times\sum C_4K_5\binom{n_{11}}{s_2}\frac{(-1)^{s_2}}{n_1!n_{11}!}\left(\frac{\eta_{23}\eta_{45}}{\eta_{24}\eta_{35}}\right)^{n_1}\left(\frac{\eta_{25}\eta_{34}}{\eta_{35}\eta_{24}}\right)^{s_2},
}
where the legs $L_5$ and the vectors $\boldsymbol{h}$ and $\boldsymbol{p}$ are defined in \cite{Fortin:2019zkm}.  For notational simplicity, we omit the indices of summation $n_1$, $n_{11}$, and $s_2$ on the sum.  Here $C_4$ and $K_5$ are given by the analog of \eqref{EqCB4} found in \cite{Fortin:2019zkm} and \eqref{EqD}, respectively, \textit{i.e.}
\eqn{
\begin{gathered}
C_4=\frac{(-h_3)_{n_1}(p_2+h_2)_{n_1}(\bar{p}_3+\bar{h}_3)_{n_1+n_{11}}(p_3)_{n_1+n_{11}}}{(\bar{p}_3+h_2)_{2n_1+n_{11}}(\bar{p}_3+h_2+1-d/2)_{n_1}},\\
K_5=\sum_{\{r_a,r_{2a},r_{ab}\}\geq0}\frac{(-h_4)_{\bar{r}_2+\bar{\bar{r}}}(-s_2)_{\bar{r}_2}(\bar{p}_4+\bar{h}_4)_{\bar{r}-\bar{\bar{r}}}}{(\bar{p}_4+\bar{h}_3)_{\bar{r}+\bar{r}_2}(\bar{p}_4+\bar{h}_3+1-d/2)_{\bar{r}_2+\bar{\bar{r}}}}\frac{(p_4-n_1)_{r_4}(\bar{p}_3+\bar{h}_3+n_1+s_2)_{r_3}}{(r_4-r_{24}-\bar{r}_4)!(r_3-r_{23}-\bar{r}_3)!}\\
\qquad\qquad\times(x_2^5)^{\bar{r}_2+\bar{\bar{r}}}(y_3^5)^{r_3-r_{23}-\bar{r}_3}(y_4^5)^{r_4-r_{24}-\bar{r}_4}\frac{(z_{24}^5)^{r_{24}}}{(r_{23})!(r_{24})!}\frac{(z_{34}^5)^{r_{34}}}{(r_{34})!}.
\end{gathered}
}

The legs and the conformal cross-ratios defined in \cite{Parikh:2019dvm} are given by
\eqna{
L_5^P&=\left(\frac{\eta_{13}}{\eta_{12}\eta_{23}}\right)^{\frac{\Delta_2}{2}}\left(\frac{\eta_{12}}{\eta_{13}\eta_{23}}\right)^{\frac{\Delta_3}{2}}\left(\frac{\eta_{12}}{\eta_{14}\eta_{24}}\right)^{\frac{\Delta_4}{2}}\left(\frac{\eta_{12}}{\eta_{15}\eta_{25}}\right)^{\frac{\Delta_5}{2}}\left(\frac{\eta_{25}}{\eta_{12}\eta_{15}}\right)^{\frac{\Delta_1}{2}}\\
&=L_5\left(\frac{\eta_{14}\eta_{25}}{\eta_{12}\eta_{45}}\right)^{\frac{\Delta_1-\Delta_5}{2}}\left(\frac{\eta_{12}\eta_{34}}{\eta_{13}\eta_{24}}\right)^{\frac{\Delta_3-\Delta_2}{2}}\left(\frac{\eta_{12}\eta_{34}\eta_{45}}{\eta_{14}\eta_{24}\eta_{35}}\right)^{\frac{\Delta_4}{2}},
}
and
\eqn{u_1^P=\frac{\eta_{14}\eta_{23}}{\eta_{13}\eta_{24}},\qquad u_2^P=\frac{\eta_{15}\eta_{24}}{\eta_{14}\eta_{25}},\qquad v_{23}^P=\frac{\eta_{12}\eta_{34}}{\eta_{13}\eta_{24}},\qquad v_{24}^P=\frac{\eta_{12}\eta_{35}}{\eta_{13}\eta_{25}},\qquad v_{34}^P=\frac{\eta_{12}\eta_{45}}{\eta_{14}\eta_{25}}.}
Hence $I_5$ becomes
\eqna{
I_5&=L_5^P(v_{23}^P)^{p_3+h_3}(v_{24}^P)^{-\bar{p}_3-\bar{h}_3}(v_{34}^P)^{-p_4}(u_1^P)^{\frac{\Delta_{k_1}}{2}}(u_2^P)^{\frac{\Delta_{k_2}}{2}}\\
&\phantom{=}\qquad\times\sum C_4K_5\binom{n_{11}}{s_2}\frac{(-1)^{s_2}}{n_1!n_{11}!}\left(u_1^P\frac{v_{34}^P}{v_{24}^P}\right)^{n_1}\left(\frac{v_{23}^P}{v_{24}^P}\right)^{s_2}.
}
With the help of the following identities [see \eqref{EqCROPE}]
\eqn{x_2^5=\frac{u_1^Pu_2^P}{v_{24}^P},\qquad1-y_3^5=\frac{1}{v_{24}^P},\qquad1-y_4^5=\frac{1}{v_{34}^P},\qquad z_{24}^5=\frac{v_{24}^P}{u_1^Pv_{34}^P},\qquad z_{34}^5=\frac{v_{23}^P}{u_1^Pv_{34}^P},}
we can express $K_5$ in terms of the conformal cross-ratios of \cite{Parikh:2019dvm}, leading to
\eqna{
I_5&=L_5^P(u_1^P)^{\frac{\Delta_{k_1}}{2}}(u_2^P)^{\frac{\Delta_{k_2}}{2}}\sum C_4\binom{n_{11}}{s_2}\frac{(-1)^{s_2}}{n_1!n_{11}!}\left(1-\frac{1}{v_{24}^P}\right)^{\sigma_3}\left(1-\frac{1}{v_{34}^P}\right)^{\sigma_4}\\
&\phantom{=}\qquad\times(u_1^P)^{n_1+r_{23}}(u_2^P)^{m_2}(v_{23}^P)^{p_3+h_3+m_2+r_{34}}(v_{24}^P)^{-\bar{p}_3-\bar{h}_3-n_1-s_2-r_{23}-r_{34}}(v_{34}^P)^{n_1-p_4-m_2+r_{23}}\\
&\phantom{=}\qquad\times\frac{(-h_4)_{m_2}(-s_2)_{m_2-r_{34}}(\bar{p}_4+\bar{h}_4)_{m_2+\sigma_3+\sigma_4}}{(\bar{p}_4+\bar{h}_3)_{2m_2+\sigma_3+\sigma_4}(\bar{p}_4+\bar{h}_3+1-d/2)_{m_2}}\\
&\phantom{=}\qquad\times\frac{(p_4-n_1)_{\sigma_4+m_2-r_{23}}(\bar{p}_3+\bar{h}_3+n_1+s_2)_{\sigma_3+r_{23}+r_{34}}}{\sigma_3!\sigma_4!r_{23}!r_{24}!r_{34}!}.
}
Thus, by combining the powers of the conformal cross-ratios, $G_5^P$ \eqref{EqG} is given by
\eqna{
G_5^P&=\sum\binom{n_{11}}{s_2}\binom{\sigma_3}{l_3}\binom{\sigma_4}{l_4}\binom{p_3+h_3+s_2+r_{34}}{m_{23}}\\
&\phantom{=}\qquad\times\binom{-l_3-\bar{p}_3-\bar{h}_3-n_1-s_2-r_{23}-r_{34}}{m_{24}}\binom{n_1-l_4-p_4-m_2+r_{23}}{m_{34}}\\
&\phantom{=}\qquad\times\frac{(-h_4)_{m_2}(-s_2)_{m_2-r_{34}}(\bar{p}_4+\bar{h}_4)_{m_2+\sigma_3+\sigma_4}}{(\bar{p}_4+\bar{h}_3)_{2m_2+\sigma_3+\sigma_4}(\bar{p}_4+\bar{h}_3+1-d/2)_{m_2}}\\
&\phantom{=}\qquad\times\frac{(p_4-n_1)_{\sigma_4+m_2-r_{23}}(\bar{p}_3+\bar{h}_3+n_1+s_2)_{\sigma_3+r_{23}+r_{34}}}{\sigma_3!\sigma_4!r_{23}!r_{24}!r_{34}!}\\
&\phantom{=}\qquad\times\frac{(-1)^{s_2+l_3+l_4+m_{23}+m_{24}+m_{34}}C_4}{n_1!n_{11}!}(u_1^P)^{n_1+r_{23}}(u_2^P)^{m_2}(1-v_{23}^P)^{m_{23}}(1-v_{24}^P)^{m_{24}}(1-v_{34}^P)^{m_{34}},
}
where all the superfluous sums must be appropriately taken care of to reach the result of \cite{Parikh:2019dvm}.

Since the terms involving $l_4$ are
\eqn{
\begin{gathered}
(-1)^{l_4}\binom{\sigma_4}{l_4}=\frac{(-\sigma_4)_{l_4}}{l_4!},\\
(-1)^{m_{34}}\binom{n_1-l_4-p_4-m_2+r_{23}}{m_{34}}=\frac{(p_4+m_2-n_1-r_{23}+m_{34})_{l_4}(p_4+m_2-n_1-r_{23})_{m_{34}}}{(p_4+m_2-n_1-r_{23})_{l_4}m_{34}!},
\end{gathered}
}
they can be re-summed with the help of the identity \eqref{Eq2F1}, and we find that $G_5^P$ becomes
\eqna{
G_5^P&=\sum\binom{n_{11}}{s_2}\binom{\sigma_3}{l_3}\binom{p_3+h_3+s_2+r_{34}}{m_{23}}\binom{-l_3-\bar{p}_3-\bar{h}_3-n_1-s_2-m_2+r_{34}}{m_{24}}(-m_{34})_{\sigma_4}\\
&\phantom{=}\qquad\times\frac{(-h_4)_{m_2}(-s_2)_{m_2-r_{34}}(\bar{p}_4+\bar{h}_4)_{m_2+\sigma_3+\sigma_4}}{(\bar{p}_4+\bar{h}_3)_{2m_2+\sigma_3+\sigma_4}(\bar{p}_4+\bar{h}_3+1-d/2)_{m_2}}\\
&\phantom{=}\qquad\times\frac{(p_4-n_1)_{m_{34}+m_2-r_{23}}(\bar{p}_3+\bar{h}_3+n_1+s_2)_{\sigma_3+r_{23}+r_{34}}}{\sigma_3!\sigma_4!r_{23}!r_{24}!r_{34}!}\\
&\phantom{=}\qquad\times\frac{(-1)^{s_2+l_3+m_{23}+m_{24}}}{n_1!n_{11}!}C_4(u_1^P)^{n_1+r_{23}}(u_2^P)^{m_2}(1-v_{23}^P)^{m_{23}}(1-v_{24}^P)^{m_{24}}\frac{(1-v_{34}^P)^{m_{34}}}{m_{34}!}.
}	
Similar steps can be performed to re-sum over $l_3$, $\sigma_3$, and finally $\sigma_4$.  The result is given by
\eqna{
G_5^P&=\sum C_4\binom{n_{11}}{s_2}\frac{(-1)^{s_2+m_{23}}}{n_1!n_{11}!}\binom{p_3+h_3+s_2+r_{34}}{m_{23}}\\
&\phantom{=}\qquad\times\frac{(-h_4)_{m_2+m_{24}+m_{34}}(-s_2)_{m_2-r_{34}}(\bar{p}_4+\bar{h}_4)_{m_2}}{(\bar{p}_4+\bar{h}_3)_{2m_2+m_{24}+m_{34}}(\bar{p}_4+\bar{h}_3+1-d/2)_{m_2}}\\
&\phantom{=}\qquad\times\frac{(p_4-n_1)_{m_{34}+m_2-r_{23}}(\bar{p}_3+\bar{h}_3+n_1+s_2)_{m_{24}+r_{23}+r_{34}}}{r_{23}!(m_2-r_{23}-r_{34})!r_{34}!}\\
&\phantom{=}\qquad\times(u_1^P)^{n_1+r_{23}}(u_2^P)^{m_2}(1-v_{23}^P)^{m_{23}}\frac{(1-v_{24}^P)^{m_{24}}}{m_{24}!}\frac{(1-v_{34}^P)^{m_{34}}}{m_{34}!}.
}

To evaluate the sum over $r_{34}$, we first change the variable as $s_2\to s_2+m_2-r_{34}$.  Then, the sum over $r_{34}$ can also be simplified using the hypergeometric type re-summation \eqref{Eq2F1}, which implies that $G_5$ can be expressed as
\eqna{
G_5^P&=\sum C_4\frac{(-1)^{s_2+m_{23}}}{n_1!}\binom{p_3+h_3+s_2+m_2}{m_{23}}\frac{(\bar{p}_3+\bar{h}_3+n_1+n_{11})_{m_2-r_{23}}}{s_2!(n_{11}-s_2-r_{23})!(m_2-r_{23})!}\\
&\phantom{=}\qquad\times\frac{(-h_4)_{m_2+m_{24}+m_{34}}(\bar{p}_4+\bar{h}_4)_{m_2}(p_4-n_1)_{m_{34}+m_2-r_{23}}}{(\bar{p}_4+\bar{h}_3)_{2m_2+m_{24}+m_{34}}(\bar{p}_4+\bar{h}_3+1-d/2)_{m_2}}\frac{(\bar{p}_3+\bar{h}_3+n_1+s_2+m_2)_{m_{24}+r_{23}}}{r_{23}!}\\
&\phantom{=}\qquad\times(u_1^P)^{n_1+r_{23}}(u_2^P)^{m_2}(1-v_{23}^P)^{m_{23}}\frac{(1-v_{24}^P)^{m_{24}}}{m_{24}!}\frac{(1-v_{34}^P)^{m_{34}}}{m_{34}!}.
}	
Using the last identity of \eqref{EqBinom} for the sum over $m_{23}$ leads to
\eqna{
G_5^P&=\sum C_4\frac{(-1)^{s_2+k_{23}+l}}{n_1!}\binom{p_3+h_3+m_2}{k_{23}}\binom{s_2}{l}\frac{(\bar{p}_3+\bar{h}_3+n_1+n_{11})_{m_2-r_{23}}}{s_2!(n_{11}-s_2-r_{23})!(m_2-r_{23})!}\\
&\phantom{=}\qquad\times\frac{(-h_4)_{m_2+m_{24}+m_{34}}(\bar{p}_4+\bar{h}_4)_{m_2}(p_4-n_1)_{m_{34}+m_2-r_{23}}}{(\bar{p}_4+\bar{h}_3)_{2m_2+m_{24}+m_{34}}(\bar{p}_4+\bar{h}_3+1-d/2)_{m_2}}\frac{(\bar{p}_3+\bar{h}_3+n_1+s_2+m_2)_{m_{24}+r_{23}}}{r_{23}!}\\
&\phantom{=}\qquad\times(u_1^P)^{n_1+r_{23}}(u_2^P)^{m_2}(1-v_{23}^P)^{k_{23}+l}\frac{(1-v_{24}^P)^{m_{24}}}{m_{24}!}\frac{(1-v_{34}^P)^{m_{34}}}{m_{34}!}.
}	
We can now make a redefinition of the variable such that $s_2\to s_2+l$ and re-sum over $s_2$ using \eqref{Eq2F1} again.  Expressing $C_4$ in terms of Pochhammer symbols, the scalar five-point conformal blocks become
\eqna{
G_5^P&=\sum\frac{(-1)^{k_{23}}}{n_1!}\binom{p_3+h_3+m_2}{k_{23}}(u_1^P)^{n_1+r_{23}}(u_2^P)^{m_2}(1-v_{23}^P)^{k_{23}+l}\frac{(1-v_{24}^P)^{m_{24}}}{m_{24}!}\frac{(1-v_{34}^P)^{m_{34}}}{m_{34}!}\\
&\phantom{=}\qquad\times\frac{(-h_4)_{m_2+m_{24}+m_{34}}(\bar{p}_4+\bar{h}_4)_{m_2}}{(\bar{p}_4+\bar{h}_3)_{2m_2+m_{24}+m_{34}}(\bar{p}_4+\bar{h}_3+1-d/2)_{m_2}}\frac{(p_4-n_1)_{m_{34}+m_2-r_{23}}}{r_{23}!}\\
&\phantom{=}\qquad\times\frac{(-h_3)_{n_1}(p_2+h_2)_{n_1}(\bar{p}_3+\bar{h}_3)_{m_2+m_{24}+n_1+r_{23}+l}}{(\bar{p}_3+h_2)_{2n_1+n_{11}}(\bar{p}_3+h_2+1-d/2)_{n_1}}\frac{(p_3)_{n_1+n_{11}}}{l!(n_{11}-r_{23}-l)!}\frac{(-m_{24}-r_{23})_{n_{11}-r_{23}-l}}{(m_2-r_{23})!}.
}
Changing variables again as in $n_{11}\to n_{11}+r_{23}+l$, we can first evaluate the summation over $n_{11}$ using \eqref{Eq2F1}, and then the summation over $l$ with the help of the second identity in \eqref{Eq3F2} after changing $k_{23}\to k_{23}-l$, and finally replace $n_1=m_1-r_{23}$, leading to
\eqna{
G_5^P&=\sum\frac{1}{(m_1-r_{23})!}(u_1^P)^{m_1}(u_2^P)^{m_2}\frac{(1-v_{23}^P)^{m_{23}}}{m_{23}!}\frac{(1-v_{24}^P)^{m_{24}}}{m_{24}!}\frac{(1-v_{34}^P)^{m_{34}}}{m_{34}!}\\
&\phantom{=}\qquad\times\frac{(-h_4)_{m_2+m_{24}+m_{34}}(\bar{p}_4+\bar{h}_4)_{m_2}}{(\bar{p}_4+\bar{h}_3)_{2m_2+m_{24}+m_{34}}(\bar{p}_4+\bar{h}_3+1-d/2)_{m_2}}\frac{(p_4-m_1+r_{23})_{m_2+m_{34}-r_{23}}}{r_{23}!}\\
&\phantom{=}\qquad\times\frac{(-h_3+m_1-m_2)_{m_{23}}(-h_3)_{m_1-r_{23}}(p_2+h_2)_{m_1+m_{24}}(\bar{p}_3+\bar{h}_3)_{m_1+m_2+m_{24}}}{(\bar{p}_3+h_2)_{2m_1+m_{23}+m_{24}}(\bar{p}_3+h_2+1-d/2)_{m_1-r_{23}}}\frac{(p_3)_{m_1}}{(m_2-r_{23})!}.
}	

At this point, we are left with only one extra sum (over $r_{23}$), as expected.  We now use the following relations,
\eqn{
\begin{gathered}
(p_4-m_1+r_{23})_{m_2+m_{34}-r_{23}}=\frac{(-1)^{m_1}(p_3)_{m_2-m_1+m_{34}}(1-p_4)_{m_1}}{(p_4-m_1)_{r_{23}}},\\
\frac{(-h_3)_{m_1-r_{23}}}{(\bar{p}_3+h_2+1-d/2)_{m_1-r_{23}}}=\frac{(-h_3)_{m_1}(-\bar{p}_3-h_2+d/2-m_1)_{r_{23}}}{(\bar{p}_3+h_2+1-d/2)_{m_1}(1+h_3-m_1)_{r_{23}}},
\end{gathered}
}
to re-sum the summation over $r_{23}$ into a ${}_3F_2$-hypergeometric function, given by
\eqn{{}_3F_{2}\left[\begin{array}{c}-m_1,-m_2,-\bar{p}_3-h_2+d/2-m_1\\p_4-m_1,1+h_3-m_1\end{array};1\right].}
With the help of the identity \eqref{Eq3F2}, this can be rewritten as
\eqn{\frac{(1+h_3)_{m_2}}{(1+h_3-m_1)_{m_2}}{}_3F_{2}\left[\begin{array}{c}-m_1,-m_2,\bar{p}_4+h_2-d/2\\p_4-m_1,-h_3-m_2\end{array};1\right],}
which translates into the result of \cite{Parikh:2019dvm}, \textit{i.e.}
\eqna{
G_5^P&=\sum_{\{m_a,m_{ab}\geq0\}}\frac{(p_2+h_2)_{m_1+m_{23}+m_{24}}(p_3)_{m_1}(-h_3)_{m_1-m_2+m_{23}}(p_4)_{m_2-m_1+m_{34}}}{(p_4)_{-m_1}(\bar{p}_3+h_2)_{2m_1+m_{23}+m_{24}}(\bar{p}_3+h_2+1-d/2)_{m_1}}\\
&\phantom{=}\qquad\times\frac{(\bar{p}_3+\bar{h}_3)_{m_1+m_2+m_{24}}(-h_4)_{m_2+m_{24}+m_{34}}(\bar{p}_4+\bar{h}_4)_{m_2}}{(-h_3)_{-m_2}(\bar{p}_4+\bar{h}_3)_{2m_2+m_{24}+m_{34}}(\bar{p}_4+\bar{h}_3+1-d/2)_{m_2}}\\
&\phantom{=}\qquad\times\frac{(u_1^P)^{m_1}}{m_1!}\frac{(u_2^P)^{m_2}}{m_2!}\frac{(1-v_{23}^P)^{m_{23}}}{m_{23}!}\frac{(1-v_{24}^P)^{m_{24}}}{m_{24}!}\frac{(1-v_{34}^P)^{m_{34}}}{m_{34}!}\\
&\phantom{=}\qquad\times{}_3F_{2}\left[\begin{array}{c}-m_1,-m_2,\bar{p}_4+h_2-d/2\\p_4-m_1,-h_3-m_2\end{array};1\right].
}[EqCB5P]
This computation shows that the results of \cite{Parikh:2019dvm} and \cite{Fortin:2019zkm} are equivalent.  Moreover, once a choice of conformal cross-ratios has been made, the OPE approach does lead to the correct result, after several re-summations.

The steps highlighted here can be repeated to obtain the scalar five-point conformal blocks in the comb channel discussed in \eqref{EqCB5}.  The proof of their equivalence at the level of \eqref{EqCB5P} and \eqref{EqCB5} follows the one for the scalar six-point conformal blocks in the snowflake channel shown in Appendix \ref{SAppCB6}.

%%%%%%%%%%%%%%%%%%%%%%%%%%%%%%%%%%%%%%%%%%%%%%%%%%
%%%%%%%%%%%%%%%%%%%%%%%%%%%%%%%%%%%%%%%%%%%%%%%%%%
	
\section{Snowflake and the OPE}\label{SAppCB6}

In this appendix we expound the proof of the scalar six-point conformal blocks in the snowflake channel.  We also present an alternative form for the scalar six-point conformal blocks in the snowflake channel, and prove that it is equivalent to the one introduced in the main text.

%%%%%%%%%%%%%%%%%%%%%%%%%%%%%%%%%%%%%%%%%%%%%%%%%%
	
\subsection{Proof of the Snowflake}

Starting from the scalar five-point correlation functions \eqref{EqCB5}, shifted such that $\mathcal{O}_{i_a}(\eta_a)\to\mathcal{O}_{i_{a-1}}(\eta_{a-1})$ with $\mathcal{O}_{i_0}(\eta_0)\equiv\mathcal{O}_{i_5}(\eta_5)$, the legs and the conformal cross-ratios transform into
\eqn{
\begin{gathered}
L_{5|\text{comb}}^{(\Delta_{i_2},\ldots,\Delta_{i_5},\Delta_{k_3})}=\left(\frac{\eta_{13}}{\eta_{12}\eta_{23}}\right)^{\frac{\Delta_{i_2}}{2}}\left(\frac{\eta_{12}}{\eta_{13}\eta_{23}}\right)^{\frac{\Delta_{i_3}}{2}}\left(\frac{\eta_{35}}{\eta_{34}\eta_{45}}\right)^{\frac{\Delta_{i_4}}{2}}\left(\frac{\eta_{34}}{\eta_{35}\eta_{45}}\right)^{\frac{\Delta_{i_5}}{2}}\left(\frac{\eta_{35}}{\eta_{13}\eta_{15}}\right)^{\frac{\Delta_{k_3}}{2}},\\
u_1^5=\frac{\eta_{15}\eta_{23}}{\eta_{12}\eta_{35}},\qquad u_2^5=\frac{\eta_{13}\eta_{45}}{\eta_{15}\eta_{34}},\qquad v_{11}^5=\frac{\eta_{13}\eta_{25}}{\eta_{12}\eta_{35}},\qquad v_{12}^5=\frac{\eta_{14}\eta_{35}}{\eta_{15}\eta_{34}},\qquad v_{22}^5=\frac{\eta_{13}\eta_{24}}{\eta_{12}\eta_{34}}.
\end{gathered}
}[EqLCR]
Here, we already substituted $\mathcal{O}_{i_1}(\eta_1)\to\mathcal{O}_{k_3}(\eta_1)$ as needed for the recurrence relation \eqref{EqIfromOPE}.  Moreover, the vectors $\boldsymbol{h}$ and $\boldsymbol{p}$ get transformed into \eqref{EqCB6hp}, which also include the new elements appearing in the recurrence relation \eqref{EqIfromOPE}, while the functional forms of $C_5$ and $F_5$ \eqref{EqCB5} remain the same [they are the same functions but of the new vectors $\boldsymbol{h}$ and $\boldsymbol{p}$ \eqref{EqCB6hp}, hence $G_5$ is the same function but of the new conformal cross-ratios and vectors].

Using the recurrence relation \eqref{EqIfromOPE} with $k=4$, $l=5$, and $m=5$, it is clear that the resulting scalar six-point correlation functions are in the snowflake channel.  To proceed, we must act with the OPE differential operator as in \eqref{EqD} on the conformal cross-ratios [see \eqref{EqCROPE}]
\eqn{
\begin{gathered}
x_5^6=\frac{\eta_{16}\eta_{45}}{\eta_{15}\eta_{46}},\qquad1-y_2^6=\frac{\eta_{12}\eta_{56}}{\eta_{15}\eta_{26}},\qquad1-y_3^6=\frac{\eta_{13}\eta_{56}}{\eta_{15}\eta_{36}},\qquad1-y_4^6=\frac{\eta_{14}\eta_{56}}{\eta_{15}\eta_{46}},\\
z_{23}^6=\frac{\eta_{23}\eta_{46}\eta_{56}}{\eta_{26}\eta_{36}\eta_{45}},\qquad z_{24}^6=\frac{\eta_{24}\eta_{56}}{\eta_{26}\eta_{45}},\qquad z_{25}^6=\frac{\eta_{25}\eta_{46}}{\eta_{26}\eta_{45}},\qquad z_{34}^6=\frac{\eta_{34}\eta_{56}}{\eta_{36}\eta_{45}},\qquad z_{35}^6=\frac{\eta_{35}\eta_{46}}{\eta_{36}\eta_{45}}.
\end{gathered}
}[Eqxyz]
This can be done easily by re-expressing \eqref{EqLCR} in terms of \eqref{Eqxyz} as
\eqn{
\begin{gathered}
u_1^5=\frac{1-y_5^6}{1-y_2^6}\frac{z_{23}^6}{z_{35}^6},\qquad u_2^5=\frac{1-y_3^6}{1-y_5^6}\frac{1}{z_{34}^6},\\
v_{11}^5=\frac{1-y_3^6}{1-y_2^6}\frac{z_{25}^6}{z_{35}^6},\qquad v_{12}^5=\frac{1-y_4^6}{1-y_5^6}\frac{z_{35}^6}{z_{34}^6},\qquad v_{22}^5=\frac{1-y_3^6}{1-y_2^6}\frac{z_{24}^6}{z_{34}^6}.
\end{gathered}
}
Defining a new set of conformal cross-ratios for the snowflake as in \eqref{EqCB6CR}, we have
\eqn{
\begin{gathered}
x_5^6=\frac{u_2^6u_3^6}{v_{33}^6},\qquad1-y_2^6=\frac{1}{v_{13}^6},\qquad1-y_3^6=\frac{1}{v_{23}^6},\qquad1-y_4^6=\frac{v_{12}^6}{v_{33}^6},\\
z_{23}^6=\frac{u_1^6v_{33}^6}{u_2^6v_{13}^6v_{23}^6},\qquad z_{24}^6=\frac{v_{22}^6}{u_2^6v_{13}^6},\qquad z_{25}^6=\frac{v_{11}^6v_{33}^6}{u_2^6v_{13}^6},\qquad z_{34}^6=\frac{1}{u_2^6v_{23}^6},\qquad z_{35}^6=\frac{v_{33}^6}{u_2^6v_{23}^6},
\end{gathered}
}
such that the action of the OPE differential operator on $I_5$ \eqref{EqCB5} gives
\eqna{
G_6&=\sum \binom{n_{11}}{s_{11}}\binom{n_{12}}{s_{12}}\binom{n_{22}}{s_{22}}\frac{(-h_5)_{m_3}(p_3-n_1+n_2+s_{12})_{\bar{r}_5}(\bar{p}_3+h_2+h_5)_{m_3+\sigma_2+\sigma_3+\sigma_4}}{(\bar{p}_3+h_2)_{2m_3+\sigma_2+\sigma_3+\sigma_4}(\bar{p}_3+h_2+1-d/2)_{m_3}}\\
&\phantom{=}\qquad\times(-1)^{s_{11}+s_{12}+s_{22}+l_2+l_3+l_4+m_{11}+m_{12}+m_{22}+m_{13}+m_{23}+m_{33}}\binom{s_{12}-r_{45}-\bar{r}_4-l_4}{m_{33}}\binom{l_4}{m_{12}}\\
&\phantom{=}\qquad\times\binom{r_{25}+s_{11}}{m_{11}}\binom{r_{24}+s_{22}}{m_{22}}\binom{\sigma_2}{l_2}\binom{\sigma_3}{l_3}\binom{\sigma_4}{l_4}\binom{h_3-n_1-s_{11}-s_{22}-r_{25}-\bar{r}_2-l_2}{m_{13}}\\
&\phantom{=}\qquad\times\binom{-p_2-\bar{h}_3+n_2+s_{11}+s_{22}-r_{35}-\bar{r}_3-l_3}{m_{23}}\\
&\phantom{=}\qquad\times\frac{(-h_3+n_1+s_{11}+s_{22})_{\sigma_2+r_{25}+\bar{r}_2}(p_2+\bar{h}_3-n_2-s_{11}-s_{22})_{\sigma_3+r_{35}+\bar{r}_3}(-s_{12})_{\sigma_4+r_{45}+\bar{r}_4}}{\sigma_2!\sigma_3!\sigma_4!}\\
&\phantom{=}\qquad\times\frac{(u_1^6)^{n_1+r_{23}}(u_2^6)^{n_2+r_{45}}(u_3^6)^{m_3}\prod_{1\leq a\leq b\leq 3}(1-v_{ab}^6)^{m_{ab}}}{r_{23}!r_{24}!r_{25}!r_{34}!r_{35}!r_{45}!n_1!n_{11}!n_{12}!n_{22}!}C_5F_5,
}[EqG6extrasums]
with the proper legs \eqref{EqCB6CR} and $m_3=\bar{r}_5+\bar{\bar{r}}$.  Here again, we omit the indices of summation on the sum for notational simplicity.  Equation \eqref{EqG6extrasums} corresponds to the scalar six-point conformal blocks in the snowflake channel, and we now aim to re-sum as many superfluous sums as possible to get to our final results \eqref{EqCB6C} and \eqref{EqCB6F}.

The sum over $l_2$ is straightforward since the terms containing $l_2$ are
\eqn{(-1)^{l_2}\binom{\sigma_2}{l_2}=\frac{(-\sigma_2)_{l_2}}{l_2!},}
and
\eqna{
&\binom{h_3-n_1-s_{11}-s_{22}-r_{25}-\bar{r}_2-l_2}{m_{13}}\\
&\qquad=(-1)^{m_{13}}\frac{(-h_3+n_1+s_{11}+s_{22}+r_{25}+\bar{r}_2)_{m_{13}}(-h_3+n_1+s_{11}+s_{22}+r_{25}+\bar{r}_2+m_{13})_{l_{2}}}{m_{13}!(-h_3+n_1+s_{11}+s_{22}+r_{25}+\bar{r}_2)_{l_{2}}}.
}
Thus, the sum over $l_2$ can be performed using \eqref{Eq2F1}.  Similarly, we can evaluate the sums over $l_3$, $l_4$, $\sigma_2$, $\sigma_3$, and $\sigma_4$, in that order,\footnote{To evaluate the sums over $l_4$ and $\sigma_4$, we must first change the variables such that $l_4\to l_4+m_{12}$ and $\sigma_4\to\sigma_4+m_{12}$, respectively.} using \eqref{Eq2F1}.  This results in
\eqna{
G_6&=\sum\binom{n_{11}}{s_{11}}\binom{n_{12}}{s_{12}}\binom{n_{22}}{s_{22}}\frac{(-h_5)_{m_3+m_{13}+m_{23}+m_{33}}(p_3-n_1+n_2+s_{12})_{\bar{r}_5}(\bar{p}_3+h_2+h_5)_{m_3+m_{12}}}{(\bar{p}_3+h_2)_{2m_3+m_{12}+m_{13}+m_{23}+m_{33}}(\bar{p}_3+h_2+1-d/2)_{m_3}}\\
&\phantom{=}\qquad\times(-1)^{s_{11}+s_{12}+s_{22}+m_{11}+m_{22}}\binom{r_{25}+s_{11}}{m_{11}}\binom{r_{24}+s_{22}}{m_{22}}(-s_{12})_{m_{12}+m_{33}+r_{45}+\bar{r}_4}\\
&\phantom{=}\qquad\times\frac{(-h_3+n_1+s_{11}+s_{22})_{m_{13}+r_{25}+\bar{r}_2}(p_2+\bar{h}_3-n_2-s_{11}-s_{22})_{m_{23}+r_{35}+\bar{r}_3}}{r_{23}!r_{24}!r_{25}!r_{34}!r_{35}!r_{45}!n_1!n_{11}!n_{12}!n_{22}!}\\
&\phantom{=}\qquad\times \frac{(u_1^6)^{n_1+r_{23}}(u_2^6)^{n_2+r_{45}}(u_3^6)^{m_3}\prod_{1\leq a\leq b\leq3}(1-v_{ab}^6)^{m_{ab}}}{m_{12}!m_{13}!m_{23}!m_{33}!}C_5F_5.
}

Since $r_{35}=m_3-r_{25}-r_{34}-r_{45}-r_{23}-r_{24}$, changing variables by $s_{12}\to s_{12}+m_{12}+m_{33}+r_{45}+r_{24}+r_{34}$ leads to
\eqna{
G_6&=\sum\binom{n_{11}}{s_{11}}\binom{n_{22}}{s_{22}}\frac{(-h_5)_{m_3+m_{13}+m_{23}+m_{33}}(\bar{p}_3+h_2+h_5)_{m_3+m_{12}}}{(\bar{p}_3+h_2)_{2m_3+m_{12}+m_{13}+m_{23}+m_{33}}(\bar{p}_3+h_2+1-d/2)_{m_3}}\\
&\phantom{=}\qquad\times(-1)^{s_{11}+s_{12}+s_{22}+m_{11}+m_{22}}\binom{r_{25}+s_{11}}{m_{11}}\binom{r_{24}+s_{22}}{m_{22}}\\
&\phantom{=}\qquad\times\frac{(p_3-n_1+n_2+s_{12}+m_{12}+m_{33}+r_{45}+r_{24}+r_{34})_{m_3-r_{23}-r_{24}-r_{34}}}{(n_{12}-s_{12}-m_{12}-m_{33}-r_{45}-r_{24}-r_{34})!(m_3-r_{25}-r_{34}-r_{45}-r_{23}-r_{24})!}\\
&\phantom{=}\qquad\times\frac{(-h_3+n_1+s_{11}+s_{22})_{m_{13}+r_{25}+r_{23}+r_{24}}(p_2+\bar{h}_3-n_2-s_{11}-s_{22})_{m_3+m_{23}-r_{24}-r_{25}-r_{45}}}{s_{12}!r_{23}!r_{24}!r_{25}!r_{34}!r_{45}!n_1!n_{11}!n_{22}!}\\
&\phantom{=}\qquad\times\frac{(u_1^6)^{n_1+r_{23}}(u_2^6)^{n_2+r_{45}}(u_3^6)^{m_3}\prod_{1\leq a\leq b\leq3}(1-v_{ab}^6)^{m_{ab}}}{m_{12}!m_{13}!m_{23}!m_{33}!}C_5F_5,
}
and we can now sum over $r_{34}$, giving [after using \eqref{EqBinom}]
\eqna{
G_6&=\sum\binom{n_{11}}{s_{11}}\binom{n_{22}}{s_{22}}\frac{(-h_5)_{m_3+m_{13}+m_{23}+m_{33}}(\bar{p}_3+h_2+h_5)_{m_3+m_{12}}}{(\bar{p}_3+h_2)_{2m_3+m_{12}+m_{13}+m_{23}+m_{33}}(\bar{p}_3+h_2+1-d/2)_{m_3}}\\
&\phantom{=}\qquad\times(-1)^{s_{11}+s_{12}+s_{22}+m_{11}+m_{22}}\binom{r_{25}}{k_{11}}\binom{s_{11}}{m_{11}-k_{11}}\binom{r_{24}}{k_{22}}\binom{s_{22}}{m_{22}-k_{22}}\\
&\phantom{=}\qquad\times\frac{(p_3-n_1+n_2+n_{12})_{m_3-r_{25}-r_{45}-r_{23}-r_{24}}(p_3-n_1+n_2+m_3+s_{12}+m_{12}+m_{33}-r_{23}-r_{25})_{r_{25}+r_{45}}}{(n_{12}-s_{12}-m_{12}-m_{33}-r_{45}-r_{24})!(m_3-r_{25}-r_{45}-r_{23}-r_{24})!}\\
&\phantom{=}\qquad\times\frac{(-h_3+n_1+s_{11}+s_{22})_{m_{13}+r_{25}+r_{23}+r_{24}}(p_2+\bar{h}_{3}-n_2-s_{11}-s_{22})_{m_3+m_{23}-r_{24}-r_{25}-r_{45}}}{s_{12}!r_{23}!r_{24}!r_{25}!r_{45}!n_1!n_{11}!n_{22}!}\\
&\phantom{=}\qquad\times\frac{(u_1^6)^{n_1+r_{23}}(u_2^6)^{n_2+r_{45}}(u_3^6)^{m_3}\prod_{1\leq a\leq b\leq3}(1-v_{ab}^6)^{m_{ab}}}{m_{12}!m_{13}!m_{23}!m_{33}!}C_5F_5.
}

We then change the variables by $r_{24}\to r_{24}+k_{22}$, and $r_{25}\to r_{25}+k_{11}$ and finally define $r_{24}=r-r_{25}$.  With these changes, we can compute the sums over $r_{25}$, $r$, $s_{12}$, and $n_{12}$, with
\eqna{
G_6&=\sum\binom{n_{11}}{s_{11}}\binom{n_{22}}{s_{22}}\frac{(-h_5)_{m_3+m_{13}+m_{23}+m_{33}}(\bar{p}_3+h_2+h_5)_{m_3+m_{12}}}{(\bar{p}_3+h_2)_{2m_3+m_{12}+m_{13}+m_{23}+m_{33}}(\bar{p}_3+h_2+1-d/2)_{m_3}}\\
&\phantom{=}\qquad\times(-1)^{s_{11}+s_{22}+m_{11}+m_{22}}\binom{s_{11}}{m_{11}-k_{11}}\binom{s_{22}}{m_{22}-k_{22}}\\
&\phantom{=}\qquad\times\frac{(p_2+h_2+n_1-n_2+m_{13}+m_{23}+2r_{23}+k_{11}+k_{22})_{m_3-r_{23}-r_{45}-k_{11}-k_{22}}}{(m_3-r_{45}-r_{23}-k_{11}-k_{22})!}\\
&\phantom{=}\qquad\times\frac{(-h_3+n_1+s_{11}+s_{22})_{m_{13}+r_{23}+k_{11}+k_{22}}(p_2+\bar{h}_3-n_2-s_{11}-s_{22})_{m_{23}+r_{23}}}{k_{11}!k_{22}!r_{23}!r_{45}!n_1!n_{11}!n_{22}!}\\
&\phantom{=}\qquad\times\frac{(-h_2)_{n_1+n_2+n_{11}+n_{22}}(-h_3)_{n_1+n_{11}+n_{22}}(p_2+h_2)_{n_1}(p_2+h_3)_{n_1}}{(p_2)_{2n_1+n_{11}+n_{22}}(p_2+1-d/2)_{n_1}}\\
&\phantom{=}\qquad\times\frac{(p_3-n_1)_{m_2+m_3+m_{12}+m_{33}-r_{23}}(p_3-h_2+h_4)_{m_2+n_{11}+k_{11}}(-h_4)_{m_2+m_{12}+m_{33}+n_{22}+k_{22}}}{(p_3-h_2)_{2m_2+m_{12}+m_{33}+n_{11}+n_{22}+k_{11}+k_{22}}(p_3-h_2+1-d/2)_{n_2}}\\
&\phantom{=}\qquad\times\frac{(u_1^6)^{n_1+r_{23}}(u_2^6)^{n_2+r_{45}}(u_3^6)^{m_3}\prod_{1\leq a\leq b\leq3}(1-v_{ab}^6)^{m_{ab}}}{m_{12}!m_{13}!m_{23}!m_{33}!}F_5,
}
where we made $C_5$ explicit.

After changing the variables by $s_{22}\to s_{22}+m_{22}-k_{22}$ and $s_{11}\to s_{11}+m_{11}-k_{11}$, and also defining $s_{22}=s-s_{11}$, we can evaluate the summation over $s_{11}$, leading to
\eqna{
G_6&=\sum\frac{(-1)^{k_{11}+k_{22}}(-n_{11}-n_{22}+m_{11}+m_{22}-k_{11}-k_{22})_{s}(-h_5)_{m_3+m_{13}+m_{23}+m_{33}}(\bar{p}_3+h_2+h_5)_{m_3+m_{12}}}{(\bar{p}_3+h_2)_{2m_3+m_{12}+m_{13}+m_{23}+m_{33}}(\bar{p}_3+h_2+1-d/2)_{m_3}}\\
&\phantom{=}\qquad\times\frac{(p_2+h_2+n_1-n_2+m_{13}+m_{23}+2r_{23}+k_{11}+k_{22})_{m_3-r_{23}-r_{45}-k_{11}-k_{22}}}{(m_3-r_{45}-r_{23}-k_{11}-k_{22})!}\\
&\phantom{=}\qquad\times\frac{(-h_3+n_1+s+m_{11}+m_{22}-k_{11}-k_{22})_{m_{13}+r_{23}+k_{11}+k_{22}}}{s!(n_{11}-m_{11}+k_{11})!(n_{22}-m_{22}+k_{22})!(m_{11}-k_{11})!(m_{22}-k_{22})!k_{11}!k_{22}!r_{23}!r_{45}!n_1!}\\
&\phantom{=}\qquad\times(p_2+\bar{h}_3-n_2-s-m_{11}-m_{22}+k_{11}+k_{22})_{m_{23}+r_{23}}\\
&\phantom{=}\qquad\times\frac{(-h_2)_{n_1+n_2+n_{11}+n_{22}}(-h_3)_{n_1+n_{11}+n_{22}}(p_2+h_2)_{n_1}(p_2+h_3)_{n_1}}{(p_2)_{2n_1+n_{11}+n_{22}}(p_2+1-d/2)_{n_1}}\\
&\phantom{=}\qquad\times\frac{(p_3-n_1)_{m_2+m_3+m_{12}+m_{33}-r_{23}}(p_3-h_2+h_4)_{m_2+n_{11}+k_{11}}(-h_4)_{m_2+m_{12}+m_{33}+n_{22}+k_{22}}}{(p_3-h_2)_{2m_2+m_{12}+m_{33}+n_{11}+n_{22}+k_{11}+k_{22}}(p_3-h_2+1-d/2)_{n_2}}\\
&\phantom{=}\qquad\times\frac{(u_1^6)^{n_1+r_{23}}(u_2^6)^{n_2+r_{45}}(u_3^6)^{m_3}\prod_{1\leq a\leq b\leq3}(1-v_{ab}^6)^{m_{ab}}}{m_{12}!m_{13}!m_{23}!m_{33}!}F_5.
}

Using the following identity,
\eqna{
&(p_2+\bar{h}_3-n_2-s-m_{11}-m_{22}+k_{11}+k_{22})_{m_{23}+r_{23}}\\
&\qquad=\sum_j\binom{m_{23}+r_{23}}{j}(p_2+\bar{h}_3-n_2-m_{11}-m_{22}+k_{11}+k_{22})_{j}(-s)_{m_{23}+r_{23}-j},
}
and changing the index of summation $s$ by $s\to s+m_{23}+r_{23}-j$, the sum over $s$ can be performed, leading to
\eqna{
G_6&=\sum\frac{(-1)^{k_{11}+k_{22}+m_{23}+r_{23}-j}(-h_5)_{m_3+m_{13}+m_{23}+m_{33}}(\bar{p}_3+h_2+h_5)_{m_3+m_{12}}}{(\bar{p}_3+h_2)_{2m_3+m_{12}+m_{13}+m_{23}+m_{33}}(\bar{p}_3+h_2+1-d/2)_{m_3}}\\
&\phantom{=}\qquad\times\frac{(p_2+h_2+n_1-n_2+m_{13}+m_{23}+2r_{23}+k_{11}+k_{22})_{m_3-r_{23}-r_{45}-k_{11}-k_{22}}}{(m_3-r_{45}-r_{23}-k_{11}-k_{22})!}\binom{m_{23}+r_{23}}{j}\\
&\phantom{=}\qquad\times\frac{(-h_3)_{m_1+m_{11}+m_{22}+m_{13}+m_{23}+r_{23}-j}(p_2+\bar{h}_3-n_2-m_{11}-m_{22}+k_{11}+k_{22})_j}{(n_{11}-m_{11}+k_{11})!(n_{22}-m_{22}+k_{22})!(m_{11}-k_{11})!(m_{22}-k_{22})!k_{11}!k_{22}!r_{23}!r_{45}!n_1!}\\
&\phantom{=}\qquad\times\frac{(-n_{11}-n_{22}-k_{11}-k_{22}+m_{11}+m_{22})_{m_{23}+r_{23}-j}(-h_2)_{n_1+n_2+n_{11}+n_{22}}(p_2+h_2)_{n_1}(p_2+h_3)_{n_1}}{(p_2)_{2n_1+n_{11}+n_{22}}(p_2+1-d/2)_{n_1}}\\
&\phantom{=}\qquad\times\frac{(p_3-n_1)_{m_2+m_3+m_{12}+m_{33}-r_{23}}(p_3-h_2+h_4)_{m_2+n_{11}+k_{11}}(-h_4)_{m_2+m_{12}+m_{33}+n_{22}+k_{22}}}{(p_3-h_2)_{2m_2+m_{12}+m_{33}+n_{11}+n_{22}+k_{11}+k_{22}}(p_3-h_2+1-d/2)_{n_2}}\\
&\phantom{=}\qquad\times(-m_{13}-r_{23}-k_{11}-k_{22})_{n_{11}+n_{22}+k_{11}+k_{22}-m_{11}-m_{22}-m_{23}-r_{23}+j}\\
&\phantom{=}\qquad\times\frac{(u_1^6)^{n_1+r_{23}}(u_2^6)^{n_2+r_{45}}(u_3^{6})^{m_3}\prod_{1\leq a\leq b\leq3}(1-v_{ab}^6)^{m_{ab}}}{m_{12}!m_{13}!m_{23}!m_{33}!}F_5.
}

With $n_{22}=n-n_{11}$, it is now possible to sum over $n_{11}$, $n$, $j$, $k_{11}$, and $k_{22}$, giving
\eqna{
G_6&=\sum\frac{(-h_5)_{m_3+m_{13}+m_{23}+m_{33}}(\bar{p}_3+h_2+h_5)_{m_3+m_{12}}}{(\bar{p}_3+h_2)_{2m_3+m_{12}+m_{13}+m_{23}+m_{33}}(\bar{p}_3+h_2+1-d/2)_{m_3}}\\
&\phantom{=}\qquad\times\frac{(p_2+h_2+m_1-t_1-m_2+t_2)_{m_3+m_{13}+m_{23}+t_1-t_2}}{(m_3-t_2-t_1)!}\frac{(-h_3)_{m_1+m_{11}+m_{22}+m_{13}}}{t_1!t_2!(m_1-t_1)!(m_2-t_2)!}\\
&\phantom{=}\qquad\times\frac{(-h_2+m_1+m_2-m_3)_{m_{11}+m_{22}}(-h_2)_{m_1-t_1+m_2-t_2}(p_2+h_2)_{m_1-t_1}(p_2+h_3)_{m_1+m_{23}}}{(p_2)_{2m_1+m_{11}+m_{22}+m_{13}+m_{23}}(p_2+1-d/2)_{m_1-t_1}}\\
&\phantom{=}\qquad\times\frac{(p_3-m_1+t_1)_{m_2+m_3+m_{12}+m_{33}-t_1}(p_3-h_2+h_4)_{m_2+m_{11}}(-h_4)_{m_2+m_{12}+m_{22}+m_{33}}}{(p_3-h_2)_{2m_2+m_{11}+m_{12}+m_{22}+m_{33}}(p_3-h_2+1-d/2)_{m_2-t_2}}\\
&\phantom{=}\qquad\times\frac{(u_1^6)^{m_1}(u_2^6)^{m_2}(u_3^6)^{m_3}\prod_{1\leq a\leq b\leq3}(1-v_{ab}^6)^{m_{ab}}}{m_{11}!m_{12}!m_{22}!m_{13}!m_{23}!m_{33}!}F_5,
}
after redefining
\eqn{r_{23}=t_1,\qquad r_{45}=t_2,\qquad n_1=m_1-t_1,\qquad n_2=m_2-t_2.}
We are thus left with three extra sums, two sums over $t_1$ and $t_2$, respectively, and one sum from $F_5$, which is given by
\eqn{F_5={}_3F_{2}\left[\begin{array}{c}-n_1,-n_2,-p_2+d/2-n_1\\1-p_2-h_2-n_1,p_3-n_1\end{array};1\right]={}_3F_{2}\left[\begin{array}{c}-m_1+t_1,-m_2+t_2,-p_2+d/2-m_1+t_1\\1-p_2-h_2-m_1+t_1,p_3-m_1+t_1\end{array};1\right].}

Using \eqref{Eq3F2} twice such that
\eqna{
&{}_3F_2\left[\begin{array}{c}-m_1+t_1,-m_2+t_2,-p_2+d/2-m_1+t_1\\1-p_2-h_2-m_1+t_1,p_3-m_1+t_1\end{array};1\right]\\
&\qquad=\frac{(p_2+h_2-m_2+t_2)_{m_1-t_1}}{(p_2+h_2)_{m_1-t_1}}{}_3F_2\left[\begin{array}{c}-m_1+t_1,-m_2+t_2,\bar{p}_3-d/2\\p_2+h_2-m_2+t_2,p_3-m_1+t_1\end{array};1\right]\\
&\qquad=\frac{(p_2+h_2-m_2+t_2)_{m_1-t_1}(-p_3+h_2+d/2-m_2+t_2)_{m_2-t_2}}{(p_2+h_2)_{m_1-t_1}(p_2+h_2-m_2+t_2)_{m_2-t_2}}\\
&\qquad\phantom{=}\qquad\times{}_3F_2\left[\begin{array}{c}p_3,-m_2+t_2,\bar{p}_3-d/2\\p_3-h_2-d/2+1,p_3-m_1+t_1\end{array};1\right],
}
and re-summing over $t_2$ using \eqref{Eq2F1} gives \eqref{EqCB6C} and \eqref{EqCB6F} (with $t_1\to t_2$ and the index of summation from the ${}_3F_2$-hypergeometric function chosen to be $t_1$).  This completes the proof of the scalar six-point conformal blocks in the snowflake channel.

To express $F_6$ in terms of Kamp\'e de F\'eriet functions \cite{exton1976multiple,srivastava1985multiple}, which are defined as
\eqn{F_{q,s,v}^{p,r,u}\left[\left.\begin{array}{c}\boldsymbol{a};\boldsymbol{c};\boldsymbol{f}\\\boldsymbol{b};\boldsymbol{d};\boldsymbol{g}\end{array}\right|x,y\right]=\sum_{m,n\geq0}\frac{(\boldsymbol{a})_{m+n}(\boldsymbol{c})_m(\boldsymbol{f})_n}{(\boldsymbol{b})_{m+n}(\boldsymbol{d})_m(\boldsymbol{g})_n}\frac{x^my^n}{m!n!},}[EqKdF]
where
\eqn{
\begin{gathered}
(\boldsymbol{a})_{m+n}=(a_1)_{m+n}\cdots(a_p)_{m+n},\qquad(\boldsymbol{b})_{m+n}=(b_1)_{m+n}\cdots(b_q)_{m+n},\\
(\boldsymbol{c})_m=(c_1)_m\cdots(c_r)_m,\qquad(\boldsymbol{d})_m=(d_1)_m\cdots(d_s)_m,\\
(\boldsymbol{f})_n=(f_1)_n\cdots(f_u)_n,\qquad(\boldsymbol{g})_n=(g_1)_n\cdots(g_v)_n,
\end{gathered}
}
we first rewrite \eqref{EqCB6F} as
\eqna{
F_6&=\frac{(-p_3+h_2+d/2-m_2)_{m_2}}{(p_3)_{-m_1}(-h_2+m_1)_{-m_3}}\sum_{t_1\geq0}\frac{(-m_2)_{t_1}(p_3)_{t_1}(\bar{p}_3-d/2)_{t_1}(-h_2+m_1)_{t_1}}{(-h_2+m_1-m_3)_{t_1}(p_3-m_1)_{t_1}(p_3-h_2+1-d/2)_{t_1}t_1!}\\
&\phantom{=}\qquad\times{}_3F_2\left[\begin{array}{c}-m_1,-m_3,-p_2+d/2-m_1\\1+h_2-m_1-t_1,p_3-m_1+t_1\end{array};1\right],
}
by re-summing over $t_2$.  Using \eqref{Eq3F2} on the ${}_3F_2$-hypergeometric function, expanding the resulting ${}_3F_2$-hypergeometric function, and combining Pochhammer symbols, we get to
\eqn{F_6=\frac{(-p_3+h_2+d/2-m_2)_{m_2}}{(p_3)_{-m_1}(-h_2)_{-m_3}}F_{2,1,0}^{1,3,2}\left[\left.\begin{array}{c}\bar{p}_3-d/2;-m_2,-h_2,p_3;-m_1,-m_3\\-h_2-m_3,p_3-m_1;p_3-h_2+1-d/2;-\end{array}\right|1,1\right],}
as stated in \eqref{EqCB6F}.

Finally, by using \eqref{Eq3F2} again on the ${}_3F_2$-hypergeometric function from the summation over $t_2$, it is possible to re-express $F_6$ in terms of a more symmetric Kamp\'e de F\'eriet function as
\eqna{
F_6&=\frac{1}{(p_3)_{-m_1}(-p_3+h_2+d/2)_{-m_2}(-\bar{p}_3-h_2+d/2)_{-m_3}}\\
&\phantom{=}\qquad\times F_{1,1,1}^{2,1,1}\left[\left.\begin{array}{c}\bar{p}_3-d/2,p_3;-m_2;-m_3\\p_3-m_1;p_3-h_2+1-d/2;\bar{p}_3+h_2+1-d/2\end{array}\right|1,1\right].
}
This form is more symmetric since it clearly shows that $F_6$ is made out of two intertwined ${}_3F_2$-hypergeometric functions, which is reminiscent of the comb channel result.

%%%%%%%%%%%%%%%%%%%%%%%%%%%%%%%%%%%%%%%%%%%%%%%%%%

\subsection{An Alternative Form}

Instead of starting from the scalar five-point correlation functions \eqref{EqCB5}, it is also possible to repeat the steps above starting from \eqref{EqCB5P}.  With the legs and the conformal cross-ratios
\eqn{
\begin{gathered}
L_6^*=\left(\frac{\eta_{13}}{\eta_{12}\eta_{23}}\right)^{\frac{\Delta_{i_2}}{2}}\left(\frac{\eta_{12}}{\eta_{13}\eta_{23}}\right)^{\frac{\Delta_{i_3}}{2}}\left(\frac{\eta_{15}}{\eta_{14}\eta_{45}}\right)^{\frac{\Delta_{i_4}}{2}}\left(\frac{\eta_{14}}{\eta_{15}\eta_{45}}\right)^{\frac{\Delta_{i_5}}{2}}\left(\frac{\eta_{12}}{\eta_{16}\eta_{26}}\right)^{\frac{\Delta_{i_6}}{2}}\left(\frac{\eta_{26}}{\eta_{12}\eta_{16}}\right)^{\frac{\Delta_{i_1}}{2}},\\
u_1^{*6}=\frac{\eta_{23}\eta_{15}\eta_{34}}{\eta_{13}\eta_{24}\eta_{35}},\qquad u_2^{*6}=\frac{\eta_{12}\eta_{34}\eta_{45}}{\eta_{14}\eta_{35}\eta_{24}},\qquad u_3^{*6}=\frac{\eta_{16}\eta_{24}\eta_{35}}{\eta_{26}\eta_{15}\eta_{34}},\\
v_{11}^{*6}=\frac{\eta_{12}\eta_{34}}{\eta_{13}\eta_{24}},\qquad v_{12}^{*6}=\frac{\eta_{34}\eta_{25}}{\eta_{24}\eta_{35}},\qquad v_{13}^{*6}=\frac{\eta_{12}\eta_{36}}{\eta_{13}\eta_{26}},\\
v_{22}^{*6}=\frac{\eta_{15}\eta_{34}}{\eta_{14}\eta_{35}},\qquad v_{23}^{*6}=\frac{\eta_{12}\eta_{46}}{\eta_{14}\eta_{26}},\qquad v_{33}^{*6}=\frac{\eta_{12}\eta_{56}}{\eta_{15}\eta_{26}},
\end{gathered}
}[EqCB6CRs]
the resulting scalar six-point conformal blocks in the snowflake channel are
\eqna{
C_6^*&=\frac{(p_3-h_2+h_4)_{m_2+m_{12}+m_{33}}}{(-h_2)_{m_1+m_2-m_3+m_{12}}}\frac{(-h_3)_{m_1+m_{12}}(p_3)_{-m_1+m_2+m_3+m_{23}+m_{33}}(p_2+h_3)_{m_1+m_{11}+m_{13}}}{(p_2)_{2m_1+m_{11}+m_{12}+m_{13}}(p_2+1-d/2)_{m_1}}\\
&\phantom{=}\qquad\times\frac{(p_3-h_2+h_4)_{m_2+m_{12}+m_{33}}(p_2+h_2)_{m_1-m_2+m_3+m_{13}}(-h_4)_{m_2+m_{22}+m_{23}}}{(p_3-h_2)_{2m_2+m_{12}+m_{22}+m_{23}+m_{33}}(p_3-h_2+1-d/2)_{m_2}}\\
&\phantom{=}\qquad\times\frac{(\bar{p}_3+h_2+h_5)_{m_3}(-h_2)_{m_1+m_2-m_3+m_{12}+m_{22}}(-h_5)_{m_3+m_{13}+m_{23}+m_{33}}}{(\bar{p}_3+h_2)_{2m_3+m_{13}+m_{23}+m_{33}}(\bar{p}_3+h_2+1-d/2)_{m_3}},
}[EqCB6Cs]
with $F_6^*$ given by \eqref{EqCB6F}.

We note here that although the alternative forms for $F_M$ are the same ($F_6^*=F_6$), the alternative forms for $C_M$ \eqref{EqCB6Cs} and \eqref{EqCB6C} are not the same function ($C_6^*\neq C_6$).  However, the two results must be equivalent as shown below.

First, it is easy to see that the conformal cross-ratios \eqref{EqCB6CRs} and \eqref{EqCB6CR} are related such that
\eqn{
\begin{gathered}
u_1^{*6}=\frac{u_1^6}{v_{22}^6},\qquad u_2^{*6}=\frac{u_2^6}{v_{12}^6v_{22}^6},\qquad u_3^{*6}=\frac{u_3^6v_{22}^6}{v_{13}^6},\\
v_{11}^{*6}=\frac{1}{v_{22}^6},\qquad v_{12}^{*6}=\frac{v_{11}^6}{v_{22}^6},\qquad v_{13}^{*6}=\frac{v_{23}^6}{v_{13}^6},\\
v_{22}^{*6}=\frac{1}{v_{12}^6},\qquad v_{23}^{*6}=\frac{v_{33}^6}{v_{12}^6v_{13}^6},\qquad v_{33}^{*6}=\frac{1}{v_{13}^6}.
\end{gathered}
}
Hence, since the scalar six-point correlation functions $I_6$ must be the same, we get the identity
\eqn{G_6=(v_{12}^6)^{h_4}(v_{13}^6)^{h_5}(v_{22}^6)^{h_2}G_6^*\equiv G_{v6}^*,}[EqCB6Id]
for the scalar six-point conformal blocks in the snowflake channel \eqref{EqG}.  To prove \eqref{EqCB6Id}, we re-express $G_{v6}^*$ in terms of the conformal cross-ratios for $G_6$, expand in the conformal cross-ratios of the latter, and evaluate the superfluous sums.

Using the fact that $F_6^*=F_6$, we first obtain
\eqna{
G_{v6}^*&=\sum\frac{(-h_2)_{m_1+m_2-m_3+m_{12}+m_{22}}(-h_5)_{m_3+m_{13}+m_{23}+m_{33}}(\bar{p}_3+h_2+h_5)_{m_3}}{(\bar{p}_3+h_2)_{2m_3+m_{13}+m_{23}+m_{33}}(\bar{p}_3+h_2+1-d/2)_{m_3}}\\
&\phantom{=}\qquad\times\frac{(p_2+h_3)_{m_1+m_{11}+m_{13}}(p_2+h_2)_{m_1-m_2+m_3+m_{13}}(p_3)_{m_2-m_1+m_3+m_{23}+m_{33}}}{(p_2)_{2m_1+m_{11}+m_{12}+m_{13}}(p_2+1-d/2)_{m_1}}\\
&\phantom{=}\qquad\times\frac{(-h_2)_{m_1+m_2-m_3+m_{11}+m_{12}}(p_3-h_2+h_4)_{m_2+m_{12}+m_{33}}}{(p_3-h_2)_{2m_2+m_{12}+m_{22}+m_{23}+m_{33}}(p_3-h_2+1-d/2)_{m_2}}\\
&\phantom{=}\qquad\times\frac{(-h_4)_{m_2+m_{22}+m_{23}}(-h_3)_{m_1+m_{12}}}{(-h_2)_{m_1+m_2-m_3+m_{12}}}\left[\prod_{1\leq a\leq3}\frac{(u_a^6)^{m_a}}{m_a!}\right](v_{12}^6)^{h_4-m_2}(v_{13}^6)^{h_5-m_3}(v_{22}^6)^{h_2+m_3-m_1-m_2}\\
&\phantom{=}\qquad\times\frac{(1-\frac{1}{v_{22}^6})^{m_{11}}}{m_{11}!}\frac{(1-\frac{v_{11}^6}{v_{22}^6})^{m_{12}}}{m_{12}!}\frac{(1-\frac{v_{23}^6}{v_{13}^6})^{m_{13}}}{m_{13}!}\frac{(1-\frac{1}{v_{12}^6})^{m_{22}}}{m_{22}!}\frac{(1-\frac{v_{33}^6}{v_{12}^6v_{13}^6})^{m_{23}}}{m_{23}!}\frac{(1-\frac{1}{v_{13}^6})^{m_{33}}}{m_{33}!}F_6,
}
in terms of the conformal cross-ratios \eqref{EqCB6CR}.  Expanding in terms of $u_a^6$ and $1-v_{ab}^6$, $G_{v6}^*$ becomes
\eqna{
G_{v6}^*&=\sum\frac{(-h_2)_{m_1+m_2-m_3+m_{12}+m_{22}}(-h_5)_{m_3+m_{13}+m_{23}+m_{33}}(\bar{p}_3+h_2+h_5)_{m_3}}{(\bar{p}_3+h_2)_{2m_3+m_{13}+m_{23}+m_{33}}(\bar{p}_3+h_2+1-d/2)_{m_3}}\\
&\phantom{=}\qquad\times\frac{(p_2+h_3)_{m_1+m_{11}+m_{13}}(p_2+h_2)_{m_1-m_2+m_3+m_{13}}(p_3)_{m_2-m_1+m_3+m_{23}+m_{33}}}{(p_2)_{2m_1+m_{11}+m_{12}+m_{13}}(p_2+1-d/2)_{m_1}}\\
&\phantom{=}\qquad\times\frac{(-h_2)_{m_1+m_2-m_3+m_{11}+m_{12}}(p_3-h_2+h_4)_{m_2+m_{12}+m_{33}}}{(p_3-h_2)_{2m_2+m_{12}+m_{22}+m_{23}+m_{33}}(p_3-h_2+1-d/2)_{m_2}}\\
&\phantom{=}\qquad\times\frac{(-h_4)_{m_2+m_{22}+m_{23}}(-h_3)_{m_1+m_{12}}}{(-h_2)_{m_1+m_2-m_3+m_{12}}}\prod_{1\leq a\leq b\leq 3}\binom{m_{ab}}{k_{ab}}\binom{k_{12}}{m^{\prime}_{11}}\binom{h_4-m_2-k_{22}-k_{23}}{m^{\prime}_{12}}\\
&\phantom{=}\qquad\times\binom{h_5-m_3-k_{13}-k_{23}-k_{33}}{m^{\prime}_{13}}\binom{h_2+m_3-m_1-m_2-k_{11}-k_{12}}{m^{\prime}_{22}}\binom{k_{13}}{m^{\prime}_{23}}\binom{k_{23}}{m^{\prime}_{33}}\\
&\phantom{=}\qquad\times(-1)^{\sum_{1\leq a\leq b\leq 3}(k_{ab}+m^{\prime}_{ab})}\prod_{1\leq a\leq3}\frac{(u_a^6)^{m_a}}{m_a!}\prod_{1\leq a\leq b\leq 3}\frac{(1-v_{ab}^6)^{m^{\prime}_{ab}}}{m_{ab}!}F_6.
}
We only need to evaluate the extra sums now to recover $C_6$ $\eqref{EqCB6C}$ and thus prove \eqref{EqCB6Id}.

We start by observing that the terms containing $k_{11}$ are given by
\eqn{(-1)^{k_{11}}\binom{m_{11}}{k_{11}}=\frac{(-m_{11})_{k_{11}}}{k_{11}!},}
and
\eqna{
&\binom{h_2+m_3-m_1-m_2-k_{11}-k_{12}}{m^{\prime}_{22}}\\
&\qquad=(-1)^{m^{\prime}_{22}}\frac{(-h_2+m_1+m_2-m_3+k_{12}+m^{\prime}_{22})_{k_{11}}(-h_2+m_1+m_2-m_3+k_{12})_{m^{\prime}_{22}}}{(-h_2+m_1+m_2-m_3+k_{12})_{k_{11}}m^{\prime}_{22}!}.
}
Thus the summation over $k_{11}$ can be done with the help of \eqref{Eq2F1}.  The same can be said for the summations over all of the $k_{ab}$, except for $k_{23}$.\footnote{To evaluate the sums over $k_{12}$ and $k_{13}$, we must first change the variables by $k_{12}\to k_{12}+m^{\prime}_{11}$ and $k_{13}\to k_{13}+m^{\prime}_{23}$, respectively.}  After these steps, $G_{v6}^*$ becomes
\eqna{
G_{v6}^*&=\sum\frac{(-h_2)_{m_1+m_2-m_3+m_{12}+m_{22}}(-h_5)_{m_3+m_{13}+m_{23}+m_{33}}(\bar{p}_3+h_2+h_5)_{m_3}}{(\bar{p}_3+h_2)_{2m_3+m_{13}+m_{23}+m_{33}}(\bar{p}_3+h_2+1-d/2)_{m_3}}\\
&\phantom{=}\qquad\times\frac{(p_2+h_3)_{m_1+m_{11}+m_{13}}(p_2+h_2)_{m_1-m_2+m_3+m_{13}}(p_3)_{m_2-m_1+m_3+m_{23}+m_{33}}}{(p_2)_{2m_1+m_{11}+m_{12}+m_{13}}(p_2+1-d/2)_{m_1}}\\
&\phantom{=}\qquad\times\frac{(-h_2)_{m_1+m_2-m_3+m^{\prime}_{11}+m^{\prime}_{22}}(p_3-h_2+h_4)_{m_2+m_{12}+m_{33}}}{(p_3-h_2)_{2m_2+m_{12}+m_{22}+m_{23}+m_{33}}(p_3-h_2+1-d/2)_{m_2}}\\
&\phantom{=}\qquad\times\frac{(-h_4)_{m_2+m_{22}+m_{23}}(-h_3)_{m_1+m_{12}}}{(-h_2)_{m_1+m_2-m_3+m_{12}}}\binom{m_{23}}{k_{23}}\binom{k_{23}}{m^{\prime}_{33}}\\
&\phantom{=}\qquad\times\frac{m_{12}!(-m^{\prime}_{22})_{m_{11}+m_{12}-m^{\prime}_{11}}(-h_5+m_3+m_{33}+m_{13}+k_{23})_{m^{\prime}_{13}+m^{\prime}_{23}-m_{13}-m_{33}}}{m^{\prime}_{11}!m^{\prime}_{12}!m^{\prime}_{13}m^{\prime}_{22}!m^{\prime}_{23}!}\\
&\phantom{=}\qquad\times\frac{m_{13}!(-m^{\prime}_{12})_{m_{22}}(-h_4+m_2+m_{22}+k_{23})_{m^{\prime}_{12}-m_{22}}(-m^{\prime}_{13})_{m_{13}+m_{33}-m^{\prime}_{23}}}{(m_{12}-m^{\prime}_{11})!(m_{13}-m^{\prime}_{23})!}\\
&\phantom{=}\qquad\times(-1)^{k_{23}+m^{\prime}_{33}}\prod_{1\leq a\leq3}\frac{(u_a^6)^{m_a}}{m_a!}\prod_{1\leq a\leq b\leq 3}\frac{(1-v_{ab}^6)^{m^{\prime}_{ab}}}{m_{ab}!}F_6.
}

We can now proceed with the summation over $m_{11}$, getting to
\eqna{
G_{v6}^*&=\sum\frac{(-h_2)_{m_1+m_2-m_3+m_{12}+m_{22}}(-h_5)_{m_3+m_{13}+m_{23}+m_{33}}(\bar{p}_3+h_2+h_5)_{m_3}}{(\bar{p}_3+h_2)_{2m_3+m_{13}+m_{23}+m_{33}}(\bar{p}_3+h_2+1-d/2)_{m_3}}\\
&\phantom{=}\qquad\times\frac{(p_2+h_3)_{m_1+m_{13}}(p_2+h_2)_{m_1-m_2+m_3+m_{13}}(p_3)_{m_2-m_1+m_3+m_{23}+m_{33}}}{(p_2)_{2m_1+m^{\prime}_{11}+m^{\prime}_{22}+m_{13}}(p_2+1-d/2)_{m_1}}\\
&\phantom{=}\qquad\times\frac{(-h_2)_{m_1+m_2-m_3+m^{\prime}_{11}+m^{\prime}_{22}}(p_3-h_2+h_4)_{m_2+m_{12}+m_{33}}}{(p_3-h_2)_{2m_2+m_{12}+m_{22}+m_{23}+m_{33}}(p_3-h_2+1-d/2)_{m_2}}\\
&\phantom{=}\qquad\times\frac{(-h_4)_{m_2+m_{22}+m_{23}}(-h_3)_{m_1+m^{\prime}_{11}+m^{\prime}_{22}}}{(-h_2)_{m_1+m_2-m_3+m_{12}}}\binom{m_{23}}{k_{23}}\binom{k_{23}}{m^{\prime}_{33}}\\
&\phantom{=}\qquad\times\frac{m_{11}!m_{12}!(-m^{\prime}_{22})_{m_{12}-m^{\prime}_{11}}(-h_5+m_3+m_{33}+m_{13}+k_{23})_{m^{\prime}_{13}+m^{\prime}_{23}-m_{13}-m_{33}}}{m^{\prime}_{11}!m^{\prime}_{12}!m^{\prime}_{13}m^{\prime}_{22}!m^{\prime}_{23}!}\\
&\phantom{=}\qquad\times\frac{m_{13}!(-m^{\prime}_{12})_{m_{22}}(-h_4+m_2+m_{22}+k_{23})_{m^{\prime}_{12}-m_{22}}(-m^{\prime}_{13})_{m_{13}+m_{33}-m^{\prime}_{23}}}{(m_{12}-m^{\prime}_{11})!(m_{13}-m^{\prime}_{23})!}\\
&\phantom{=}\qquad\times(-1)^{k_{23}+m^{\prime}_{33}}\prod_{1\leq a\leq3}\frac{(u_a^6)^{m_a}}{m_a!}\prod_{1\leq a\leq b\leq 3}\frac{(1-v_{ab}^6)^{m^{\prime}_{ab}}}{m_{ab}!}F_6.
}

To evaluate the summation over $k_{23}$, we first change the variable by $k_{23}\to k_{23}+m^{\prime}_{33}$.  Then, using the following identity
\eqna{
&(-h_4+m_2+m_{22}+m^{\prime}_{33}+k_{23})_{m^{\prime}_{12}-m_{22}}\\
&\qquad=\sum_j\binom{m^{\prime}_{12}-m_{22}}{j}(-h_4+m_2+m_{22}+m_{23})_{m^{\prime}_{12}-m_{22}-j}(-m_{23}+m^{\prime}_{33}+k_{23})_j,
}
we can re-sum over $k_{23}$, leading to
\eqna{
G_{v6}^*&=\sum\frac{(-h_2)_{m_1+m_2-m_3+m_{12}+m_{22}}(\bar{p}_3+h_2+h_5)_{m_3}}{(\bar{p}_3+h_2)_{2m_3+m_{13}+m_{23}+m_{33}}(\bar{p}_3+h_2+1-d/2)_{m_3}}\\
&\phantom{=}\qquad\times\frac{(p_2+h_3)_{m_1+m_{13}}(p_2+h_2)_{m_1-m_2+m_3+m_{13}}(p_3)_{m_2-m_1+m_3+m_{23}+m_{33}}}{(p_2)_{2m_1+m^{\prime}_{11}+m^{\prime}_{22}+m_{13}}(p_2+1-d/2)_{m_1}}\\
&\phantom{=}\qquad\times\frac{(-h_2)_{m_1+m_2-m_3+m^{\prime}_{11}+m^{\prime}_{22}}(p_3-h_2+h_4)_{m_2+m_{12}+m_{33}}}{(p_3-h_2)_{2m_2+m_{12}+m_{22}+m_{23}+m_{33}}(p_3-h_2+1-d/2)_{m_2}}\\
&\phantom{=}\qquad\times\frac{(-h_4)_{m_2+m^{\prime}_{12}+m_{23}-j}(-h_3)_{m_1+m^{\prime}_{11}+m^{\prime}_{22}}}{(-h_5+m_3+m_{13}+m_{23}+m_{33})_{-j}(-h_2)_{m_1+m_2-m_3+m_{12}}}\\
&\phantom{=}\qquad\times\frac{m_{11}!m_{12}!(-m^{\prime}_{22})_{m_{12}-m^{\prime}_{11}}(-h_5)_{m_3+m^{\prime}_{13}+m^{\prime}_{23}+m^{\prime}_{33}}}{m^{\prime}_{11}!m^{\prime}_{12}!m^{\prime}_{13}m^{\prime}_{22}!m^{\prime}_{23}!m^{\prime}_{33}!}\\
&\phantom{=}\qquad\times\frac{m_{13}!m_{23}!(-m^{\prime}_{12})_{m_{22}+j}(-m^{\prime}_{13}-m^{\prime}_{23}+m_{13}+m_{33})_{m_{23}-m^{\prime}_{33}-j}(-m^{\prime}_{13})_{m_{13}+m_{33}-m^{\prime}_{23}}}{j!(m_{12}-m^{\prime}_{11})!(m_{23}-m^{\prime}_{33}-j)!(m_{13}-m^{\prime}_{23})!}\\
&\phantom{=}\qquad\times\prod_{1\leq a\leq3}\frac{(u_a^6)^{m_a}}{m_a!}\prod_{1\leq a\leq b\leq 3}\frac{(1-v_{ab}^6)^{m^{\prime}_{ab}}}{m_{ab}!}F_6.
}

At this point, we sum over $m_{22}$ and $m_{12}$ after we change the variable such that $m_{12}\to m_{12}+m^{\prime}_{11}$.  This gives
\eqna{
G_{v6}^*&=\sum\frac{(\bar{p}_3+h_2+h_5)_{m_3}}{(\bar{p}_3+h_2)_{2m_3+m_{13}+m_{23}+m_{33}}(\bar{p}_3+h_2+1-d/2)_{m_3}}\\
&\phantom{=}\qquad\times\frac{(p_2+h_3)_{m_1+m_{13}}(p_2+h_2)_{m_1-m_2+m_3+m_{13}}(p_3)_{m_2-m_1+m_3+m^{\prime}_{12}+m_{23}+m_{33}-j}}{(p_2)_{2m_1+m^{\prime}_{11}+m^{\prime}_{22}+m_{13}}(p_2+1-d/2)_{m_1}}\\
&\phantom{=}\qquad\times\frac{(-h_2)_{m_1+m_2-m_3+m^{\prime}_{11}+m^{\prime}_{22}}(p_3-h_2+h_4)_{m_2+m^{\prime}_{11}+m_{33}}}{(p_3-h_2)_{2m_2+m^{\prime}_{11}+m^{\prime}_{12}+m^{\prime}_{22}+m_{23}+m_{33}-j}(p_3-h_2+1-d/2)_{m_2}}\\
&\phantom{=}\qquad\times\frac{(-h_5)_{m_3+m^{\prime}_{13}+m^{\prime}_{23}+m^{\prime}_{33}}(-h_4)_{m_2+m^{\prime}_{12}+m^{\prime}_{22}+m_{23}-j}(-h_3)_{m_1+m^{\prime}_{11}+m^{\prime}_{22}}}{m_{33}!(-h_5+m_3+m_{13}+m_{23}+m_{33})_{-j}}\\
&\phantom{=}\qquad\times\frac{(-m^{\prime}_{12})_{j}(-m^{\prime}_{13}-m^{\prime}_{23}+m_{13}+m_{33})_{m_{23}-m^{\prime}_{33}-j}(-m^{\prime}_{13})_{m_{13}+m_{33}-m^{\prime}_{23}}}{j!(m_{23}-m^{\prime}_{33}-j)!(m_{13}-m^{\prime}_{23})!}\\
&\phantom{=}\qquad\times\prod_{1\leq a\leq3}\frac{(u_a^6)^{m_a}}{m_a!}\prod_{1\leq a\leq b\leq 3}\frac{(1-v_{ab}^6)^{m^{\prime}_{ab}}}{m^{\prime}_{ab}!}F_6.
}

Changing the variable by $m_{23}\to m_{23}+m^{\prime}_{33}+j$ and then redefining $m_{23}=m-m_{33}$, we can finally perform the re-summations over $m_{33}$, $j$, $m$, and $m_{13}$ (where we first make the substitution $m_{13}\to m_{13}+m^{\prime}_{23}$).  Redefining $m^{\prime}_{ab}$ by $m_{ab}$, we are thus left with $G_{v6}^*=G_6$, completing the proof.

%%%%%%%%%%%%%%%%%%%%%%%%%%%%%%%%%%%%%%%%%%%%%%%%%%
%%%%%%%%%%%%%%%%%%%%%%%%%%%%%%%%%%%%%%%%%%%%%%%%%%
	
\section{Symmetry Properties}\label{SAppSym}

This appendix presents the proofs of the symmetry properties of the scalar six-point conformal blocks in the snowflake channel, which are generated by rotations, reflections and dendrite permutations as in \eqref{EqSymRot}, \eqref{EqSymRef}, and \eqref{EqSymPerm}, respectively.  Collectively, these symmetries generate the $(\mathbb{Z}_2)^3\rtimes D_3$ group discussed in Section~\ref{SecSymmetry}.

%%%%%%%%%%%%%%%%%%%%%%%%%%%%%%%%%%%%%%%%%%%%%%%%%%

\subsection{Rotations of the Triangle}

To prove invariance under the rotation generator $R$ \eqref{EqSymRot}, it is only necessary to check that \eqref{EqCB6F} verifies
\eqn{F_{6|\text{snowflake}}^{(d,h_2,h_3,h_4,h_5;p_2,p_3,p_4,p_5,p_6)}(m_1,m_2,m_3)=F_{6|\text{snowflake}}^{(d,-p_3,h_4,h_5,h_3;p_3-h_2,p_2+h_2,p_5,p_6,p_4)}(m_2,m_3,m_1).}[EqSymRotF]
To simplify the notation, we denote \eqref{EqSymRotF} as $F_6=F_{6R}$.

First, we rewrite \eqref{EqCB6F} as
\eqna{
F_6&=\frac{(-p_3+h_2+d/2-m_2)_{m_2}}{(p_3)_{-m_1}(-h_2+m_1+m_2)_{-m_3}}\sum_{t_1,t_2\geq0}\frac{(-m_1)_{t_2}(-m_2)_{t_1}(-m_3)_{t_2}}{(p_3-h_2-d/2+1)_{t_1}}\\
&\phantom{=}\qquad\times\frac{(p_3)_{t_1}(\bar{p}_3-d/2)_{t_1}(-p_2+d/2-m_1)_{t_2}(1+h_2-m_1+m_3-m_2)_{m_2-t_1}}{(1+h_2-m_1-m_2)_{m_2-t_1+t_2}(p_3-m_1)_{t_1+t_2}t_1!t_2!},
}
and then use the identity
\eqna{
\frac{(1+h_2-m_1+m_3-m_2)_{m_2-t_1}}{(1+h_2-m_1-m_2)_{m_2-t_1+t_2}}&=\frac{(1+h_2-m_1+m_3-m_2)_{m_2-t_1}}{(1+h_2-m_1-m_2+t_2)_{m_2-t}(1+h_2-m_1-m_2)_{t_2}}\\
&=\sum_{t_3\geq0}\frac{(-m_3+t_2)_{t_3}(-m_2+t_1)_{t_3}}{(1+h_2-m_1-m_2+t_2)_{t_3}(1+h_2-m_1-m_2)_{t_2}t_3!},
}
to express $F_6$ as
\eqna{
F_6&=\frac{(-p_3+h_2+d/2-m_2)_{m_2}}{(p_3)_{-m_1}(-h_2+m_1+m_2)_{-m_3}}\sum_{t_1,t_2,t_3\geq0}\frac{(-m_1)_{t_2}(-m_2)_{t_1+t_3}(-m_3)_{t_2+t_3}}{(p_3-h_2-d/2+1)_{t_1}}\\
&\phantom{=}\qquad\times\frac{(p_3)_{t_1}(\bar{p}_3-d/2)_{t_1}(-p_2+d/2-m_1)_{t_2}}{(1+h_2-m_1-m_2)_{t_2+t_3}(p_3-m_1)_{t_1+t_2}t_1!t_2!t_3!}.
}
We now modify the sum over $t_1$ with the help of \eqref{Eq3F2},
\eqna{
{}_3F_2\left[\begin{array}{c}-m_2+t_3,\bar{p}_3-d/2,p_3\\p_3-h_2-d/2+1,p_3-m_1+t_2\end{array};1\right]&=\frac{(p_2+h_2-m_2+t_3)_{m_2-t_3}}{(-p_3+h_2+d/2-m_2+t_3)_{m_2-t_3}}\\
&\phantom{=}\qquad\times{}_3F_2\left[\begin{array}{c}-m_2+t_3,\bar{p}_3-d/2,-m_1+t_2\\p_2+h_2-m_2+t_3,p_3-m_1+t_2\end{array};1\right],
}
to get
\eqna{
F_6&=\sum_{t_2,t_3\geq0}\frac{(-m_1)_{t_2}(-m_2)_{t_3}(-m_3)_{t_2+t_3}}{(p_3)_{-m_1}(p_2+h_2)_{-m_2}(-h_2+m_1+m_2)_{-m_3}t_2!t_3!}\\
&\phantom{=}\qquad\times\frac{(-p_2+d/2-m_1)_{t_2}(-p_3+h_2+d/2-m_2)_{t_3}}{(p_3-m_1)_{t_2}(p_2+h_2-m_2)_{t_3}(1+h_2-m_1-m_2)_{t_2+t_3}}\\
&\phantom{=}\qquad\times{}_3F_2\left[\begin{array}{c}-m_1+t_2,\bar{p}_3-d/2,-m_2+t_3\\p_3-m_1+t_2,p_2+h_2-m_2+t_3\end{array};1\right],
}
or
\eqna{
F_6&=\sum_{t_1,t_2,t_3\geq0}\frac{(-m_1)_{t_1+t_2}(-m_2)_{t_1+t_3}(-m_3)_{t_2+t_3}}{(p_3)_{-m_1}(p_2+h_2)_{-m_2}(-h_2+m_1+m_2)_{-m_3}t_1!t_2!t_3!}\\
&\phantom{=}\qquad\times\frac{(\bar{p}_3-d/2)_{t_1}(-p_2+d/2-m_1)_{t_2}(-p_3+h_2+d/2-m_2)_{t_3}}{(p_3-m_1)_{t_1+t_2}(p_2+h_2-m_2)_{t_1+t_3}(1+h_2-m_1-m_2)_{t_2+t_3}}.
}[EqF6]
This is the simplest form to check the invariance of $F_6$ under the reflection generator \eqref{EqSymRef}.  We can use \eqref{Eq3F2} again to obtain
\eqna{
F_6&=\sum_{t_2,t_3\geq0}\frac{(-m_1)_{t_2}(-m_2)_{t_3}(-m_3)_{t_2+t_3}}{(-p_2+d/2)_{-m_1}(p_2+h_2)_{-m_2}(-h_2+m_1+m_2)_{-m_3}t_2!t_3!}\\
&\phantom{=}\qquad\times\frac{(-p_3+h_2+d/2-m_2)_{t_3}}{(p_2+h_2-m_2)_{t_3}(1+h_2-m_1-m_2)_{t_2+t_3}}{}_3F_2\left[\begin{array}{c}-m_1+t_2,\bar{p}_3-d/2,p_2+h_2\\p_2+1-d/2,p_2+h_2-m_2+t_3\end{array};1\right].
}
It is now possible to expand the ${}_3F_2$-hypergeometric function as a summation over $t_1$ and evaluate the sum over $t_2$ with the help of \eqref{Eq2F1}, leading to
\eqna{
F_6&=\sum_{t_1,t_3\geq0}\frac{(-m_1)_{t_1}(-m_2)_{t_3}(-m_3)_{t_3}}{(-p_2+d/2)_{-m_1}(p_2+h_2)_{-m_2}(-h_2+m_1+m_2)_{-m_3}t_1!t_3!}\\
&\phantom{=}\qquad\times\frac{(1+h_2-m_1-m_2+m_3)_{m_1-t_1}(\bar{p}_3-d/2)_{t_1}(p_2+h_2)_{t_1}(-p_3+h_2+d/2-m_2)_{t_3}}{(p_2+1-d/2)_{t_1}(p_2+h_2-m_2)_{t_1+t_3}(1+h_2-m_1-m_2)_{m_1-t_1+t_3}}.
}
Expressing the sum over $t_3$ as a ${}_3F_2$-hypergeometric function, we get
\eqna{
F_6&=\sum_{t_1\geq0}\frac{(-h_2+m_2-m_3+t_1)_{m_3}}{(-p_2+d/2)_{-m_1}(p_2+h_2)_{-m_2}}\frac{(-m_1)_{t_1}(\bar{p}_3-d/2)_{t_1}(p_2+h_2)_{t_1}}{(p_2+1-d/2)_{t_1}(p_2+h_2-m_2)_{t_1}t_1!}\\
&\phantom{=}\qquad\times{}_3F_2\left[\begin{array}{c}-m_2,-m_3,-p_3+h_2+d/2-m_2\\1+h_2-m_2-t_1,p_2+h_2-m_2+t_1\end{array};1\right].
}
Using \eqref{Eq3F2} once more leads to
\eqna{
F_6&=\sum_{t_1\geq0}\frac{(-h_2-m_3+t_1)_{m_3}}{(-p_2+d/2)_{-m_1}(p_2+h_2)_{-m_2}}\frac{(-m_1)_{t_1}(\bar{p}_3-d/2)_{t_1}(p_2+h_2)_{t_1}}{(p_2+1-d/2)_{t_1}(p_2+h_2-m_2)_{t_1}t_1!}\\
&\phantom{=}\qquad\times{}_3F_2\left[\begin{array}{c}-m_2,-m_3,\bar{p}_3-d/2+t_1\\-h_2-m_3+t_1,p_2+h_2-m_2+t_1\end{array};1\right]\\
&=\sum_{t_1\geq0}\frac{(-h_2-m_3+t_1)_{m_3}}{(-p_2+d/2)_{-m_1}(p_2+h_2)_{-m_2}}\frac{(-m_1)_{t_1}(\bar{p}_3-d/2)_{t_1}(p_2+h_2)_{t_1}}{(p_2+1-d/2)_{t_1}(p_2+h_2-m_2)_{t_1}t_1!}\\
&\phantom{=}\qquad\times{}_3F_2\left[\begin{array}{c}-m_2,-m_3,\bar{p}_3-d/2+t_1\\p_2+h_2-m_2+t_1,-h_2-m_3+t_1\end{array};1\right],
}
where in the last equality we simply changed the order of the two bottom parameters of the ${}_3F_2$-hypergeometric function.  With this change, we can finally re-use \eqref{Eq3F2} to get
\eqna{
F_6&=\sum_{t_1\geq0}\frac{(p_2+h_2-m_2+m_3+t_1)_{m_2}(-h_2-m_3+t_1)_{m_3}}{(-p_2+d/2)_{-m_1}(p_2+h_2)_{-m_2}}\frac{(-m_1)_{t_1}(\bar{p}_3-d/2)_{t_1}(p_2+h_2)_{t_1}}{(p_2+1-d/2)_{t_1}(p_2+h_2-m_2)_{m_2+t_1}t_1!}\\
&\phantom{=}\qquad\times{}_3F_2\left[\begin{array}{c}-m_2,-m_3,-\bar{p}_3-h_2+d/2-m_3\\1-p_2-h_2-m_3-t_1,-h_2-m_3+t_1\end{array};1\right],
}
which is nothing else than $F_6=F_{6R}$, completing the proof of \eqref{EqSymRotF} and \eqref{EqSymRot}.

%%%%%%%%%%%%%%%%%%%%%%%%%%%%%%%%%%%%%%%%%%%%%%%%%%

\subsection{Reflections of the Triangle}

Invariance under the reflection generator $S$ implies the identity \eqref{EqSymRef}, which we rewrite as $G_6=G_{6S}$ to simplify the notation.  Before proceeding, we observe that, under the reflection generator, $F_6\to F_6$ from the definition \eqref{EqF6}.  Expressing $G_{6S}$ in terms of the original conformal cross-ratios and expanding, we get
\eqna{
G_{6S}&=\sum\frac{(p_3-h_2+h_4)_{m^{\prime}_{2}+m_{23}}(p_2+h_3)_{m^{\prime}_{1}+m_{11}}(\bar{p}_3+h_2+h_5)_{m^{\prime}_3+m_{12}}}{(\bar{p}_3+h_2)_{2m^{\prime}_3+m_{12}+m_{13}+m_{23}+m_{33}}(\bar{p}_3+h_2+1-d/2)_{m^{\prime}_3}}\\
&\phantom{=}\qquad\times\frac{(p_2+h_2)_{m^{\prime}_{1}+m^{\prime}_3-m^{\prime}_2+m_{12}+m_{33}}(p_3)_{m^{\prime}_2-m^{\prime}_1+m^{\prime}_3+m_{13}+m_{23}}(-h_2)_{m^{\prime}_1+m^{\prime}_2-m^{\prime}_3+m_{11}+m_{22}}}{(p_3-h_2)_{2m^{\prime}_{2}+m_{11}+m_{22}+m_{13}+m_{23}}(p_3-h_2+1-d/2)_{m^{\prime}_2}}\\
&\phantom{=}\qquad\times\frac{(-h_4)_{m^{\prime}_2+m_{11}+m_{22}+m_{13}}(-h_3)_{m^{\prime}_{1}+m_{12}+m_{22}+m_{33}}(-h_5)_{m^{\prime}_3+m_{13}+m_{23}+m_{33}}}{(p_2)_{2m^{\prime}_{1}+m_{11}+m_{12}+m_{22}+m_{33}}(p_2+1-d/2)_{m^{\prime}_1}}\\
&\phantom{=}\qquad\times\binom{m_{11}}{k_{11}}\binom{m_{12}}{k_{12}}\binom{m_{13}}{k_{13}}\binom{m_{22}}{k_{22}}\binom{m_{23}}{k_{23}}\binom{m_{33}}{k_{33}}\binom{h_3-m^{\prime}_1-k_{12}-k_{22}-k_{33}}{m^{\prime}_{11}}\\
&\phantom{=}\qquad\times\binom{h_4-m^{\prime}_{2}-k_{11}-k_{13}-k_{22}}{m^{\prime}_{12}}\binom{k_{33}}{m^{\prime}_{13}}\binom{k_{22}}{m^{\prime}_{22}}\binom{h_5-m^{\prime}_{3}-k_{13}-k_{23}-k_{33}}{m^{\prime}_{23}}\binom{k_{13}}{m^{\prime}_{33}}\\
&\phantom{=}\qquad\times(-1)^{\sum_{1\leq a\leq b\leq 3}(k_{ab}+m^{\prime}_{ab})}F_6\prod_{1\leq a\leq3}\frac{(u_a^6)^{m^{\prime}_a}}{m^{\prime}_a!}\prod_{1\leq a\leq b\leq3}\frac{(1-v_{ab}^6)^{m^{\prime}_{ab}}}{m_{ab}!},
}
where we must re-sum all the extra sums.

We start by evaluating the summations over $k_{11}$, $k_{12}$, and $k_{23}$ using \eqref{Eq2F1}, getting to
\eqna{
G_{6S}&=\sum\frac{(p_3-h_2+h_4)_{m^{\prime}_{2}+m_{23}}(p_2+h_3)_{m^{\prime}_{1}+m_{11}}(\bar{p}_3+h_2+h_5)_{m^{\prime}_3+m_{12}}}{(\bar{p}_3+h_2)_{2m^{\prime}_3+m_{12}+m_{13}+m_{23}+m_{33}}(\bar{p}_3+h_2+1-d/2)_{m^{\prime}_3}}\\
&\phantom{=}\qquad\times\frac{(p_2+h_2)_{m^{\prime}_{1}+m^{\prime}_3-m^{\prime}_2+m_{12}+m_{33}}(p_3)_{m^{\prime}_2-m^{\prime}_1+m^{\prime}_3+m_{13}+m_{23}}(-h_2)_{m^{\prime}_1+m^{\prime}_2-m^{\prime}_3+m_{11}+m_{22}}}{(p_3-h_2)_{2m^{\prime}_{2}+m_{11}+m_{22}+m_{13}+m_{23}}(p_3-h_2+1-d/2)_{m^{\prime}_2}}\\
&\phantom{=}\qquad\times\frac{(-h_4)_{m^{\prime}_2+m_{11}+m_{22}+m_{13}}(-h_3)_{m^{\prime}_{1}+m_{12}+m_{22}+m_{33}}(-h_5)_{m^{\prime}_3+m_{13}+m_{23}+m_{33}}}{(p_2)_{2m^{\prime}_{1}+m_{11}+m_{12}+m_{22}+m_{33}}(p_2+1-d/2)_{m^{\prime}_1}}\\
&\phantom{=}\qquad\times\binom{m_{13}}{k_{13}}\binom{m_{22}}{k_{22}}\binom{m_{33}}{k_{33}}\frac{(-m^{\prime}_{11})_{m_{12}}(-h_3+m^{\prime}_1+m_{12}+k_{22}+k_{33})_{m^{\prime}_{11}-m_{12}}}{m^{\prime}_{11}!}\\
&\phantom{=}\qquad\times\frac{(-m^{\prime}_{12})_{m_{11}}(-h_4+m^{\prime}_{2}+m_{11}+k_{13}+k_{22})_{m^{\prime}_{12}-m_{11}}}{m^{\prime}_{12}!}\binom{k_{33}}{m^{\prime}_{13}}\binom{k_{22}}{m^{\prime}_{22}}\binom{k_{13}}{m^{\prime}_{33}}\\
&\phantom{=}\qquad\times\frac{(-m^{\prime}_{23})_{m_{23}}(-h_5+m^{\prime}_{3}+m_{23}+k_{13}+k_{33})_{m^{\prime}_{23}-m_{23}}}{m^{\prime}_{23}!}(-1)^{k_{13}+k_{22}+k_{33}+m^{\prime}_{13}+m^{\prime}_{22}+m^{\prime}_{33}}\\
&\phantom{=}\qquad\times F_6\prod_{1\leq a\leq3}\frac{(u_a^6)^{m^{\prime}_a}}{m^{\prime}_a!}\prod_{1\leq a\leq b\leq3}\frac{(1-v_{ab}^6)^{m^{\prime}_{ab}}}{m_{ab}!}.
}

We then change the variables by $k_{13}\to k_{13}+m^{\prime}_{33}$, $k_{22}\to k_{22}+m^{\prime}_{22}$, and $k_{33}\to k_{33}+m^{\prime}_{13}$, and use the identity
\eqna{
&(-h_4+m^{\prime}_{2}+m^{\prime}_{22}+m^{\prime}_{33}+m_{11}+k_{13}+k_{22})_{m^{\prime}_{12}-m_{11}}\\
&\qquad=\sum_{j_1}\binom{m^{\prime}_{12}-m_{11}}{j_1}(-h_4+m^{\prime}_{2}+m^{\prime}_{22}+m_{11}+m_{13}+k_{22})_{m^{\prime}_{12}-m_{11}-j_1}(-m_{13}+m^{\prime}_{33}+k_{13})_{j_1},
}
to re-sum over $k_{13}$, leading to
\eqna{
G_{6S}&=\sum\frac{(p_3-h_2+h_4)_{m^{\prime}_{2}+m_{23}}(p_2+h_3)_{m^{\prime}_{1}+m_{11}}(\bar{p}_3+h_2+h_5)_{m^{\prime}_3+m_{12}}}{(\bar{p}_3+h_2)_{2m^{\prime}_3+m_{12}+m_{13}+m_{23}+m_{33}}(\bar{p}_3+h_2+1-d/2)_{m^{\prime}_3}}\\
&\phantom{=}\qquad\times\frac{(p_2+h_2)_{m^{\prime}_{1}+m^{\prime}_3-m^{\prime}_2+m_{12}+m_{33}}(p_3)_{m^{\prime}_2-m^{\prime}_1+m^{\prime}_3+m_{13}+m_{23}}(-h_2)_{m^{\prime}_1+m^{\prime}_2-m^{\prime}_3+m_{11}+m_{22}}}{(p_3-h_2)_{2m^{\prime}_{2}+m_{11}+m_{22}+m_{13}+m_{23}}(p_3-h_2+1-d/2)_{m^{\prime}_2}}\\
&\phantom{=}\qquad\times\frac{(-h_4)_{m^{\prime}_2+m_{11}+m_{22}+m_{13}}(-h_3)_{m^{\prime}_{1}+m_{12}+m_{22}+m_{33}}(-h_5)_{m^{\prime}_3+m_{13}+m_{23}+m_{33}}}{(p_2)_{2m^{\prime}_{1}+m_{11}+m_{12}+m_{22}+m_{33}}(p_2+1-d/2)_{m^{\prime}_1}}\\
&\phantom{=}\qquad\times\frac{(-m^{\prime}_{11})_{m_{12}}(-h_3+m^{\prime}_1+m^{\prime}_{13}+m^{\prime}_{22}+m_{12}+k_{22}+k_{33})_{m^{\prime}_{11}-m_{12}}}{m_{11}!(m_{13}-m^{\prime}_{33}-j_{1})!}\\
&\phantom{=}\qquad\times\frac{(-m^{\prime}_{12})_{m_{11}+j_1}(-h_4+m^{\prime}_{2}+m^{\prime}_{22}+m_{11}+m_{13}+k_{22})_{m^{\prime}_{12}-m_{11}-j_1}}{k_{33}!m_{12}!(m_{22}-m^{\prime}_{22}-k_{22})!}\\
&\phantom{=}\qquad\times\frac{(-m^{\prime}_{23})_{m_{13}+m_{23}-m^{\prime}_{33}-j_1}(-h_5+m^{\prime}_{3}+m^{\prime}_{13}+m_{13}+m_{23}+k_{33}-j_1)_{m^{\prime}_{23}+m^{\prime}_{33}-m_{13}-m_{23}+j_1}}{j_1!k_{22}!m_{23}!(m_{33}-m^{\prime}_{13}-k_{33})!}\\
&\phantom{=}\qquad\times(-1)^{k_{22}+k_{33}}F_6\prod_{1\leq a\leq3}\frac{(u_a^6)^{m^{\prime}_a}}{m^{\prime}_a!}\prod_{1\leq a\leq b\leq3}\frac{(1-v_{ab}^6)^{m^{\prime}_{ab}}}{m^{\prime}_{ab}!}.
}

Similarly, using the identity
\eqna{
&(-h_3+m^{\prime}_1+m^{\prime}_{13}+m^{\prime}_{22}+m_{12}+k_{22}+k_{33})_{m^{\prime}_{11}-m_{12}}\\
&\qquad=\sum_{j_2}\binom{m^{\prime}_{11}-m_{12}}{j_2}(-h_3+m^{\prime}_1+m^{\prime}_{22}+m_{12}+m_{33}+k_{22})_{m^{\prime}_{11}-m_{12}-j_2}(-m_{33}+m^{\prime}_{13}+k_{33})_{j_2},
}
we can compute the sum over $k_{33}$ and we get
\eqna{
G_{6S}&=\sum\frac{(p_3-h_2+h_4)_{m^{\prime}_{2}+m_{23}}(p_2+h_3)_{m^{\prime}_{1}+m_{11}}(\bar{p}_3+h_2+h_5)_{m^{\prime}_3+m_{12}}}{(\bar{p}_3+h_2)_{2m^{\prime}_3+m_{12}+m_{13}+m_{23}+m_{33}}(\bar{p}_3+h_2+1-d/2)_{m^{\prime}_3}}\\
&\phantom{=}\qquad\times\frac{(p_2+h_2)_{m^{\prime}_{1}+m^{\prime}_3-m^{\prime}_2+m_{12}+m_{33}}(p_3)_{m^{\prime}_2-m^{\prime}_1+m^{\prime}_3+m_{13}+m_{23}}(-h_2)_{m^{\prime}_1+m^{\prime}_2-m^{\prime}_3+m_{11}+m_{22}}}{(p_3-h_2)_{2m^{\prime}_{2}+m_{11}+m_{22}+m_{13}+m_{23}}(p_3-h_2+1-d/2)_{m^{\prime}_2}}\\
&\phantom{=}\qquad\times\frac{(-h_4)_{m^{\prime}_2+m_{11}+m_{22}+m_{13}}(-h_3)_{m^{\prime}_{1}+m_{12}+m_{22}+m_{33}}(-h_5)_{m^{\prime}_3+m^{\prime}_{13}+m^{\prime}_{23}+m^{\prime}_{33}}}{(p_2)_{2m^{\prime}_{1}+m_{11}+m_{12}+m_{22}+m_{33}}(p_2+1-d/2)_{m^{\prime}_1}}\\
&\phantom{=}\qquad\times\frac{(-m^{\prime}_{11})_{m_{12}+j_2}(-h_3+m^{\prime}_1+m^{\prime}_{22}+m_{12}+m_{33}+k_{22})_{m^{\prime}_{11}-m_{12}-j_2}}{m_{11}!(m_{13}-m^{\prime}_{33}-j_{1})!}\\
&\phantom{=}\qquad\times\frac{(-m^{\prime}_{12})_{m_{11}+j_1}(-h_4+m^{\prime}_{2}+m^{\prime}_{22}+m_{11}+m_{13}+k_{22})_{m^{\prime}_{12}-m_{11}-j_1}}{m_{12}!(m_{22}-m^{\prime}_{22}-k_{22})!}\\
&\phantom{=}\qquad\times\frac{(-m^{\prime}_{23})_{m_{13}+m_{23}+m_{33}-m^{\prime}_{13}-m^{\prime}_{33}-j_1-j_2}}{(-h_5+m^{\prime}_{3}+m_{13}+m_{23}+m_{33})_{-j_1-j_2}j_1!j_2!k_{22}!m_{23}!(m_{33}-m^{\prime}_{13}-j_{2})!}\\
&\phantom{=}\qquad\times(-1)^{k_{22}}F_6\prod_{1\leq a\leq3}\frac{(u_a^6)^{m^{\prime}_a}}{m^{\prime}_a!}\prod_{1\leq a\leq b\leq3}\frac{(1-v_{ab}^6)^{m^{\prime}_{ab}}}{m^{\prime}_{ab}!}.
}

Once again, we introduce
\eqna{
&(-h_3+m^{\prime}_1+m^{\prime}_{22}+m_{12}+m_{33}+k_{22})_{m^{\prime}_{11}-m_{12}-j_2}\\
&\qquad=\sum_{j_3}\binom{m^{\prime}_{11}-m_{12}-j_2}{j_3}(-h_3+m^{\prime}_1+m_{12}+m_{22}+m_{33})_{m^{\prime}_{11}-m_{12}-j_2-j_3}(-m_{22}+m^{\prime}_{22}+k_{22})_{j_3},
}
and sum over $k_{22}$ to obtain
\eqna{
G_{6S}&=\sum\frac{(p_3-h_2+h_4)_{m^{\prime}_{2}+m_{23}}(p_2+h_3)_{m^{\prime}_{1}+m_{11}}(\bar{p}_3+h_2+h_5)_{m^{\prime}_3+m_{12}}}{(\bar{p}_3+h_2)_{2m^{\prime}_3+m_{12}+m_{13}+m_{23}+m_{33}}(\bar{p}_3+h_2+1-d/2)_{m^{\prime}_3}}\\
&\phantom{=}\qquad\times\frac{(p_2+h_2)_{m^{\prime}_{1}+m^{\prime}_3-m^{\prime}_2+m_{12}+m_{33}}(p_3)_{m^{\prime}_2-m^{\prime}_1+m^{\prime}_3+m_{13}+m_{23}}(-h_2)_{m^{\prime}_1+m^{\prime}_2-m^{\prime}_3+m_{11}+m_{22}}}{(p_3-h_2)_{2m^{\prime}_{2}+m_{11}+m_{22}+m_{13}+m_{23}}(p_3-h_2+1-d/2)_{m^{\prime}_2}}\\
&\phantom{=}\qquad\times\frac{(-h_4)_{m^{\prime}_2+m^{\prime}_{12}+m^{\prime}_{22}+m_{13}-j_1}(-h_3)_{m^{\prime}_{1}+m^{\prime}_{11}+m_{22}+m_{33}-j_2-j_3}(-h_5)_{m^{\prime}_3+m^{\prime}_{13}+m^{\prime}_{23}+m^{\prime}_{33}}}{(p_2)_{2m^{\prime}_{1}+m_{11}+m_{12}+m_{22}+m_{33}}(p_2+1-d/2)_{m^{\prime}_1}}\\
&\phantom{=}\qquad\times\frac{(-m^{\prime}_{12})_{m_{11}+m_{22}-m^{\prime}_{22}+j_1-j_3}}{(-h_4+m^{\prime}_{2}+m_{22}+m_{11}+m_{13})_{-j_3}m_{12}!(m_{22}-m^{\prime}_{22}-j_{3})!}\\
&\phantom{=}\qquad\times\frac{(-m^{\prime}_{23})_{m_{13}+m_{23}+m_{33}-m^{\prime}_{13}-m^{\prime}_{33}-j_1-j_2}}{(-h_5+m^{\prime}_{3}+m_{13}+m_{23}+m_{33})_{-j_1-j_2}j_1!j_2!j_3!m_{23}!(m_{33}-m^{\prime}_{13}-j_{2})!}\\
&\phantom{=}\qquad\times\frac{(-m^{\prime}_{11})_{m_{12}+j_2+j_3}}{m_{11}!(m_{13}-m^{\prime}_{33}-j_{1})!}F_6\prod_{1\leq a\leq3}\frac{(u_a^6)^{m^{\prime}_a}}{m^{\prime}_a!}\prod_{1\leq a\leq b\leq3}\frac{(1-v_{ab}^6)^{m^{\prime}_{ab}}}{m^{\prime}_{ab}!}.
}

At this point, we define $m_{22}=m-m_{11}$ and evaluate the summation over $m_{11}$, always using \eqref{Eq2F1}, to find
\eqna{
G_{6S}&=\sum\frac{(p_3-h_2+h_4)_{m^{\prime}_{2}+m_{23}}(p_2+h_3)_{m^{\prime}_{1}}(\bar{p}_3+h_2+h_5)_{m^{\prime}_3+m_{12}}}{(\bar{p}_3+h_2)_{2m^{\prime}_3+m_{12}+m_{13}+m_{23}+m_{33}}(\bar{p}_3+h_2+1-d/2)_{m^{\prime}_3}}\\
&\phantom{=}\qquad\times\frac{(p_2+h_2)_{m^{\prime}_{1}+m^{\prime}_3-m^{\prime}_2+m_{12}+m_{33}}(p_3)_{m^{\prime}_2-m^{\prime}_1+m^{\prime}_3+m_{13}+m_{23}}(-h_2)_{m^{\prime}_1+m^{\prime}_2-m^{\prime}_3+m}}{(p_3-h_2)_{2m^{\prime}_{2}+m+m_{13}+m_{23}}(p_3-h_2+1-d/2)_{m^{\prime}_2}}\\
&\phantom{=}\qquad\times\frac{(-h_4)_{m^{\prime}_2+m^{\prime}_{12}+m^{\prime}_{22}+m_{13}-j_1}(-h_3)_{m^{\prime}_{1}+m^{\prime}_{11}+m^{\prime}_{22}+m_{33}-j_2}(-h_5)_{m^{\prime}_3+m^{\prime}_{13}+m^{\prime}_{23}+m^{\prime}_{33}}}{(p_2)_{2m^{\prime}_{1}+m+m_{12}+m_{33}}(p_2+1-d/2)_{m^{\prime}_1}}\\
&\phantom{=}\qquad\times\frac{(p_2+2m^{\prime}_1+m^{\prime}_{11}+m^{\prime}_{22}+m_{33}-j_2)_{m-m^{\prime}_{22}-j_3}(-m^{\prime}_{12})_{m-m^{\prime}_{22}+j_1-j_3}}{(-h_4+m^{\prime}_{2}+m+m_{13})_{-j_3}m_{12}!(m-m^{\prime}_{22}-j_{3})!}\\
&\phantom{=}\qquad\times\frac{(-m^{\prime}_{23})_{m_{13}+m_{23}+m_{33}-m^{\prime}_{13}-m^{\prime}_{33}-j_1-j_2}}{(-h_5+m^{\prime}_{3}+m_{13}+m_{23}+m_{33})_{-j_1-j_2}m_{23}!(m_{33}-m^{\prime}_{13}-j_{2})!}\\
&\phantom{=}\qquad\times\frac{(-m^{\prime}_{11})_{m_{12}+j_2+j_3}}{(m_{13}-m^{\prime}_{33}-j_{1})!}F_6\prod_{1\leq a\leq3}\frac{(u_a^6)^{m^{\prime}_a}}{m^{\prime}_a!}\prod_{1\leq a\leq b\leq3}\frac{(1-v_{ab}^6)^{m^{\prime}_{ab}}}{m^{\prime}_{ab}!}.
}

We now express the summation over $j_3$ as a ${}_3F_2$-hypergeometric function, use \eqref{Eq3F2}, and re-expand the ${}_3F_2$-hypergeometric function as a sum over $j_3$ to rewrite $G_{6S}$ as
\eqna{
G_{6S}&=\sum\frac{(p_3-h_2+h_4)_{m^{\prime}_{2}+m_{23}}(p_2+h_3)_{m^{\prime}_{1}}(\bar{p}_3+h_2+h_5)_{m^{\prime}_3+m_{12}}}{(\bar{p}_3+h_2)_{2m^{\prime}_3+m_{12}+m_{13}+m_{23}+m_{33}}(\bar{p}_3+h_2+1-d/2)_{m^{\prime}_3}}\\
&\phantom{=}\qquad\times\frac{(p_2+h_2)_{m^{\prime}_{1}+m^{\prime}_3-m^{\prime}_2+m_{12}+m_{33}}(p_3)_{m^{\prime}_2-m^{\prime}_1+m^{\prime}_3+m_{13}+m_{23}}(-h_2)_{m^{\prime}_1+m^{\prime}_2-m^{\prime}_3+m}}{(p_3-h_2)_{2m^{\prime}_{2}+m+m_{13}+m_{23}}(p_3-h_2+1-d/2)_{m^{\prime}_2}}\\
&\phantom{=}\qquad\times\frac{(-h_4)_{m^{\prime}_2+m^{\prime}_{12}+m^{\prime}_{22}+m_{13}-j_1+j_3}(-h_3)_{m^{\prime}_{1}+m^{\prime}_{11}+m^{\prime}_{22}+m_{33}-j_2}(-h_5)_{m^{\prime}_3+m^{\prime}_{13}+m^{\prime}_{23}+m^{\prime}_{33}}}{(p_2)_{2m^{\prime}_{1}+m^{\prime}_{22}+m_{12}+m_{33}+j_3}(p_2+1-d/2)_{m^{\prime}_1}}\\
&\phantom{=}\qquad\times\frac{(-m^{\prime}_{12})_{m-m^{\prime}_{22}+j_1-j_3}(-m^{\prime}_{23})_{m_{13}+m_{23}+m_{33}-m^{\prime}_{13}-m^{\prime}_{33}-j_1-j_2}}{(-h_5+m^{\prime}_{3}+m_{13}+m_{23}+m_{33})_{-j_1-j_2}j_1!j_2!j_3!m_{12}!(m-m^{\prime}_{22}-j_{3})!m_{23}!(m_{33}-m^{\prime}_{13}-j_{2})!}\\
&\phantom{=}\qquad\times\frac{(-m^{\prime}_{11})_{m_{12}+j_2+j_3}}{(m_{13}-m^{\prime}_{33}-j_{1})!}F_6\prod_{1\leq a\leq3}\frac{(u_a^6)^{m^{\prime}_a}}{m^{\prime}_a!}\prod_{1\leq a\leq b\leq3}\frac{(1-v_{ab}^6)^{m^{\prime}_{ab}}}{m^{\prime}_{ab}!}.
}

After changing $m$ by $m\to m+m^{\prime}_{22}+j_3$, we evaluate the summation over $m$, leading to
\eqna{
G_{6S}&=\sum\frac{(p_3-h_2+h_4)_{m^{\prime}_{2}+m_{23}}(p_2+h_3)_{m^{\prime}_{1}}(\bar{p}_3+h_2+h_5)_{m^{\prime}_3+m_{12}}}{(\bar{p}_3+h_2)_{2m^{\prime}_3+m_{12}+m_{13}+m_{23}+m_{33}}(\bar{p}_3+h_2+1-d/2)_{m^{\prime}_3}}\\
&\phantom{=}\qquad\times\frac{(p_2+h_2)_{m^{\prime}_{1}+m^{\prime}_3-m^{\prime}_2+m_{12}+m_{33}}(p_3)_{m^{\prime}_2-m^{\prime}_1+m^{\prime}_3+m^{\prime}_{12}+m_{13}+m_{23}-j_1}(-h_2)_{m^{\prime}_1+m^{\prime}_2-m^{\prime}_3+m^{\prime}_{22}+j_3}}{(p_3-h_2)_{2m^{\prime}_{2}+m^{\prime}_{12}+m^{\prime}_{22}+m_{13}+m_{23}-j_1+j_3}(p_3-h_2+1-d/2)_{m^{\prime}_2}}\\
&\phantom{=}\qquad\times\frac{(-h_4)_{m^{\prime}_2+m^{\prime}_{12}+m^{\prime}_{22}+m_{13}-j_1+j_3}(-h_3)_{m^{\prime}_{1}+m^{\prime}_{11}+m^{\prime}_{22}+m_{33}-j_2}(-h_5)_{m^{\prime}_3+m^{\prime}_{13}+m^{\prime}_{23}+m^{\prime}_{33}}}{(p_2)_{2m^{\prime}_{1}+m^{\prime}_{22}+m_{12}+m_{33}+j_3}(p_2+1-d/2)_{m^{\prime}_1}}\\
&\phantom{=}\qquad\times\frac{(-m^{\prime}_{12})_{j_1}(-m^{\prime}_{23})_{m_{13}+m_{23}+m_{33}-m^{\prime}_{13}-m^{\prime}_{33}-j_1-j_2}}{(-h_5+m^{\prime}_{3}+m_{13}+m_{23}+m_{33})_{-j_1-j_2}j_1!j_2!j_3!m_{12}!m_{23}!(m_{33}-m^{\prime}_{13}-j_{2})!}\\
&\phantom{=}\qquad\times\frac{(-m^{\prime}_{11})_{m_{12}+j_2+j_3}}{(m_{13}-m^{\prime}_{33}-j_{1})!}F_6\prod_{1\leq a\leq3}\frac{(u_a^6)^{m^{\prime}_a}}{m^{\prime}_a!}\prod_{1\leq a\leq b\leq3}\frac{(1-v_{ab}^6)^{m^{\prime}_{ab}}}{m^{\prime}_{ab}!}.
}

We then change the variable $m_{13}$ by $m_{13}\to m_{13}+m^{\prime}_{33}+j_1$ and compute the summation over $j_1$.  As a result, we get
\eqna{
G_{6S}&=\sum\frac{(p_3-h_2+h_4)_{m^{\prime}_{2}+m_{23}}(p_2+h_3)_{m^{\prime}_{1}}(\bar{p}_3+h_2+h_5)_{m^{\prime}_3+m^{\prime}_{12}+m_{12}}}{(\bar{p}_3+h_2)_{2m^{\prime}_3+m^{\prime}_{12}+m^{\prime}_{33}+m_{12}+m_{13}+m_{23}+m_{33}}(\bar{p}_3+h_2+1-d/2)_{m^{\prime}_3}}\\
&\phantom{=}\qquad\times\frac{(p_2+h_2)_{m^{\prime}_{1}+m^{\prime}_3-m^{\prime}_2+m_{12}+m_{33}}(p_3)_{m^{\prime}_2-m^{\prime}_1+m^{\prime}_3+m^{\prime}_{12}+m^{\prime}_{33}+m_{13}+m_{23}}(-h_2)_{m^{\prime}_1+m^{\prime}_2-m^{\prime}_3+m^{\prime}_{22}+j_3}}{(p_3-h_2)_{2m^{\prime}_{2}+m^{\prime}_{12}+m^{\prime}_{22}+m^{\prime}_{33}+m_{13}+m_{23}+j_3}(p_3-h_2+1-d/2)_{m^{\prime}_2}}\\
&\phantom{=}\qquad\times\frac{(-h_4)_{m^{\prime}_2+m^{\prime}_{12}+m^{\prime}_{22}+m^{\prime}_{33}+m_{13}+j_3}(-h_3)_{m^{\prime}_{1}+m^{\prime}_{11}+m^{\prime}_{22}+m_{33}-j_2}(-h_5)_{m^{\prime}_3+m^{\prime}_{13}+m^{\prime}_{23}+m^{\prime}_{33}}}{(p_2)_{2m^{\prime}_{1}+m^{\prime}_{22}+m_{12}+m_{33}+j_3}(p_2+1-d/2)_{m^{\prime}_1}}\\
&\phantom{=}\qquad\times\frac{(-m^{\prime}_{23})_{m_{13}+m_{23}+m_{33}-m^{\prime}_{13}-j_2}}{(-h_5+m^{\prime}_{3}+m^{\prime}_{33}+m_{13}+m_{23}+m_{33})_{-j_2}j_2!j_3!m_{12}!m_{23}!(m_{33}-m^{\prime}_{13}-j_{2})!}\\
&\phantom{=}\qquad\times\frac{(-m^{\prime}_{11})_{m_{12}+j_2+j_3}}{(m_{13})!}F_6\prod_{1\leq a\leq3}\frac{(u_a^6)^{m^{\prime}_a}}{m^{\prime}_a!}\prod_{1\leq a\leq b\leq3}\frac{(1-v_{ab}^6)^{m^{\prime}_{ab}}}{m^{\prime}_{ab}!}.
}

We now redefine $m_{23}=n-m_{13}$ and evaluate the summation over $m_{13}$ with \eqref{Eq2F1}, which implies
\eqna{
G_{6S}&=\sum\frac{(p_3-h_2+h_4)_{m^{\prime}_{2}}(p_2+h_3)_{m^{\prime}_{1}}(\bar{p}_3+h_2+h_5)_{m^{\prime}_3+m^{\prime}_{12}+m_{12}}}{(\bar{p}_3+h_2)_{2m^{\prime}_3+m^{\prime}_{12}+m^{\prime}_{33}+m_{12}+n+m_{33}}(\bar{p}_3+h_2+1-d/2)_{m^{\prime}_3}}\\
&\phantom{=}\qquad\times\frac{(p_2+h_2)_{m^{\prime}_{1}+m^{\prime}_3-m^{\prime}_2+m_{12}+m_{33}}(p_3)_{m^{\prime}_2-m^{\prime}_1+m^{\prime}_3+m^{\prime}_{12}+m^{\prime}_{33}+n}(-h_2)_{m^{\prime}_1+m^{\prime}_2-m^{\prime}_3+m^{\prime}_{22}+j_3}}{(p_3-h_2)_{2m^{\prime}_{2}+m^{\prime}_{12}+m^{\prime}_{22}+m^{\prime}_{33}+j_3}(p_3-h_2+1-d/2)_{m^{\prime}_2}}\\
&\phantom{=}\qquad\times\frac{(-h_4)_{m^{\prime}_2+m^{\prime}_{12}+m^{\prime}_{22}+m^{\prime}_{33}+j_3}(-h_3)_{m^{\prime}_{1}+m^{\prime}_{11}+m^{\prime}_{22}+m_{33}-j_2}(-h_5)_{m^{\prime}_3+m^{\prime}_{13}+m^{\prime}_{23}+m^{\prime}_{33}}}{(p_2)_{2m^{\prime}_{1}+m^{\prime}_{22}+m_{12}+m_{33}+j_3}(p_2+1-d/2)_{m^{\prime}_1}}\\
&\phantom{=}\qquad\times\frac{(-m^{\prime}_{23})_{n+m_{33}-m^{\prime}_{13}-j_2}}{(-h_5+m^{\prime}_{3}+m^{\prime}_{33}+n+m_{33})_{-j_2}j_2!j_3!m_{12}!(m_{33}-m^{\prime}_{13}-j_{2})!}\\
&\phantom{=}\qquad\times\frac{(-m^{\prime}_{11})_{m_{12}+j_2+j_3}}{n!}F_6\prod_{1\leq a\leq3}\frac{(u_a^6)^{m^{\prime}_a}}{m^{\prime}_a!}\prod_{1\leq a\leq b\leq3}\frac{(1-v_{ab}^6)^{m^{\prime}_{ab}}}{m^{\prime}_{ab}!}.
}

We can proceed with the summation over $j_2$, which gives a ${}_3F_2$-hypergeometric function, use \eqref{Eq3F2} once more, and re-expand with the same index of summation to rewrite $G_{6S}$ as
\eqna{
G_{6S}&=\sum\frac{(p_3-h_2+h_4)_{m^{\prime}_{2}}(p_2+h_3)_{m^{\prime}_{1}}(\bar{p}_3+h_2+h_5)_{m^{\prime}_3+m^{\prime}_{12}+m_{12}}}{(\bar{p}_3+h_2)_{2m^{\prime}_3+m^{\prime}_{12}+m^{\prime}_{33}+m_{12}+n+m_{33}}(\bar{p}_3+h_2+1-d/2)_{m^{\prime}_3}}\\
&\phantom{=}\qquad\times\frac{(p_2+h_2)_{m^{\prime}_{1}+m^{\prime}_3-m^{\prime}_2+m_{12}+m_{33}}(p_3)_{m^{\prime}_2-m^{\prime}_1+m^{\prime}_3+m^{\prime}_{12}+m^{\prime}_{33}+n}(-h_2)_{m^{\prime}_1+m^{\prime}_2-m^{\prime}_3+m^{\prime}_{22}+j_3}}{(p_3-h_2)_{2m^{\prime}_{2}+m^{\prime}_{12}+m^{\prime}_{22}+m^{\prime}_{33}+j_3}(p_3-h_2+1-d/2)_{m^{\prime}_2}}\\
&\phantom{=}\qquad\times\frac{(-h_4)_{m^{\prime}_2+m^{\prime}_{12}+m^{\prime}_{22}+m^{\prime}_{33}+j_3}(-h_3)_{m^{\prime}_{1}+m^{\prime}_{11}+m^{\prime}_{13}+m^{\prime}_{22}}(-h_5)_{m^{\prime}_3+m^{\prime}_{13}+m^{\prime}_{23}+m^{\prime}_{33}+j_2}}{(p_2)_{2m^{\prime}_{1}+m^{\prime}_{22}+m_{12}+m_{33}+j_3}(p_2+1-d/2)_{m^{\prime}_1}}\\
&\phantom{=}\qquad\times\frac{(-h_3+m^{\prime}_1+m^{\prime}_{13}+m^{\prime}_{22}+m_{12}+j_2+j_3)_{m_{33}-m^{\prime}_{13}-j_2}(-m^{\prime}_{23})_{m_{33}-m^{\prime}_{13}-j_2}}{j_2!j_3!m_{12}!(m_{33}-m^{\prime}_{13}-j_{2})!}\\
&\phantom{=}\qquad\times\frac{(-m^{\prime}_{11})_{m_{12}+j_2+j_3}}{n!}F_6\prod_{1\leq a\leq3}\frac{(u_a^6)^{m^{\prime}_a}}{m^{\prime}_a!}\prod_{1\leq a\leq b\leq3}\frac{(1-v_{ab}^6)^{m^{\prime}_{ab}}}{m^{\prime}_{ab}!}.
}

This allows us to sum over $n$, leading to
\eqna{
G_{6S}&=\sum\frac{(p_3-h_2+h_4)_{m^{\prime}_{2}}(p_2+h_3)_{m^{\prime}_{1}}(\bar{p}_3+h_2+h_5)_{m^{\prime}_3+m^{\prime}_{12}+m_{12}}}{(\bar{p}_3+h_2)_{2m^{\prime}_3+m^{\prime}_{12}+m^{\prime}_{13}+m^{\prime}_{23}+m^{\prime}_{33}+m_{12}+j_2}(\bar{p}_3+h_2+1-d/2)_{m^{\prime}_3}}\\
&\phantom{=}\qquad\times\frac{(p_2+h_2)_{m^{\prime}_{1}+m^{\prime}_3-m^{\prime}_2+m^{\prime}_{13}+m^{\prime}_{23}+m_{12}+j_{2}}(p_3)_{m^{\prime}_2-m^{\prime}_1+m^{\prime}_3+m^{\prime}_{12}+m^{\prime}_{33}}(-h_2)_{m^{\prime}_1+m^{\prime}_2-m^{\prime}_3+m^{\prime}_{22}+j_3}}{(p_3-h_2)_{2m^{\prime}_{2}+m^{\prime}_{12}+m^{\prime}_{22}+m^{\prime}_{33}+j_3}(p_3-h_2+1-d/2)_{m^{\prime}_2}}\\
&\phantom{=}\qquad\times\frac{(-h_4)_{m^{\prime}_2+m^{\prime}_{12}+m^{\prime}_{22}+m^{\prime}_{33}+j_3}(-h_3)_{m^{\prime}_{1}+m^{\prime}_{11}+m^{\prime}_{13}+m^{\prime}_{22}}(-h_5)_{m^{\prime}_3+m^{\prime}_{13}+m^{\prime}_{23}+m^{\prime}_{33}+j_2}}{(p_2)_{2m^{\prime}_{1}+m^{\prime}_{22}+m_{12}+m_{33}+j_3}(p_2+1-d/2)_{m^{\prime}_1}}\\
&\phantom{=}\qquad\times\frac{(-h_3+m^{\prime}_1+m^{\prime}_{13}+m^{\prime}_{22}+m_{12}+j_2+j_3)_{m_{33}-m^{\prime}_{13}-j_2}(-m^{\prime}_{23})_{m_{33}-m^{\prime}_{13}-j_2}}{(m_{33}-m^{\prime}_{13}-j_{2})!}\\
&\phantom{=}\qquad\times\frac{(-m^{\prime}_{11})_{m_{12}+j_2+j_3}}{j_2!j_3!m_{12}!}F_6\prod_{1\leq a\leq3}\frac{(u_a^6)^{m^{\prime}_a}}{m^{\prime}_a!}\prod_{1\leq a\leq b\leq3}\frac{(1-v_{ab}^6)^{m^{\prime}_{ab}}}{m^{\prime}_{ab}!}.
}

We then change $m_{33}$ by $m_{33}\to m_{33}+m^{\prime}_{13}+j_2$ and sum over $m_{33}$ to get
\eqna{
G_{6S}&=\sum\frac{(p_3-h_2+h_4)_{m^{\prime}_{2}}(p_2+h_3)_{m^{\prime}_{1}+m^{\prime}_{23}}(\bar{p}_3+h_2+h_5)_{m^{\prime}_3+m^{\prime}_{12}+m_{12}}}{(\bar{p}_3+h_2)_{2m^{\prime}_3+m^{\prime}_{12}+m^{\prime}_{13}+m^{\prime}_{23}+m^{\prime}_{33}+m_{12}+j_2}(\bar{p}_3+h_2+1-d/2)_{m^{\prime}_3}}\\
&\phantom{=}\qquad\times\frac{(p_2+h_2)_{m^{\prime}_{1}+m^{\prime}_3-m^{\prime}_2+m^{\prime}_{13}+m^{\prime}_{23}+m_{12}+j_{2}}(p_3)_{m^{\prime}_2-m^{\prime}_1+m^{\prime}_3+m^{\prime}_{12}+m^{\prime}_{33}}(-h_2)_{m^{\prime}_1+m^{\prime}_2-m^{\prime}_3+m^{\prime}_{22}+j_3}}{(p_3-h_2)_{2m^{\prime}_{2}+m^{\prime}_{12}+m^{\prime}_{22}+m^{\prime}_{33}+j_3}(p_3-h_2+1-d/2)_{m^{\prime}_2}}\\
&\phantom{=}\qquad\times\frac{(-h_4)_{m^{\prime}_2+m^{\prime}_{12}+m^{\prime}_{22}+m^{\prime}_{33}+j_3}(-h_3)_{m^{\prime}_{1}+m^{\prime}_{11}+m^{\prime}_{13}+m^{\prime}_{22}}(-h_5)_{m^{\prime}_3+m^{\prime}_{13}+m^{\prime}_{23}+m^{\prime}_{33}+j_2}}{(p_2)_{2m^{\prime}_{1}+m^{\prime}_{13}+m^{\prime}_{22}+m^{\prime}_{23}+m_{12}+j_2+j_3}(p_2+1-d/2)_{m^{\prime}_1}}\\
&\phantom{=}\qquad\times\frac{(-m^{\prime}_{11})_{m_{12}+j_2+j_3}}{j_2!j_3!m_{12}!}F_6\prod_{1\leq a\leq3}\frac{(u_a^6)^{m^{\prime}_a}}{m^{\prime}_a!}\prod_{1\leq a\leq b\leq3}\frac{(1-v_{ab}^6)^{m^{\prime}_{ab}}}{m^{\prime}_{ab}!}.
}

We now sum over $m_{12}$, leading to a ${}_3F_2$-hypergeometric function, and use \eqref{Eq3F2} one last time to rewrite $G_{6S}$ as
\eqna{
G_{6S}&=\sum\frac{(p_3-h_2+h_4)_{m^{\prime}_{2}}(p_2+h_3)_{m^{\prime}_{1}+m^{\prime}_{23}}(\bar{p}_3+h_2+h_5)_{m^{\prime}_3+m^{\prime}_{12}}}{(\bar{p}_3+h_2)_{2m^{\prime}_3+m^{\prime}_{12}+m^{\prime}_{13}+m^{\prime}_{23}+m^{\prime}_{33}+m_{12}+j_2}(\bar{p}_3+h_2+1-d/2)_{m^{\prime}_3}}\\
&\phantom{=}\qquad\times\frac{(p_2+h_2)_{m^{\prime}_{1}+m^{\prime}_3-m^{\prime}_2+m^{\prime}_{13}+m^{\prime}_{23}+m_{12}+j_{2}}(p_3)_{m^{\prime}_2-m^{\prime}_1+m^{\prime}_3+m^{\prime}_{12}+m^{\prime}_{33}}(-h_2)_{m^{\prime}_1+m^{\prime}_2-m^{\prime}_3+m^{\prime}_{11}+m^{\prime}_{22}-m_{12}-j_2}}{(p_3-h_2)_{2m^{\prime}_{2}+m^{\prime}_{12}+m^{\prime}_{22}+m^{\prime}_{33}+j_3}(p_3-h_2+1-d/2)_{m^{\prime}_2}}\\
&\phantom{=}\qquad\times\frac{(-1)^{m_{12}}(-h_4)_{m^{\prime}_2+m^{\prime}_{12}+m^{\prime}_{22}+m^{\prime}_{33}+j_3}(-h_3)_{m^{\prime}_{1}+m^{\prime}_{11}+m^{\prime}_{13}+m^{\prime}_{22}}(-h_5)_{m^{\prime}_3+m^{\prime}_{13}+m^{\prime}_{23}+m^{\prime}_{33}+m_{12}+j_2}}{(p_2)_{2m^{\prime}_{1}+m^{\prime}_{11}+m^{\prime}_{13}+m^{\prime}_{22}+m^{\prime}_{23}}(p_2+1-d/2)_{m^{\prime}_1}}\\
&\phantom{=}\qquad\times\frac{(-m^{\prime}_{11})_{m_{12}+j_2+j_3}}{j_2!j_3!m_{12}!}F_6\prod_{1\leq a\leq3}\frac{(u_a^6)^{m^{\prime}_a}}{m^{\prime}_a!}\prod_{1\leq a\leq b\leq3}\frac{(1-v_{ab}^6)^{m^{\prime}_{ab}}}{m^{\prime}_{ab}!}.
}

This transformation allows us to sum over $j_3$ following \eqref{Eq2F1}, which leads to
\eqna{
G_{6S}&=\sum\frac{(p_3-h_2+h_4)_{m^{\prime}_{2}+m^{\prime}_{11}-m_{12}-j_2}(p_2+h_3)_{m^{\prime}_{1}+m^{\prime}_{23}}(\bar{p}_3+h_2+h_5)_{m^{\prime}_3+m^{\prime}_{12}}}{(\bar{p}_3+h_2)_{2m^{\prime}_3+m^{\prime}_{12}+m^{\prime}_{13}+m^{\prime}_{23}+m^{\prime}_{33}+m_{12}+j_2}(\bar{p}_3+h_2+1-d/2)_{m^{\prime}_3}}\\
&\phantom{=}\qquad\times\frac{(p_2+h_2)_{m^{\prime}_{1}+m^{\prime}_3-m^{\prime}_2+m^{\prime}_{13}+m^{\prime}_{23}+m_{12}+j_{2}}(p_3)_{m^{\prime}_2-m^{\prime}_1+m^{\prime}_3+m^{\prime}_{12}+m^{\prime}_{33}}(-h_2)_{m^{\prime}_1+m^{\prime}_2-m^{\prime}_3+m^{\prime}_{11}+m^{\prime}_{22}-m_{12}-j_2}}{(p_3-h_2)_{2m^{\prime}_{2}+m^{\prime}_{11}+m^{\prime}_{12}+m^{\prime}_{22}+m^{\prime}_{33}-m_{12}-j_2}(p_3-h_2+1-d/2)_{m^{\prime}_2}}\\
&\phantom{=}\qquad\times\frac{(-1)^{m_{12}}(-h_4)_{m^{\prime}_2+m^{\prime}_{12}+m^{\prime}_{22}+m^{\prime}_{33}}(-h_3)_{m^{\prime}_{1}+m^{\prime}_{11}+m^{\prime}_{13}+m^{\prime}_{22}}(-h_5)_{m^{\prime}_3+m^{\prime}_{13}+m^{\prime}_{23}+m^{\prime}_{33}+m_{12}+j_2}}{(p_2)_{2m^{\prime}_{1}+m^{\prime}_{11}+m^{\prime}_{13}+m^{\prime}_{22}+m^{\prime}_{23}}(p_2+1-d/2)_{m^{\prime}_1}}\\
&\phantom{=}\qquad\times\frac{(-m^{\prime}_{11})_{m_{12}+j_2}}{j_2!m_{12}!}F_6\prod_{1\leq a\leq3}\frac{(u_a^6)^{m^{\prime}_a}}{m^{\prime}_a!}\prod_{1\leq a\leq b\leq3}\frac{(1-v_{ab}^6)^{m^{\prime}_{ab}}}{m^{\prime}_{ab}!}.
}

We are thus left with two extra sums (over $m_{12}$ and $j_2$).  However, they are both trivial.  Indeed, by redefining $j_2\to j_2-m_{12}$ and using the binomial identity
\eqn{\sum_{m_{12}}(-1)^{m_{12}}\frac{1}{m_{12}!(j_2-m_{12})!}=\frac{1}{j_2!}(1-1)^{j_2},}
we find that $j_2=0$ and thus
\eqna{
G_{6S}&=\sum\frac{(p_3-h_2+h_4)_{m^{\prime}_{2}+m^{\prime}_{11}}(p_2+h_3)_{m^{\prime}_{1}+m^{\prime}_{23}}(\bar{p}_3+h_2+h_5)_{m^{\prime}_3+m^{\prime}_{12}}}{(\bar{p}_3+h_2)_{2m^{\prime}_3+m^{\prime}_{12}+m^{\prime}_{13}+m^{\prime}_{23}+m^{\prime}_{33}}(\bar{p}_3+h_2+1-d/2)_{m^{\prime}_3}}\\
&\phantom{=}\qquad\times\frac{(p_2+h_2)_{m^{\prime}_{1}+m^{\prime}_3-m^{\prime}_2+m^{\prime}_{13}+m^{\prime}_{23}}(p_3)_{m^{\prime}_2-m^{\prime}_1+m^{\prime}_3+m^{\prime}_{12}+m^{\prime}_{33}}(-h_2)_{m^{\prime}_1+m^{\prime}_2-m^{\prime}_3+m^{\prime}_{11}+m^{\prime}_{22}}}{(p_3-h_2)_{2m^{\prime}_{2}+m^{\prime}_{11}+m^{\prime}_{12}+m^{\prime}_{22}+m^{\prime}_{33}}(p_3-h_2+1-d/2)_{m^{\prime}_2}}\\
&\phantom{=}\qquad\times\frac{(-h_4)_{m^{\prime}_2+m^{\prime}_{12}+m^{\prime}_{22}+m^{\prime}_{33}}(-h_3)_{m^{\prime}_{1}+m^{\prime}_{11}+m^{\prime}_{13}+m^{\prime}_{22}}(-h_5)_{m^{\prime}_3+m^{\prime}_{13}+m^{\prime}_{23}+m^{\prime}_{33}}}{(p_2)_{2m^{\prime}_{1}+m^{\prime}_{11}+m^{\prime}_{13}+m^{\prime}_{22}+m^{\prime}_{23}}(p_2+1-d/2)_{m^{\prime}_1}}\\
&\phantom{=}\qquad\times F_6\prod_{1\leq a\leq3}\frac{(u_a^6)^{m^{\prime}_a}}{m^{\prime}_a!}\prod_{1\leq a\leq b\leq3}\frac{(1-v_{ab}^6)^{m^{\prime}_{ab}}}{m^{\prime}_{ab}!},
}
which implies $G_{6S}=G_6$ and proves \eqref{EqSymRef}.

%%%%%%%%%%%%%%%%%%%%%%%%%%%%%%%%%%%%%%%%%%%%%%%%%%

\subsection{Permutations of the Dendrites}

Finally, we present the proof of the invariance of the scalar six-point correlation functions under dendrite permutation $P$ \eqref{EqSymPerm}.  For this proof, we use the alternative form \eqref{EqCB6Cs} for which $F_6^*=F_6$ \eqref{EqCB6F}, where $F_6^*\to F_6^*$ trivially under $P$.  Equation \eqref{EqSymPerm} thus becomes
\eqna{
&G_{6|\text{snowflake}}^{*(d,h_2,h_3,h_4,h_5;p_2,p_3,p_4,p_5,p_6)}(u_1^{*6},u_2^{*6},u_3^{*6};v_{11}^{*6},v_{12}^{*6},v_{13}^{*6},v_{22}^{*6},v_{23}^{*6},v_{33}^{*6})\\
&\qquad=(v_{11}^{*6})^{h_2}(v_{12}^{*6})^{h_2}(v_{13}^{*6})^{h_5}\\
&\qquad\phantom{=}\qquad\times G_{6|\text{snowflake}}^{*(d,h_2,-p_2-h_3,h_4,h_5;p_2,p_3,p_4,p_5,p_6)}\left(\frac{u_1^{*6}}{v_{11}^{*6}v_{12}^{*6}},\frac{u_2^{*6}}{v_{11}^{*6}v_{12}^{*6}},\frac{u_3^{*6}v_{11}^{*6}v_{12}^{*6}}{v_{13}^{*6}};\frac{1}{v_{11}^{*6}},\frac{1}{v_{12}^{*6}},\frac{1}{v_{13}^{*6}},\frac{v_{22}^{*6}}{v_{12}^{*6}},\frac{v_{23}^{*6}}{v_{13}^{*6}},\frac{v_{33}^{*6}}{v_{13}^{*6}}\right).
}[EqSymPermG]
For notational simplicity, we rewrite \eqref{EqSymPermG} as $G_6^*=G_{6P}^*$.

First, by expanding $G_{6P}^*$ in terms of the initial conformal cross-ratios, we obtain
\eqna{
G_{6P}^*&=\sum\frac{(-h_2)_{m_1+m_2-m_3+m_{12}+m_{22}}(-h_5)_{m_3+m_{13}+m_{23}+m_{33}}(\bar{p}_3+h_2+h_5)_{m_3}}{(\bar{p}_3+h_2)_{2m_3+m_{13}+m_{23}+m_{33}}(\bar{p}_3+h_2+1-d/2)_{m_3}}\\
&\phantom{=}\qquad\times\frac{(-h_3)_{m_1+m_{11}+m_{13}}(p_2+h_2)_{m_1-m_2+m_3+m_{13}}(p_3)_{m_2-m_1+m_3+m_{23}+m_{33}}}{(p_2)_{2m_1+m_{11}+m_{12}+m_{13}}(p_2+1-d/2)_{m_1}}\\
&\phantom{=}\qquad\times\frac{(-h_2)_{m_1+m_2-m_3+m_{11}+m_{12}}(p_3-h_2+h_4)_{m_2+m_{12}+m_{33}}}{(p_3-h_2)_{2m_2+m_{12}+m_{22}+m_{23}+m_{33}}(p_3-h_2+1-d/2)_{m_2}}\frac{(-h_4)_{m_2+m_{22}+m_{23}}(p_2+h_3)_{m_1+m_{12}}}{(-h_2)_{m_1+m_2-m_3+m_{12}}}\\
&\phantom{=}\qquad\times\binom{m_{11}}{k_{11}}\binom{m_{12}}{k_{12}}\binom{m_{13}}{k_{13}}\binom{m_{22}}{k_{22}}\binom{m_{23}}{k_{23}}\binom{m_{33}}{k_{33}}\binom{h_2+m_3-m_1-m_2-k_{11}}{m^{\prime}_{11}}\\
&\phantom{=}\qquad\times\binom{h_2+m_3-m_1-m_2-k_{12}-k_{22}}{m^{\prime}_{12}}\binom{h_5-m_3-k_{13}-k_{23}-k_{33}}{m^{\prime}_{13}}\binom{k_{22}}{m^{\prime}_{22}}\binom{k_{23}}{m^{\prime}_{23}}\binom{k_{33}}{m^{\prime}_{33}}\\
&\phantom{=}\qquad\times(-1)^{\sum_{1\leq a\leq b\leq3}(k_{ab}+m^{\prime}_{ab})}F_6^*\prod_{1\leq a\leq3}\frac{(u_a^{*6})^{m_a}}{m_a!}\prod_{1\leq a\leq b\leq3}\frac{(1-v_{ab}^{*6})^{m^{\prime}_{ab}}}{m_{ab}!},
}
where all superfluous sums must be evaluated.

After evaluating the summations over all the $k_{ab}$ (with change of variables $k_{ab}\to k_{ab}+m^{\prime}_{ab}$ for $2\leq a\leq b\leq 3$), we find
\eqna{
G_{6P}^*&=\sum\frac{(-h_2)_{m_1+m_2-m_3+m^{\prime}_{12}+m^{\prime}_{22}}(-h_5)_{m_3+m^{\prime}_{13}+m^{\prime}_{23}+m^{\prime}_{33}}(\bar{p}_3+h_2+h_5)_{m_3}}{(\bar{p}_3+h_2)_{2m_3+m_{13}+m_{23}+m_{33}}(\bar{p}_3+h_2+1-d/2)_{m_3}}\\
&\phantom{=}\qquad\times\frac{(-h_3)_{m_1+m_{11}+m_{13}}(p_2+h_2)_{m_1-m_2+m_3+m_{13}}(p_3)_{m_2-m_1+m_3+m_{23}+m_{33}}}{(p_2)_{2m_1+m_{11}+m_{12}+m_{13}}(p_2+1-d/2)_{m_1}}\\
&\phantom{=}\qquad\times\frac{(-h_2)_{m_1+m_2-m_3+m_{11}+m_{12}}(p_3-h_2+h_4)_{m_2+m_{12}+m_{33}}}{(p_3-h_2)_{2m_2+m_{12}+m_{22}+m_{23}+m_{33}}(p_3-h_2+1-d/2)_{m_2}}\frac{(-h_4)_{m_2+m_{22}+m_{23}}(p_2+h_3)_{m_1+m_{12}}}{(-h_2)_{m_1+m_2-m_3+m_{12}}}\\
&\phantom{=}\qquad\times\frac{(-m^{\prime}_{11})_{m_{11}}(-h_2-m_3+m_1+m_2+m_{11})_{m^{\prime}_{11}-m_{11}}}{m_{11}!(m_{22}-m_{22}^{\prime})!(m_{23}-m_{23}^{\prime})!(m_{33}-m_{33}^{\prime})!}\frac{(-m^{\prime}_{12})_{m_{12}+m_{22}-m^{\prime}_{22}}}{m_{12}!}\\
&\phantom{=}\qquad\times\frac{(-m^{\prime}_{13})_{m_{13}+m_{23}+m_{33}-m^{\prime}_{23}-m^{\prime}_{33}}}{m_{13}!}F^*_6\prod_{1\leq a\leq3}\frac{(u_a^{*6})^{m_a}}{m_a!}\prod_{1\leq a\leq b\leq3}\frac{(1-v_{ab}^{*6})^{m^{\prime}_{ab}}}{m^{\prime}_{ab}!}.
}

Re-summing over $m_{22}$, after the change the variable $m_{22}\to m_{22}+m^{\prime}_{22}$, we obtain
\eqna{
G_{6P}^*&=\sum\frac{(-h_2)_{m_1+m_2-m_3+m^{\prime}_{12}+m^{\prime}_{22}}(-h_5)_{m_3+m^{\prime}_{13}+m^{\prime}_{23}+m^{\prime}_{33}}(\bar{p}_3+h_2+h_5)_{m_3}}{(\bar{p}_3+h_2)_{2m_3+m_{13}+m_{23}+m_{33}}(\bar{p}_3+h_2+1-d/2)_{m_3}}\\
&\phantom{=}\qquad\times\frac{(-h_3)_{m_1+m_{11}+m_{13}}(p_2+h_2)_{m_1-m_2+m_3+m_{13}}(p_3)_{m_2-m_1+m_3+m_{23}+m_{33}}}{(p_2)_{2m_1+m_{11}+m_{12}+m_{13}}(p_2+1-d/2)_{m_1}}\\
&\phantom{=}\qquad\times\frac{(-h_2)_{m_1+m_2-m_3+m_{11}+m_{12}}(p_3-h_2+h_4)_{m_2+m^{\prime}_{12}+m_{33}}}{(p_3-h_2)_{2m_2+m^{\prime}_{12}+m^{\prime}_{22}+m_{23}+m_{33}}(p_3-h_2+1-d/2)_{m_2}}\frac{(-h_4)_{m_2+m^{\prime}_{22}+m_{23}}(p_2+h_3)_{m_1+m_{12}}}{(-h_2)_{m_1+m_2-m_3+m_{12}}}\\
&\phantom{=}\qquad\times\frac{(-m^{\prime}_{11})_{m_{11}}(-h_2-m_3+m_1+m_2+m_{11})_{m^{\prime}_{11}-m_{11}}}{m_{11}!(m_{23}-m_{23}^{\prime})!(m_{33}-m_{33}^{\prime})!}\frac{(-m^{\prime}_{12})_{m_{12}}}{m_{12}!}\\
&\phantom{=}\qquad\times\frac{(-m^{\prime}_{13})_{m_{13}+m_{23}+m_{33}-m^{\prime}_{23}-m^{\prime}_{33}}}{m_{13}!}F_6^*\prod_{1\leq a\leq3}\frac{(u_a^{*6})^{m_a}}{m_a!}\prod_{1\leq a\leq b\leq3}\frac{(1-v_{ab}^{*6})^{m^{\prime}_{ab}}}{m^{\prime}_{ab}!}.
}

Now, we redefine variables such that $m_{23}\to m_{23}+m^{\prime}_{23}$ and $m_{33}\to m_{33}+m^{\prime}_{33}$, and we define $m_{33}=m-m_{23}$ to evaluate the sums over $m_{23}$, $m$, and $m_{13}$, always with the help of \eqref{Eq2F1},
\eqna{
G_{6P}^*&=\sum\frac{(-h_2)_{m_1+m_2-m_3+m^{\prime}_{12}+m^{\prime}_{22}}(-h_5)_{m_3+m^{\prime}_{13}+m^{\prime}_{23}+m^{\prime}_{33}}(\bar{p}_3+h_2+h_5)_{m_3}}{(\bar{p}_3+h_2)_{2m_3+m^{\prime}_{13}+m^{\prime}_{23}+m^{\prime}_{33}}(\bar{p}_3+h_2+1-d/2)_{m_3}}\\
&\phantom{=}\qquad\times\frac{(-h_3)_{m_1+m_{11}}(p_2+h_2)_{m_1-m_2+m_3+m^{\prime}_{13}}(p_3)_{m_2-m_1+m_3+m^{\prime}_{23}+m^{\prime}_{33}}}{(p_2)_{2m_1+m_{11}+m_{12}+m^{\prime}_{13}}(p_2+1-d/2)_{m_1}}\\
&\phantom{=}\qquad\times\frac{(-h_2)_{m_1+m_2-m_3+m_{11}+m_{12}}(p_3-h_2+h_4)_{m_2+m^{\prime}_{12}+m^{\prime}_{33}}}{(p_3-h_2)_{2m_2+m^{\prime}_{12}+m^{\prime}_{22}+m^{\prime}_{23}+m^{\prime}_{33}}(p_3-h_2+1-d/2)_{m_2}}\frac{(-h_4)_{m_2+m^{\prime}_{22}+m^{\prime}_{23}}(p_2+h_3)_{m_1+m_{12}+m^{\prime}_{13}}}{(-h_2)_{m_1+m_2-m_3+m_{12}}}\\
&\phantom{=}\qquad\times\frac{(-m^{\prime}_{11})_{m_{11}}(-h_2-m_3+m_1+m_2+m_{11})_{m^{\prime}_{11}-m_{11}}}{m_{11}!}\frac{(-m^{\prime}_{12})_{m_{12}}}{m_{12}!}\\
&\phantom{=}\qquad\times F_6^*\prod_{1\leq a\leq3}\frac{(u_a^{*6})^{m_a}}{m_a!}\prod_{1\leq a\leq b\leq3}\frac{(1-v_{ab}^{*6})^{m^{\prime}_{ab}}}{m^{\prime}_{ab}!}.
}

The summation over $m_{11}$ corresponds to a ${}_3F_2$-hypergeometric function which can be transformed with the help of \eqref{Eq3F2}, leading to
\eqna{
&{}_3F_2\left[\begin{array}{c}-m^{\prime}_{11},-h_3+m_1,-h_2+m_1+m_2-m_3+m_{12}\\p_2+2m_1+m_{12}+m^{\prime}_{13},-h_2+m_1+m_2-m_3\end{array};1\right]=\frac{(p_2+h_3+m_1+m_{12}+m^{\prime}_{13})_{m^{\prime}_{11}}}{(p_2+2m_1+m_{12}+m^{\prime}_{13})_{m^{\prime}_{11}}}\\
&\qquad\times{}_3F_2\left[\begin{array}{c}-m^{\prime}_{11},-h_3+m_1,-m_{12}\\1-p_2-h_3-m_1-m_{12}-m^{\prime}_{11}-m^{\prime}_{13},-h_2+m_1+m_2-m_3\end{array};1\right].
}
Hence, we find that
\eqna{
G_{6P}^*&=\sum\frac{(-h_2)_{m_1+m_2-m_3+m^{\prime}_{12}+m^{\prime}_{22}}(-h_5)_{m_3+m^{\prime}_{13}+m^{\prime}_{23}+m^{\prime}_{33}}(\bar{p}_3+h_2+h_5)_{m_3}}{(\bar{p}_3+h_2)_{2m_3+m^{\prime}_{13}+m^{\prime}_{23}+m^{\prime}_{33}}(\bar{p}_3+h_2+1-d/2)_{m_3}}\\
&\phantom{=}\qquad\times\frac{(-h_3)_{m_1+m_{11}}(p_2+h_2)_{m_1-m_2+m_3+m^{\prime}_{13}}(p_3)_{m_2-m_1+m_3+m^{\prime}_{23}+m^{\prime}_{33}}}{(p_2)_{2m_1+m^{\prime}_{11}+m_{12}+m^{\prime}_{13}}(p_2+1-d/2)_{m_1}}\\
&\phantom{=}\qquad\times\frac{(-1)^{m_{11}}(-m_{12})_{m_{11}}(-h_4)_{m_2+m^{\prime}_{22}+m^{\prime}_{23}}(p_2+h_3)_{m_1+m_{12}-m_{11}+m^{\prime}_{11}+m^{\prime}_{13}}(p_3-h_2+h_4)_{m_2+m^{\prime}_{12}+m^{\prime}_{33}}}{(p_3-h_2)_{2m_2+m^{\prime}_{12}+m^{\prime}_{22}+m^{\prime}_{23}+m^{\prime}_{33}}(p_3-h_2+1-d/2)_{m_2}}\\
&\phantom{=}\qquad\times\frac{(-m^{\prime}_{11})_{m_{11}}(-h_2-m_3+m_1+m_2+m_{11})_{m^{\prime}_{11}-m_{11}}}{m_{11}!}\frac{(-m^{\prime}_{12})_{m_{12}}}{m_{12}!}\\
&\phantom{=}\qquad\times F_6^*\prod_{1\leq a\leq3}\frac{(u_a^{*6})^{m_a}}{m_a!}\prod_{1\leq a\leq b\leq3}\frac{(1-v_{ab}^{*6})^{m^{\prime}_{ab}}}{m^{\prime}_{ab}!}.
}

After evaluating the summation over $m_{12}$ (with $m_{12}\to m_{12}+m_{11}$ first) with the help of \eqref{Eq2F1}, the result becomes
\eqna{
G_{6P}^*&=\sum\frac{(-h_2)_{m_1+m_2-m_3+m^{\prime}_{12}+m^{\prime}_{22}}(-h_5)_{m_3+m^{\prime}_{13}+m^{\prime}_{23}+m^{\prime}_{33}}(\bar{p}_3+h_2+h_5)_{m_3}}{(\bar{p}_3+h_2)_{2m_3+m^{\prime}_{13}+m^{\prime}_{23}+m^{\prime}_{33}}(\bar{p}_3+h_2+1-d/2)_{m_3}}\\
&\phantom{=}\qquad\times\frac{(-h_3)_{m_1+m^{\prime}_{12}}(p_2+h_2)_{m_1-m_2+m_3+m^{\prime}_{13}}(p_3)_{m_2-m_1+m_3+m^{\prime}_{23}+m^{\prime}_{33}}}{(p_2)_{2m_1+m^{\prime}_{11}+m^{\prime}_{12}+m^{\prime}_{13}}(p_2+1-d/2)_{m_1}}\\
&\phantom{=}\qquad\times\frac{(-h_4)_{m_2+m^{\prime}_{22}+m^{\prime}_{23}}(p_2+h_3)_{m_1+m^{\prime}_{11}+m^{\prime}_{13}}(p_3-h_2+h_4)_{m_2+m^{\prime}_{12}+m^{\prime}_{33}}}{(p_3-h_2)_{2m_2+m^{\prime}_{12}+m^{\prime}_{22}+m^{\prime}_{23}+m^{\prime}_{33}}(p_3-h_2+1-d/2)_{m_2}}\\
&\phantom{=}\qquad\times\frac{(-m^{\prime}_{12})_{m_{11}}(-m^{\prime}_{11})_{m_{11}}(-h_2-m_3+m_1+m_2+m_{11})_{m^{\prime}_{11}-m_{11}}}{m_{11}!}\\
&\phantom{=}\qquad\times F_6^*\prod_{1\leq a\leq3}\frac{(u_a^{*6})^{m_a}}{m_a!}\prod_{1\leq a\leq b\leq3}\frac{(1-v_{ab}^{*6})^{m^{\prime}_{ab}}}{m^{\prime}_{ab}!}.
}

Finally, we evaluate the summation over $m_{11}$ and the result is 
\eqna{
G_{6P}^*&=\sum\frac{(-h_2)_{m_1+m_2-m_3+m^{\prime}_{12}+m^{\prime}_{22}}(-h_5)_{m_3+m^{\prime}_{13}+m^{\prime}_{23}+m^{\prime}_{33}}(\bar{p}_3+h_2+h_5)_{m_3}}{(\bar{p}_3+h_2)_{2m_3+m^{\prime}_{13}+m^{\prime}_{23}+m^{\prime}_{33}}(\bar{p}_3+h_2+1-d/2)_{m_3}}\\
&\phantom{=}\qquad\frac{(p_3-h_2+h_4)_{m_2+m^{\prime}_{12}+m^{\prime}_{33}}(-h_3)_{m_1+m^{\prime}_{12}}(p_2+h_2)_{m_1-m_2+m_3+m^{\prime}_{13}}(p_3)_{m_2-m_1+m_3+m^{\prime}_{23}+m^{\prime}_{33}}}{(p_2)_{2m_1+m^{\prime}_{11}+m^{\prime}_{12}+m^{\prime}_{13}}(p_2+1-d/2)_{m_1}}\\
&\phantom{=}\qquad\frac{(-h_2)_{m_1+m_2-m_3+m^{\prime}_{11}+m^{\prime}_{12}}(-h_4)_{m_2+m^{\prime}_{22}+m^{\prime}_{23}}(p_2+h_3)_{m_1+m^{\prime}_{11}+m^{\prime}_{13}}}{(-h_2)_{m_1+m_2-m_3+m^{\prime}_{12}}p_3-h_2)_{2m_2+m^{\prime}_{12}+m^{\prime}_{22}+m^{\prime}_{23}+m^{\prime}_{33}}(p_3-h_2+1-d/2)_{m_2}}\\
&\phantom{=}\qquad\times F_6^*\prod_{1\leq a\leq3}\frac{(u_a^{*6})^{m_a}}{m_a!}\prod_{1\leq a\leq b\leq3}\frac{(1-v_{ab}^{*6})^{m^{\prime}_{ab}}}{m^{\prime}_{ab}!},
}
proving \eqref{EqSymPermG} and thus \eqref{EqSymPerm}.

%%%%%%%%%%%%%%%%%%%%%%%%%%%%%%%%%%%%%%%%%%%%%%%%%%
%%%%%%%%%%%%%%%%%%%%%%%%%%%%%%%%%%%%%%%%%%%%%%%%%%

\bibliography{Snowflake}

\end{document}